\documentclass[doublespace,nopageskip]{VTthesis} % nopageskip - Removes arbitrary blank pages.

\usepackage{lipsum}
\usepackage{cleveref}
\usepackage{nomencl}
\usepackage{float}
\usepackage{graphicx}

\title{Generative AI Framework for 3D Object Generation in Augmented Reality}

\keywords{Generative AI, Augmented reality, Natural language processing, Image processing}

\author{Majid Behravan}

\program{Computer Science and Application} 

\degree{Master of Science} 

\submitdate{December 6, 2024} 

\principaladvisor{Denis Gra{\v{c}}anin}
\firstreader{Christopher L North}
\secondreader{Xuan Wang}
\dedication{To the beautiful eyes of someone I love with all my heart.}
\acknowledge{
I am deeply grateful to my advisor, Dr. Denis Gra{\v{c}}anin, for his exceptional support, guidance, and encouragement throughout this journey.
Special thanks to Dr. Mohamed Azab for his collaboration which significantly supported the projects.
Thank you to the lab members for their support and collaboration: Arezoo, Elham, Shiva, Nikitha, Poorvesh, Anthony and Sunday.
Your help and encouragement mean so much to me.
Thank you all!}

\abstract{This thesis presents a framework that integrates state-of-the-art generative AI models for real-time creation of three-dimensional (3D) objects in augmented reality (AR) environments. The primary goal is to convert diverse inputs, such as images and speech, into accurate 3D models, enhancing user interaction and immersion. Key components include advanced object detection algorithms, user-friendly interaction techniques, and robust AI models like Shap-E for 3D generation. Leveraging Vision Language Models (VLMs) and Large Language Models (LLMs), the system captures spatial details from images and processes textual information to generate comprehensive 3D objects, seamlessly integrating virtual objects into real-world environments.
The framework demonstrates applications across industries such as gaming, education, retail, and interior design. It allows players to create personalized in-game assets, customers to see products in their environments before purchase, and designers to convert real-world objects into 3D models for real-time visualization. A significant contribution is democratizing 3D model creation, making advanced AI tools accessible to a broader audience, fostering creativity and innovation. The framework addresses challenges like handling multilingual inputs, diverse visual data, and complex environments, improving object detection and model generation accuracy, as well as loading 3D models in AR space in real-time.
In conclusion, this thesis integrates generative AI and AR for efficient 3D model generation, enhancing accessibility and paving the way for innovative applications and improved user interactions in AR environments.}
\abstractgenaud{This thesis explores how advanced artificial intelligence (AI) can create realistic three-dimensional (3D) objects in real-time within augmented reality (AR) environments. AR is a technology that overlays digital content onto the real world through AR glasses. Our primary goal is to transform different types of inputs, such as pictures and speech, into precise 3D models, enhancing the user's experience and interaction with their surroundings.
The framework includes advanced techniques to detect and process objects from images and text, using powerful AI models. These models, called Vision Language Models (VLMs) and Large Language Models (LLMs), help the system analyze the inputs accurately and provide suggestions for creating objects that fit the environment and the user's needs. The integration of these technologies with text-to-3D and image-to-3D models allows virtual objects to blend seamlessly into the real world, creating an immersive experience.
This technology has practical uses in various fields. In gaming, it allows players to design and interact with custom game items. In retail, it enables customers to see how products would look in their own space before buying them. For interior design, it allows designers to create 3D models of real-world objects for planning and visualization.
A key achievement of this research is making 3D model creation more accessible to everyone, not just experts. This democratization fosters creativity and innovation, allowing more people to benefit from AR technology. The framework also addresses technical challenges, such as understanding multiple languages and complex environments, improving the accuracy and quality of the generated 3D models.
}
 \makenomenclature
 
\begin{document}
% The following lines set up the front matter of your thesis and are required to ensure proper formatting per the VT ETD standards. 
  \frontmatter
  \maketitle
  \tableofcontents

% The list of figures and tables are now optional per the official ETD standards.  Unless you have a very good reason for removing them, you should leave these lists in the document. Comment them out to remove them.
	\listoffigures
	\listoftables
    \printnomenclature %Creates a list of abbreviations. Comment out to remove it. 

% sample text for abbreviations:
% NLP is a field of computer science, artificial intelligence, and linguistics concerned with the interactions between computers and human (natural) languages.

% \printnomenclature[2cm]
 
% $\sigma$ is the eighteenth letter of the Greek alphabet, and carries the 's' sound. In the system of Greek numerals, it has a value of 200. 
 
% \nomenclature{$\sigma$}{The total mass of angels per unit area}

% The following sets up the document for the main part of the thesis. Do not comment out or remove this line.
	\mainmatter

% ==================================Chapter 1
    \chapter{Introduction} \label{ch:introduction}

    \section{Problem Statement} \label{se:Problem_Statement}

	 \nomenclature{NLP}{Natural Language Processing}
  \nomenclature{AI}{Artificial Intelligence}
  \nomenclature{LLMs}{Large Language Models}
  \nomenclature{AR}{Augmented Reality}
  \nomenclature{XR}{Extended Reality}
\nomenclature{NeRF}{Neural Radiance Fields}
  \nomenclature{STT}{Speech-to-Text}
\nomenclature{TTS}{Text-to-Speech}
\nomenclature{VLMs}{Visual Language Models}
\nomenclature{3D}{Three-dimensional}

The rapid advancements in generative artificial intelligence (AI) have enabled remarkable capabilities in various domains, including natural language processing (NLP), image and video generation, and complex 3D model creation~\cite{10649569}.  LLMs such as those described by Brown et al.~\cite{Brown2020} play a pivotal role in facilitating such developments. These models have revolutionized how we interact with technology by offering more immersive experiences, as demonstrated by the integration of generative AI in creating 3D objects for augmented reality (AR) and extended reality (XR) environments.

Despite significant progress, generating consistent and efficient 3D models from natural language or image inputs remains a challenge. Notable models like DreamFusion, which is built on diffusion processes for text-to-3D generation, have been influential but come with significant limitations, such as high latency and computational demands~\cite{poole2023dreamfusion}. Specifically, DreamFusion’s generation time ranges from 30 to 40 minutes, rendering it impractical for real-time AR applications~\cite{Weng2024}. Furthermore, as highlighted by Li et al.~\cite{Li2023GenerativeAI3D}, models leveraging Neural Radiance Fields (NeRF) and related techniques require substantial computational resources, limiting their scalability and effectiveness in dynamic AR settings.

Another issue associated with diffusion-based generative models is inconsistency in output. Users can often experience varied results from identical prompts due to inherent stochastic elements of these models, as reported by Rombach et al.~\cite{rombach2022high}. This inconsistency undermines user experience and trust, especially when uniformity and reliability are critical for professional and commercial applications.

To address these challenges, we introduce the \textit{Matrix} framework, which streamlines the text-to-3D generation process by significantly reducing the generation time to under 50 seconds. This innovation is achieved through the integration of the Shap-E model~\cite{jun2023shape} and targeted optimizations that ensure compatibility with AR devices, facilitating faster loading times and improved real-time user interaction.

The \textit{Matrix} framework offers not only rapid transformation of spoken words into detailed 3D models but also enhances inclusivity by supporting multilingual speech-to-text (STT) and text-to-speech (TTS) capabilities. Current generative AI solutions primarily support English, as noted in the literature~\cite{Touvron2023}, limiting their accessibility in global contexts. By incorporating multilingual translation layers, our system broadens the usability of text-to-3D generation in diverse linguistic settings, overcoming the noted challenges of translation accuracy and performance when extending beyond English~\cite{radford2021learning}.

Efficient GPU utilization is another focal point. Generative AI models for 3D creation typically require significant GPU resources, which can be costly and impractical for real-time applications~\cite{gao2022get3d}. To mitigate this, we integrate a pre-generated objects repository that allows users to select from previously created 3D models, reducing the computational load and enhancing the system's responsiveness. This approach also addresses the challenge of result consistency by providing reliable, reusable options, which tackles the issue highlighted by existing literature on diffusion-based models~\cite{rombach2022high}.

The \textit{Matrix} framework leverages LLMs for recommending related objects based on user input, enriching user interactions by providing contextually relevant enhancements and suggestions. This capability enables a more dynamic and personalized user experience in AR environments.

Visual Language Models (VLMs) play a critical role in detecting and analyzing the user’s environment, facilitating the recommendation of appropriate 3D objects. By processing spatial arrangements, existing elements, and aesthetic attributes, VLMs support the seamless integration of generated objects that align with the context of the AR space.

Moreover, the framework incorporates advanced image-to-3D conversion techniques, enabling users to transform captured images into 3D models efficiently. This feature bridges the gap between physical and digital realms, enhancing real-time interactions and providing users with a robust toolkit for AR content creation.

The optimization of 3D model output size is a further technical challenge. Large, detailed 3D objects can create latency issues, especially on AR devices, as file sizes can exceed 2.5MB, resulting in loading times exceeding 5 minutes~\cite{chan2022efficient}. By applying the quadric edge collapse algorithm to reduce the number of vertices, our framework ensures that objects maintain high visual fidelity at a reduced computational cost, achieving an optimal balance at approximately 1,000 vertices. This adjustment improves the loading and performance of AR applications, particularly on devices like Microsoft HoloLens.

Additionally, the reliability of STT systems is paramount in speech-driven 3D model generation. Real-world AR applications must contend with factors such as background noise and varied accents, which can lead to transcription errors~\cite{nash2022}. Our system includes a visual confirmation interface that allows users to verify detected objects before rendering, minimizing errors and optimizing GPU usage.

In conclusion, the integration of generative AI with AR remains fraught with challenges related to latency, consistency, and multilingual support. Our \textit{Matrix} framework addresses these issues by leveraging optimized AI models, enhancing GPU efficiency, supporting multiple languages, and minimizing transcription errors. This comprehensive approach provides a pathway toward scalable, real-time AR solutions that bridge current technological gaps and expand the practical applications of generative AI.	 
    
    \section{Research Objectives} \label{se:Research_Objectives}
    
    This research aims to address the existing challenges in the field of generative AI for AR environments by developing a comprehensive framework. The specific objectives of this study are:

\begin{enumerate}

\item To develop an optimized generative AI framework that reduces the generation time of 3D objects, enabling real-time interactions in AR.
 
\item To leverage LLMs for recognizing spoken input and recommending contextually relevant objects, enriching user interaction.

\item To establish a pre-generated object repository for efficient GPU usage and consistent outputs.

\item To integrate VLMs to detect areas and recommend objects based on spatial analysis, improving contextual relevance.

\item To implement image-to-3D conversion for seamless real-world object integration into AR environments.

\end{enumerate}

    \section{Research Questions} \label{se:Research_Questions}  
    % \section{Research Questions or Hypotheses} \label{se:Research_Questions}
This study aims to answer the following research questions:

\begin{enumerate}
\item How can LLMs and VLMs be employed to enhance the recommendation of contextually relevant objects in AR while supporting 3D object generation?
 
\item How can image-to-3D and text-to-3D object generation be optimized for near real-time integration into AR while maintaining accuracy and visual fidelity?

 \item What is the usability of using 3D object generation and AR integration while maintaining accuracy and visual fidelity?
 
\end{enumerate}

    \section{Significance of the Study} \label{se:Significance_Study}
The significance of this study lies in its potential to overcome the key limitations of current generative AI systems for AR environments. By reducing generation time, enhancing GPU efficiency, and integrating multilingual capabilities, this research contributes to the advancement of real-time and inclusive AR experiences. Furthermore, the incorporation of LLMs and VLMs for object recognition and context-based recommendations fosters dynamic interactions, enhancing both user engagement and practical applications in industries such as design, education, and retail. The development of an efficient, secure, and customizable framework broadens the accessibility and applicability of AR technologies.

  \section{Contribution} \label{se:Contribution}
    This thesis presents several key contributions to the field of generative AI for AR applications, particularly in the dynamic and real-time generation of 3D models from diverse input formats. These contributions are highlighted below:

\begin{itemize}
    \item \textbf{Framework Development for Real-Time 3D Generation:} A framework was designed and implemented to integrate generative AI, AR, and vision language models (VLMs), achieving real-time 3D model creation from textual, visual, and speech-based inputs.
    
    \item \textbf{Advanced Object Detection and Context-Aware Recommendations:} Leveraging VLMs, the system analyzes the user’s environment, recommends contextually relevant objects, and seamlessly integrates virtual objects with real-world surroundings, enhancing AR experience through personalization and contextual relevance.

    \item \textbf{Enhanced User Interaction and Immersive Experiences:} By employing Shap-E models and seamless interaction controls, the framework democratizes 3D model generation, enabling users across industries—such as education, gaming, and retail—to customize AR content without specialized knowledge.
\end{itemize}

Through these advancements, this thesis provides a novel AI-driven system that bridges the gap between generative AI and AR applications, addressing latency, object consistency, and context-awareness issues. This work extends the capabilities presented in our related paper, \emph{Generative Multi-Modal Artificial Intelligence for Dynamic Real-Time Context-Aware Content Creation in Augmented Reality}~\cite{behravan2024generative}, by demonstrating enhanced accuracy in object recommendations, reduction in latency for 3D model rendering, and improved user interaction in real-time AR environments.

    \section{Methodology Overview} \label{se:Methodology_Overview} 
    The research employs a multifaceted methodology comprising the following key components:

\begin{itemize}
\item \textbf{Framework Development:} Utilizing the Shap-E model~\cite{jun2023shape} as the base for text-to-3D generation, with modifications tailored for AR applications to optimize loading and interaction times.

\item \textbf{LLM and VLM Integration:} Leveraging LLMs to recognize spoken input and recommend contextually relevant objects, and VLMs to analyze images for spatial understanding and object placement.
\item \textbf{Pre-Generated Repository:} Developing a repository of pre-generated 3D objects to ensure consistency, reduce GPU usage, and expedite generation time.
\item \textbf{Multilingual Integration:} Implementing AI-driven translation models for speech-to-text and text-to-speech, ensuring seamless interaction in multiple languages.
\item \textbf{Performance Analysis:} Testing the system's performance on AR devices, including generation time, model output size, and user experience metrics.

\end{itemize}

     \section{Thesis Structure} \label{se:Thesis_Structure}
The structure of this thesis is as follows:

\begin{enumerate}
    \item \textbf{Chapter 1: Introduction} \\
    This chapter outlines the problem statement, research objectives, research questions, significance of the study, methodology overview, and thesis structure.
    
    \item \textbf{Chapter 2: Literature Review} \\
    This chapter reviews the current state of research in generative AI, AR, VLMs, and LLMs.
    
    \item \textbf{Chapter 3: Methodology and Implementation} \\
    This chapter details the algorithms and methodologies developed for 3D model generation and describes the integration of the developed algorithms into the AR framework.
    
    \item \textbf{Chapter 4: Evaluation and Results} \\
    Presents the findings from the system's performance tests and user interaction studies, accompanied by data analysis.
    
    \item \textbf{Chapter 5: Discussion} \\
    % This chapter analyzes the findings and their implications for various industries.
    Interprets the results in the context of research objectives and questions, addressing the implications and limitations.

    \item \textbf{Chapter 6: Conclusion and Future Work} \\
    Summarizes the research contributions, discusses future research directions, and potential enhancements.
\end{enumerate}

    \section{Summary} \label{se:Summary}

    This chapter has provided an introduction to the research topic, highlighting the problem statement, research objectives, and questions guiding this study. The significance of the research was outlined, emphasizing the potential impact on real-time AR experiences and generative AI applications. A brief overview of the methodology was provided, along with a summary of the thesis structure. This foundation sets the stage for an in-depth exploration of related literature and the methodologies employed in this research.
% ==================================Chapter 2
    \chapter{Review of Literature} \label{ch:lit_review}

\nomenclature{ASR}{Automatic Speech Recognition}
\nomenclature{MONA}{Multimodal Orofacial Neural Audio}
\nomenclature{NER}{Named Entity Recognition}

 \nomenclature{CLIP}{Contrastive Language-Image Pretraining}

 \nomenclature{FLAVA}{Foundational Language and Vision Alignment}
 
\section{Advances in AI-Driven Interactive Technologies}

We delve into three critical areas of technological advancement that significantly enhance interactive technologies and digital design:

\begin{itemize}
\item 
Speech-to-Text and Text-to-Speech technologies.
\item
The application of LLMs for object detection and recommendation.
\item
The innovation of Text-to-3D and Image-to-3D conversion.
\item
Vision language model.
 
\end{itemize}

The corresponding subsections provide an in-depth analysis of the current research and developments in these fields, illustrating their transformative impacts across various industries such as communication, healthcare, and digital content creation.
These technologies not only demonstrate the integration of AI in enhancing user experiences but also underscore the synergy between advanced computational models and practical applications in real-world scenarios.

\subsection{speech-to-text and Text-to-Speech}

STT technologies convert spoken languages into written text.
STT has various applications such as voice user interfaces, transcribing meetings, and accessibility features for those unable to type.
Significant advancements in AI, particularly in speech processing and language modeling, have facilitated new capabilities in multilingual and multimodal applications.
Hono et al. explored the integration of pre-trained speech and language models for end-to-end speech recognition, demonstrating how these models leverage large amounts of data and sophisticated algorithms to perform complex tasks such as Automatic Speech Recognition (ASR) without the need for extensive training from scratch~\cite{Hono2022}.

Furthermore, the work by~\cite{SeamlessM4T2023} on SeamlessM4T highlights the integration of multimodal translation systems that support STT and TTS functionalities, encompassing up to 100 languages.
This model is based on a unified approach using a vast corpus of aligned speech translations, which significantly enhances the translation accuracy and robustness against various speech variations. Its adoption in multilingual applications, such as the mobile platform discussed, highlights its capability to address linguistic challenges effectively in disaster scenarios, ensuring accurate and robust cross-linguistic communication~\cite{behravan2024multilingual}.

Additionally, Hu et al. introduced WavLLM, a robust and adaptive speech large language model that combines dual encoders and a two-stage curriculum learning approach.
This model exemplifies the next step in evolving LLMs to handle complex auditory tasks and improve generalization across varied contexts~\cite{WavLLM2024}.

Benster et al. presented an innovative approach to silent speech recognition with their Multimodal Orofacial Neural Audio (MONA) system, which significantly improves recognition accuracy using cross-modal alignment techniques.
This breakthrough reduces the word error rate substantially, showcasing the potential of non-invasive interfaces for soundless verbal communication~\cite{Benster2024}.

In the context of mental health, Zhang et al. developed a method to integrate acoustic speech information into LLMs for depression detection.
Their approach, which utilizes acoustic landmarks, represents a significant advancement in how speech patterns can be analyzed to detect and analyze depressive states, demonstrating the broad applicability of LLMs beyond traditional text-based tasks~\cite{Zhang2024}.

Lastly, Hu et al. in another significant study, introduced GenTranslate, which adopts a generative paradigm for multilingual speech and machine translation.
This new model enhances the quality of translations by integrating the diverse information present in multiple translation hypotheses, which is a step forward in optimizing language model outputs for complex multilingual environments~\cite{GenTranslate2024}.

STT and TTS technologies are crucial for improving accessibility in AR environments. They allow users to interact with AR systems through voice, which is particularly beneficial for individuals with ADHD who may struggle with traditional input methods like typing or touch interfaces~\cite{ADHDAR2024}.

These studies collectively demonstrate the diverse applications of AI in enhancing speech recognition, translation, and healthcare diagnostics, aligning closely with the objectives of frameworks focusing on real-time environment customization through speech-driven interfaces.

\subsection{LLMs for Object Detection and Recommendation}

The integration of LLMs into applications requiring precise object detection and recommendation systems has been a focal area of recent research, as demonstrated in several groundbreaking studies.
The advancements in LLMs have significantly contributed to enhancing NER capabilities and recommendation systems, essential components of our AR framework.

Yan et al.~\cite{yan2024ltner} developed the LTNER framework, which leverages the GPT-3.5 model to improve the accuracy of LLMs in NER tasks significantly.
Their approach uses a novel Contextualized Entity Marking Gen Method, highlighting LLMs' potential to comprehend and process context more effectively than traditional methods.

Similarly, Ashok and Lipton~\cite{ashok2023promptner} introduced PromptNER, a state-of-the-art algorithm for few-shot Named Entity Recognition (NER).
This method employs prompt-based heuristics to enable LLMs to identify and explain potential entities based on pre-defined types, demonstrating the adaptability of LLMs in handling varied NER tasks without extensive training.

In the realm of recommendations, Kim et al.~\cite{kim2024allmrec} proposed the A-LLMRec system, an efficient all-round LLM-based recommender system that integrates collaborative filtering knowledge directly into LLMs.
Their system showcases the flexibility of LLMs in utilizing existing datasets and knowledge bases to provide accurate recommendations, even in scenarios with limited user-item interactions.

LLMs are used in various ways within virtual environments, including personalized e-learning systems where they adapt content to learners' emotional and cognitive states, enhancing engagement and fostering deeper interaction and comprehension~\cite{LLMELearning2024}.

Further extending the capabilities of LLMs in recommendations, Zhang et al.~\cite{zhang2024language} discussed the adaptations of LLM training paradigms for recommender systems.
Their tutorial emphasized the potential of LLMs to transfer knowledge across domains, improving the effectiveness and trustworthiness of recommendations.

Moreover, the work of Cao et al.~\cite{cao2024knowledge} on integrating recommendation-specific knowledge into LLMs addresses the knowledge gap between traditional recommender systems and LLM capabilities.
Their method enhances LLMs' ability to generate more relevant and personalized recommendations by incorporating data samples that reflect user preferences and item correlations.

\subsection{Text-to-3D}

Text-to-3D conversion involves generating three-dimensional models from textual descriptions.
This technology is particularly interesting in fields like XR, where text-based inputs can be used to create immersive environments or objects dynamically.

The translation of textual descriptions into 3D models has been enriched by a variety of advanced methodologies, exploring autoregressive models, diffusion processes, and deep learning techniques.
Among these,~\cite{mittal2022autosdf}'s ``AutoSDF'' and \cite{fu2022shapecrafter}'s ``ShapeCrafter'' have demonstrated how recursive algorithms and shape priors can innovatively generate shapes from textual descriptions~\cite{mittal2022autosdf,fu2022shapecrafter}.
 
Significant developments also include voxelized diffusion models as demonstrated by~\cite{cheng2023sdfusion} in ``SDFusion'' and \cite{li2023diffusionsdf} in ``Diffusion-SDF,'' which significantly enhance 3D shape reconstruction and generation capabilities from complex textual inputs~\cite{cheng2023sdfusion,li2023diffusionsdf}.
Moreover, \cite{zhao2023michelangelo} introduce a unique alignment of shape, image, and text representations in ``Michelangelo,'' facilitating highly conditional 3D shape generation \cite{zhao2023michelangelo}.

A central piece of our framework is the use of the Shap-E model by~\cite{jun2023shape}, which generates detailed 3D objects by interpreting complex textual data, showcasing the potent capabilities of LLMs in 3D modeling \cite{jun2023shape}.
This model's efficiency in creating detailed and contextually appropriate 3D representations significantly contributes to the advancement of user interaction in XR spaces.

Additionally, the integration of structure-aware generative models is notably advanced through ``ShapeScaffolder'' by~\cite{tian2023shapescaffolder} and ``SALAD'' by \cite{koo2023salad}, which emphasize generating realistic and structurally sound 3D models from textual descriptions, enhancing both the aesthetic and functional aspects of generated objects~\cite{tian2023shapescaffolder,koo2023salad}.
These developments underline the increasing sophistication in generating 3D content that is not only visually accurate but also contextually appropriate to user input.

This comprehensive synthesis of technologies underscores our framework's reliance on cutting-edge advancements in AI, specifically leveraging the intricate processing capabilities of LLMs and innovative generative techniques to create immersive XR experiences.
Our approach aims to seamlessly integrate these technologies to deliver a dynamic, responsive, and highly customizable user experience in AR environments.

\subsection{VLMs}

VLMs, such as CLIP (Contrastive Language-Image Pretraining) and FLAVA (Foundational Language and Vision Alignment), have been pivotal in bridging the gap between visual and textual data.
CLIP, introduced in~\cite{radford2021learning}, demonstrated remarkable zero-shot classification capabilities by training on 400 million image-caption pairs, using a contrastive loss to align image and text representations in a shared space.
FLAVA, developed by~\cite{singh2022flava}, employs masking strategies to enhance multimodal learning, achieving state-of-the-art performance across various benchmarks.

% Contrastive training leverages pairs of positive and negative examples, training the VLM to predict similar representations for positive pairs and different representations for negative pairs.
% Masking involves reconstructing masked image patches given some unmasked text, or vice versa, training the VLM to reconstruct masked words or image patches.

\nomenclature{CoCa}{Contrastive Captioner}

Generative models, such as CoCa (Contrastive Captioner) and CM3leon, have further advanced the field by enabling the generation of both text and images.
CoCa, proposed in~\cite{yu2022coca}, integrates a generative loss with contrastive learning, allowing the model to perform tasks like visual question answering without additional multimodal fusion modules.
CM3leon, introduced by the Chameleon Team~\cite{team2024cm3leon}, leverages a cross-attention mechanism to handle interleaved text and image data, enhancing its capabilities in text-to-image and image-to-text generation tasks.

\nomenclature{BLIP}{Bootstrapping Language-Image Pre-training}

Generative VLMs can generate images or captions and are typically the most expensive to train.
Improving training data quality through synthetic data generation and augmentation techniques has shown significant benefits.
Bootstrapping Language-Image Pre-training (BLIP)~\cite{li2022blip} generates synthetic captions to enhance data consistency and completeness, leading to better model performance.
Nguyen et a.~\cite{nguyen2023improving} demonstrated that mixing real and synthetic captions during pretraining improves alignment and overall model effectiveness.

\nomenclature{LLaVA}{Large Language-and-Vision Assistant}

Leveraging pretrained language models, such as LLaVA (Large Language-and-Vision Assistant) and MiniGPT, has become a popular approach to reduce computational costs while maintaining high performance~\cite{liu2023visual}.
Pre-trained backbones use models like Llama to learn a mapping between an image encoder and the LLM.
Frozen, introduced in~\cite{tsimpoukelli2021frozen}, connects vision encoders to pretrained language models using a lightweight mapping network, enabling rapid adaptation to new tasks.
MiniGPT-4, described in~\cite{zhu2023minigpt}, employs a linear projection layer to align image representations with a language model, significantly reducing training time and resources.

In addition to these models, recent advancements have explored the integration of video data into VLMs.
This extension is crucial for applications requiring temporal context, such as video summarization and real-time interaction in AR environments.
Models like Flamingo, which process interleaved text and video frames, have demonstrated enhanced performance in tasks involving dynamic visual content.
These developments highlight the importance of multimodal data that includes both static images and video sequences for comprehensive vision language understanding.

Another significant area of progress is the improvement of grounding techniques in VLMs.
Grounding involves associating textual descriptions with specific visual elements, which is essential for tasks like object detection and scene understanding in AR.
Methods that incorporate bounding box annotations and negative captioning have been particularly effective.
These techniques ensure that models not only generate accurate descriptions but also correctly identify and localize objects within a scene, thereby improving the practical utility of AR applications.

The use of large-scale pretraining datasets has also been a focal point in recent research.
Datasets such as LAION-400M, which consist of hundreds of millions of image-text pairs, have been instrumental in training robust VLMs. However, the quality and diversity of these datasets are critical.
Efforts to curate and balance these datasets, such as by removing duplicates and ensuring a wide range of concepts, have been shown to enhance the performance of models in zero-shot learning scenarios.

Finally, there has been a growing emphasis on the ethical considerations and societal impacts of VLMs.
Issues such as bias in training data, the potential for generating harmful content, and the need for responsible AI development are increasingly being addressed.
Researchers are developing frameworks for assessing and mitigating biases in VLMs and are advocating for transparent and inclusive practices in the creation and deployment of these technologies.
As generative multi-modal AI continues to evolve, ensuring its ethical use will be paramount to gaining public trust and maximizing its positive impact on society.
Incorporating context-aware content in AR improves user experience and task performance by providing relevant, situational information that enhances efficiency and reduces errors~\cite{Dasgupta-2020-a}.
Moreover, advancements in integrating physiological signals with LLMs demonstrate potential for creating more empathetic and context-aware AI systems~\cite{dongre2024}.

By building on these advancements, future research in generative multi-modal AI for AR can continue to push the boundaries of what is possible, creating more interactive, intuitive, and context-aware digital environments.

\subsection{Image-to-3D Generation}

The field of image-to-3D generation using generative AI has seen substantial progress, with numerous methodologies contributing to its rapid development.
We describe the key works and models that have advanced this domain, particularly in the context of single-view 3D reconstruction.

\nomenclature{GANs}{Generative Adversarial Networks}

Early advancements in 3D generative modeling largely focused on the development and application of Generative Adversarial Networks (GANs)~\cite{goodfellow2014generative}.
This approach was pivotal in shaping the early landscape of 3D generation, enabling the creation of models from representations like point clouds~\cite{achlioptas2018learning} and implicit representations~\cite{chen2019learning}.
Building on this foundation, recent studies have integrated GANs with differentiable rendering methods, which utilize multiple rendered views as the basis for the loss signal, significantly enhancing the quality of generated 3D models~\cite{chan2022efficient}.

\nomenclature{VQ-VAE}{Vector-Quantized Variational Autoencoder}

The advent of diffusion models marked a significant leap forward in high-quality image generation, with growing interest in extending these models to the 3D domain.
Typically, this involves training a Vector-Quantized Variational Autoencoder (VQ-VAE) on 3D representations such as triplane~\cite{shue2023triplane} and point cloud~\cite{zeng2022latent}, followed by applying the diffusion model in the latent space.
Although direct training on 3D representations is less common, it is gaining traction, particularly with point clouds~\cite{luo2021diffusion}, voxels \cite{zheng2023locally}, and neural wavelet coefficients~\cite{hui2022neural}.

Among the innovative methods, ShapeClipper stands out by reconstructing 3D object shapes from single-view RGB images using CLIP-based shape consistency and geometric constraints.
This approach leverages the semantic consistency of CLIP embeddings to enhance 3D shape learning, showing superior performance on datasets like Pix3D, Pascal3D+, and OpenImages~\cite{huang2023shapeclipper}.

Similarly, Shap-E employs autoencoders trained on explicit 3D representations combined with a diffusion model in the latent space, offering a flexible approach to representing complex 3D structures. 
This method addresses the limitations of fixed-resolution explicit outputs, providing a scalable solution for 3D model generation~\cite{achlioptas2018learning}.

\nomenclature{MCC}{Multiview Compressive Coding}
\nomenclature{CNNs}{convolutional neural networks}

Multiview Compressive Coding (MCC) advances the field by learning a general-purpose 3D representation from RGB-D views of diverse real-world objects.
MCC utilizes an attention mechanism on encoded appearance and geometric cues, outperforming traditional global-feature strategies.
Its robustness in handling diverse viewpoints and occlusions is particularly advantageous for real-world applications~\cite{fu2021auto}.

Voxel-based methods have also seen significant advancements.
These approaches benefit from the regularity of voxel representation and the application of 2D convolutional neural networks (CNNs).
However, they often face a trade-off between resolution and computational efficiency.
For example, voxel-grid methods offer better accuracy than bounding boxes but at the cost of increased computational resources~\cite{zamir2018taskonomy}.
SparseFusion addresses this by integrating sparse voxel grids with neural networks, achieving high-fidelity 3D reconstructions from sparse input views, suitable for real-time AR and VR applications~\cite{fu2021auto}.

Mesh-based techniques, exemplified by Mesh R-CNN, reconstruct object meshes by deforming a template such as a unit sphere.
Despite their efficiency, these methods are constrained by the fixed topology of the initial mesh.
Innovations like Total3DUnderstanding improve on this by integrating room layout, object bounding boxes, and meshes into a unified framework, achieving state-of-the-art results in indoor scene reconstruction on datasets such as SUN RGB-D and Pix3~\cite{nie2020total3dunderstanding}.

Implicit surface methods, including those utilizing NeRF, have gained prominence for their ability to implicitly represent complex geometries.
Techniques like PixelNeRF and CodeNeRF employ lightweight networks to model arbitrary topologies, proving effective for scene generalization and novel-view synthesis~\cite{yu2021pixelnerf}.

% InstPIFu enhances pixel-aligned feature methods in cluttered indoor scenes by introducing an instance-aligned attention module. This innovation effectively decouples mixed local features, leading to more accurate and detailed 3D object reconstructions from single images \cite{zhang2021holistic}.

\nomenclature{InstPIFu}{instance-aligned implicit function}

In the realm of holistic scene understanding, approaches such as Towards High-Fidelity Single-view Holistic Reconstruction use an instance-aligned implicit function (InstPIFu) for detailed object reconstruction and an implicit representation for complex room geometries.
These methods outperform existing techniques on datasets like SUN RGB-D, Pix3D, 3D-FUTURE, and 3D-FRONT~\cite{liu2023towards}.

Bai and Li provide a comprehensive overview of advancements in 3D generative AI, highlighting significant progress in 3D object creation, character and motion generation, and the development of neural network-based 3D rendering models such as NeRF and 3D Gaussian Splatting~\cite{bai2024progress}.

The Make-A-Shape model by Autodesk represents a substantial advancement in the field, training a large-scale 3D generative model using a vast dataset of 10 million publicly available shapes.
This model introduces novel techniques such as the wavelet-tree representation for compact encoding and the subband adaptive training strategy, enabling effective learning of both coarse and detailed features.
Make-A-Shape demonstrates superior performance in generating high-quality 3D shapes across diverse input modalities~\cite{hui2024make}.

Lastly, the GET3D model by NVIDIA leverages recent advances in differentiable surface modeling and rendering, along with 2D Generative Adversarial Networks, to train a model that generates explicit textured 3D meshes.
This model excels in producing high-fidelity 3D shapes with detailed geometry and complex topology from 2D image collections, making it highly applicable across various real-world applications, from gaming and entertainment to e-commerce and design~\cite{gao2022get3d}.

\section{Gaps in the Literature}
\label{se:Gaps in the Literature}
Despite advancements in developing robust and user-friendly AR applications, significant gaps remain, particularly in speech-to-3D translation and image-to-3D conversion. While previous works have LLMs and AI-driven methods, there are still notable areas that require further exploration:

\begin{itemize}
\item \textbf{Output Size and Latency}: Generative models often produce 3D outputs with large file sizes, leading to latency in AR devices. This affects the real-time nature of the user experience, making interactions slower and less fluid.

\item \textbf{Image-to-3D Conversion Limitations}: Current models struggle with complex backgrounds and lighting variations in images captured through AR headsets, leading to inaccurate 3D reconstructions. This impacts the quality and realism of AR applications, hindering user immersion and satisfaction.

\item \textbf{Contextual Awareness}: LLMs lack real-time environmental context, which limits their ability to create AR objects that match the surrounding space in terms of color, style, and other visual elements. This gap reduces the relevance and visual coherence of AR content.

\item \textbf{Multilingual Support in Open-Source LLMs}: Most open-source LLMs are trained predominantly on English datasets, limiting their effective application in multilingual contexts. This gap restricts the usability and inclusivity of AR applications on a global scale, impacting their effectiveness in diverse linguistic settings and limiting global user engagement.

\nomenclature{UI}{User Interface}

\item \textbf{Speech-to-Text  Accuracy}: Current STT systems face issues such as transcription inaccuracies due to background noise, diverse accents, and speech nuances. These errors can impede user interaction, leading to misinterpretations of user commands and decreased efficiency in AR environments. If not properly addressed in the user interface (UI), these inaccuracies can increase the GPU load of the system, resulting in higher operational costs.

\item \textbf{Efficient GPU Utilization}: High computational demands of GPU usage remain a challenge, especially when generating new 3D models in real-time. This impacts the performance and cost-effectiveness of AR systems, affecting their scalability and responsiveness.

\item \textbf{Integration of 3D Models into AR Environments}: Ensuring accurate scaling and alignment of generated 3D models for seamless integration into real-world views remains an area that needs refinement. This affects the realism and usability of AR applications, diminishing the overall user experience.
\end{itemize}

   \chapter{Automatic 3D object creation in AR} \label{ch:Problem Definition}

\section{Introduction}
AR enhances real-world experiences by overlaying digital objects in physical environments. A significant advancement in this domain is the automatic creation of 3D objects from various inputs such as text, images, or speech, offering users more immersive, interactive experiences. However, real-time and contextually accurate 3D object generation remains challenging due to processing demands, data requirements, and the complexity of real-world spatial adaptation.

\section{Technologies for 3D Object Generation}
\begin{enumerate}
    \item \textbf{Generative AI Models}: Models such as Shap-E leverage AI to convert textual and visual inputs into 3D models. These models use deep learning frameworks to understand input prompts and generate object shapes, textures, and orientations.
    \item \textbf{Large Language Models (LLMs)}: LLMs play a crucial role in interpreting and processing natural language inputs, allowing users to specify desired object characteristics through voice commands or text prompts.
    \item \textbf{Vision Language Models (VLMs)}: VLMs, such as CLIP and FLAVA, bridge visual and textual data, enabling the AR system to interpret surroundings accurately and recommend context-appropriate objects.
    \item \textbf{Image-to-3D Conversion Techniques}: These approaches convert 2D images to 3D objects using techniques like voxel-based models, mesh reconstruction, and neural radiance fields (NeRF). They play a central role in adapting real-world images into AR.
\end{enumerate}

\section{Key Challenges in Automatic 3D Object Creation}
\begin{enumerate}
    \item \textbf{Latency and Computation Load}: Generating 3D models from high-complexity data (e.g., high-resolution images or detailed textual descriptions) requires significant GPU resources, affecting real-time performance.
    \item \textbf{Data Consistency and Accuracy}: Variability in outputs due to stochastic elements in generative models impacts user experience, particularly when identical prompts yield different results.
    \item \textbf{Contextual Relevance}: Models must account for the context and spatial layout of the AR environment to ensure objects fit naturally. This includes understanding user environments, spatial relations, and aesthetic coherence.
    \item \textbf{Multilingual and Multimodal Challenges}: Ensuring accuracy across different languages and input types is critical for global usability, requiring robust, multilingual processing layers for speech and text inputs.
    \item \textbf{Resource Management in Real-Time Applications}: Large 3D models can create high memory and GPU usage, impacting device performance, especially on mobile AR hardware.
\end{enumerate}

\section{Approaches to Address Challenges}
\begin{enumerate}
    \item \textbf{Efficient Frameworks}: Implementing frameworks that optimize model loading and caching reduces latency and improves real-time interactions, as seen in methods like pre-generating object libraries.
    \item \textbf{Incorporation of Pre-Generated Object Repositories}: By utilizing libraries of pre-existing 3D models, the system reduces generation time and computational load, enhancing both speed and reliability.
    \item \textbf{Multimodal Integration with VLMs and LLMs}: Leveraging VLMs for spatial awareness and LLMs for context-aware recommendations improves user interaction by suggesting relevant objects and layouts.
    \item \textbf{Object Size Optimization Techniques}: Using algorithms such as the quadric edge collapse for vertex reduction in complex models balances visual fidelity with manageable file sizes, optimizing AR device performance.
\end{enumerate}

\section{Summary}
Automatic 3D object creation in AR combines multiple AI techniques, addressing a crucial gap in enhancing user experience through realistic digital overlays. Addressing latency, accuracy, and contextual challenges remains key to realizing responsive and immersive AR systems, making automatic 3D generation accessible to wider audiences and practical for applications in design, retail, and education. The following chapter will detail the methodology used to implement these solutions, outlining technical processes and integration into the AR environment.

% ==================================Chapter 4
    \chapter{Methodology and Implementation} \label{ch:Methodology and Implementation}
\begin{figure}
    \centering
    \includegraphics[width=0.7\linewidth]{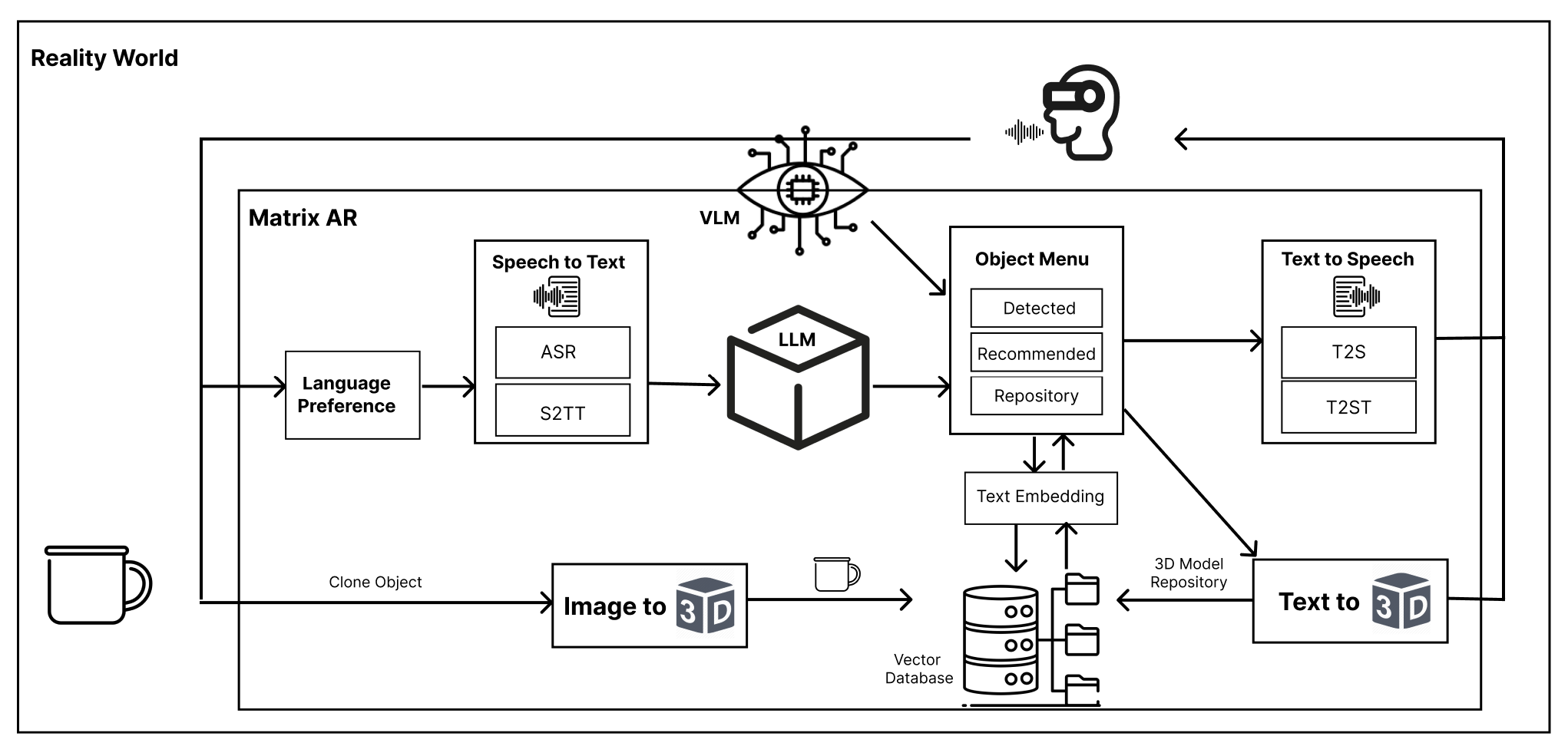}
    \caption{Conceptual Model of the Framework for AR Interaction – This model illustrates the integration of the three main subsystems within the Matrix AR environment: Speech-to-3D, Context-Aware Object Recommendation, and Image-to-3D. The workflow demonstrates how user inputs (speech or images) are processed through ASR , LLM , and VLM to generate, recommend, and render 3D objects in real-time. Text embedding supports efficient retrieval from the vector database and 3D model repository, ensuring seamless user interaction and object customization in the AR space.}
    \label{fig:Framework overview}
\end{figure}

This framework comprises three main subsystems designed to enhance interactive AR experiences: 1) Speech-to-3D, 2) Context-Aware Object Recommendation, and 3) Image-to-3D Generation. Each subsystem addresses specific challenges associated with real-time, user-centric content generation and integration within AR environments. Figure \ref{fig:Framework overview} illustrates the conceptual model of this framework, providing an overview of the interconnected subsystems. The following sections delve into the detailed implementation and functionalities of each subsystem.

\section{Speech-to-3D Subsystem}

\textit{Matrix} framework functionality is meticulously designed to ensure an intuitive, inclusive, and interactive user experience.
By harnessing state-of-the-art technologies and advanced AI models, we facilitate a seamless transition from verbal commands to visual outputs in real-time, thereby enhancing the capability of users to interact dynamically with the AR environment.

We outline the Speech-to-3D Subsystem, which is used to interpret user input, generate 3D models, and allow for user-driven customization of the AR environment.

\begin{figure} [ht]  
\centering  % Centers the image
\includegraphics[width=0.7\linewidth]{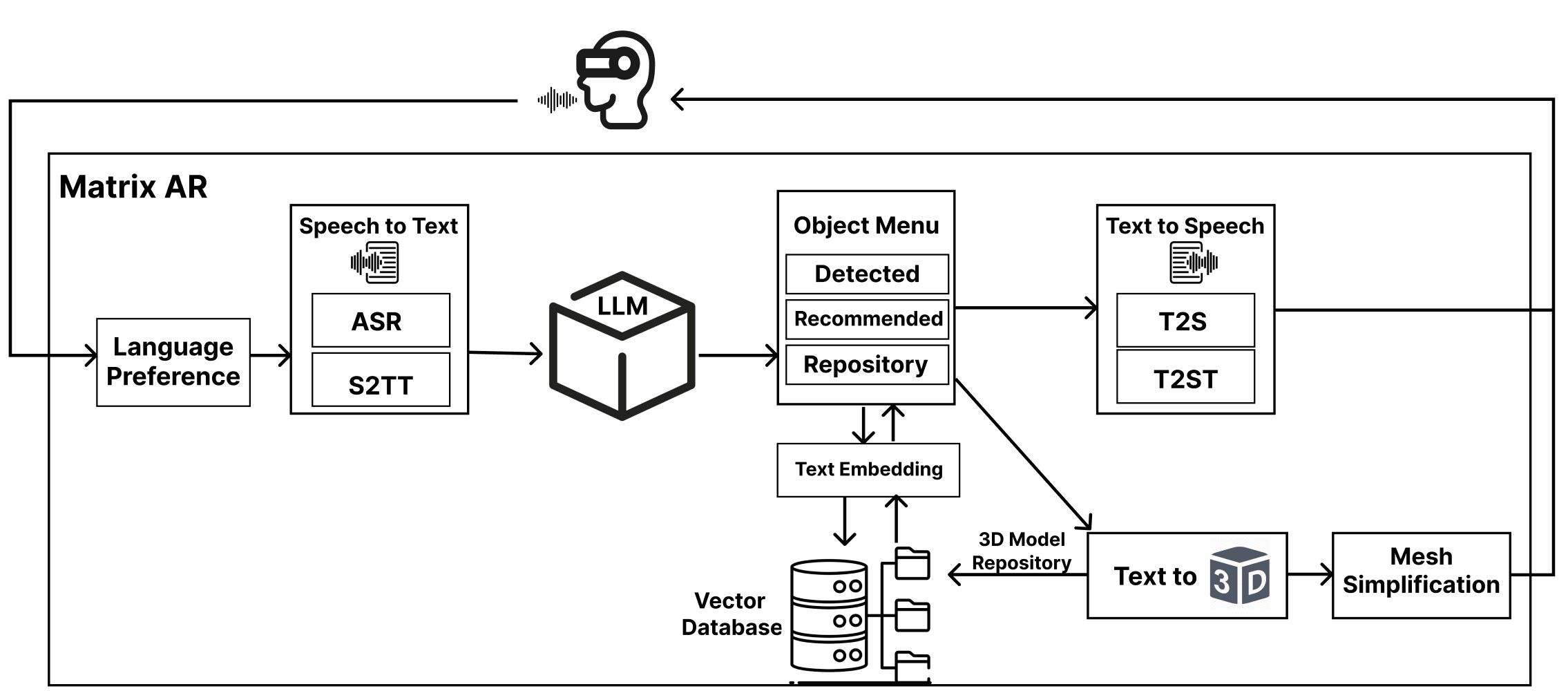}  % Adjusts the image's width to 80% of the text width
\caption{The overview of the developed speech-to-3D subsystem.}  % Optional caption
\label{fig:Research Framework}  % Optional label for referencing
\end{figure}

We use an AR application to illustrate the implemented \textit{Matrix} framework service and functionality (\cref{fig:Research Framework}), from initial language selection to the final customization of the interactive environment.
The goal is to maximize user engagement and  responsiveness, ensuring that the technology remains accessible and efficient across various languages and functional demands.

\nomenclature{S2TT}{Speech-to-Text Translation}

\nomenclature{T2ST}{Text-to-Speech Translation}

\nomenclature{T2S}{Text-to-Speech}

\textit{Matrix} Speech-to-3D Subsystem \cref{fig:Research Framework}, includes several key components and processes that are labeled with abbreviations to denote specific functions.
S2TT is a Speech-to-Text Translation, indicating the conversion process from spoken language to textual format.
T2S and T2ST denote Text-to-Speech and Text-to- Speech Translation, respectively.
Each component plays a crucial role in ensuring that the functions work seamlessly across various stages of user interaction.

\textbf{Language Selection}

\begin{figure} [ht]  
\centering  % Centers the image
\includegraphics[width=0.5\linewidth]{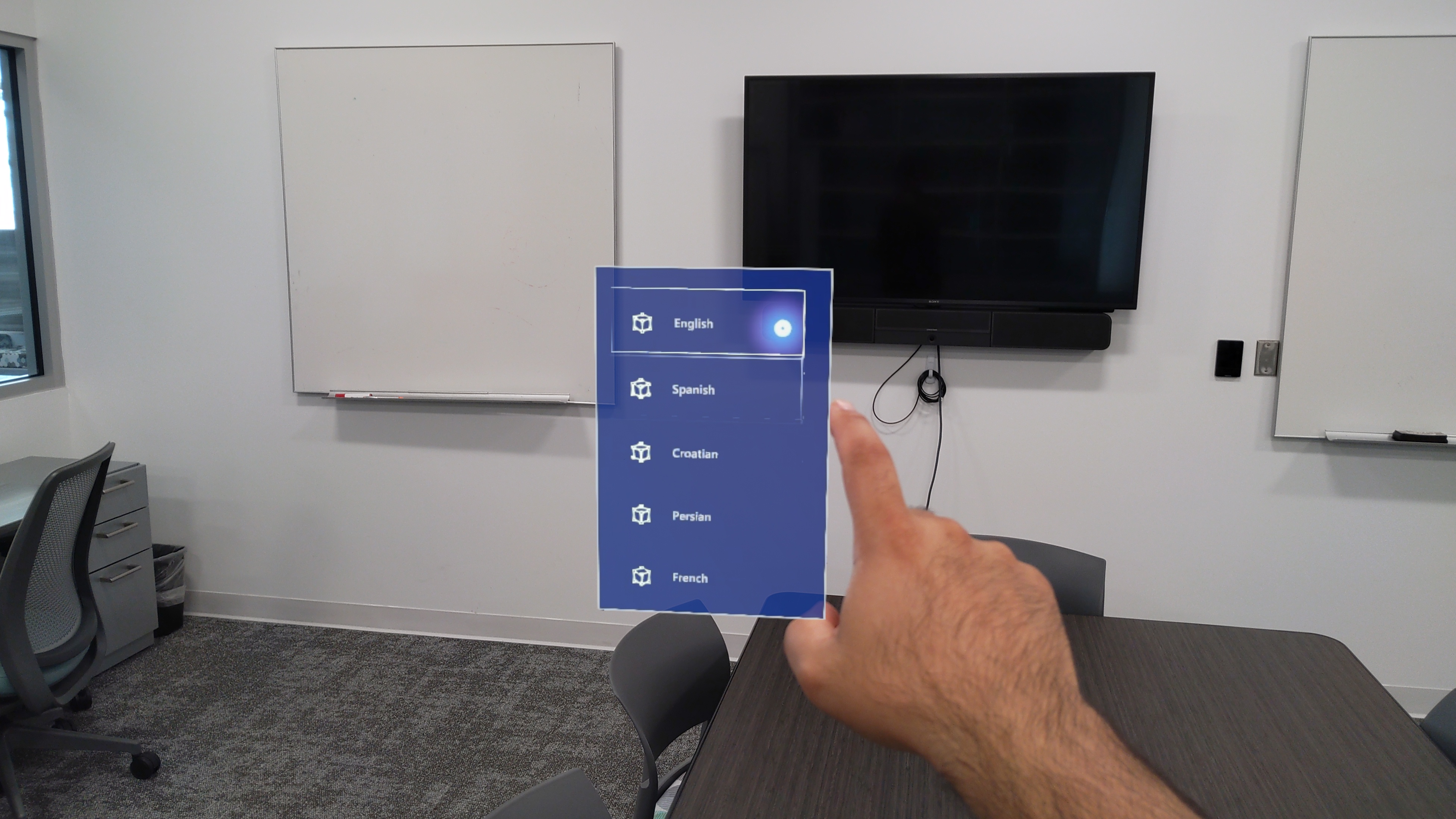}  % Adjusts the image's width to 80% of the text width
\caption{The illustrative AR application example: the language selection menu.}  % Optional caption
\label{fig:Language Selection Menu}  % Optional label for referencing
\end{figure}

Upon entering the AR environment, users are greeted with a welcoming voice message, initiating a user-friendly introduction to the system.
Following this, they are prompted to select their preferred language from a menu.
This critical step ensures that all subsequent interactions between the user and the system are conducted in the chosen language, thereby tailoring the experience to each individual’s linguistic preference.
The system covers around 100 languages, supported by Meta's SeamlessM4T technology, ensuring inclusivity and accessibility for a diverse user base.
A short list of these languages is presented in the \cref{fig:Language Selection Menu} for reference.

\textbf{Command Initiation}

The AR application utilizes the capabilities of a Microsoft HoloLens 2 device for command voice, setting a high standard for responsiveness and interaction.
To start recording, the application requires the utterance of the wake word, ``MATRIX.''
Upon this command, the application begins recording, and it continues until the user says ``STOP.''
This method allows the user to control when their input is being processed, enhancing privacy and user control over the interaction.

\textbf{Translation and Transcription}

Using \textit{Matrix}  framework, an application can listen and detect user speech and support interactive communication with the user, responding to their input language and selected objects.
For non-English speech, we have integrated the open-source SeamlessM4T model to translate the speech into English.
This is crucial for ensuring that applications can understand and process inputs from a wide range of languages efficiently. %(\cref{fig:Speech to Text Flow}).

For English inputs, the audio is transcribed directly using the ASR capabilities of SeamlessM4T, which are specifically fine-tuned for high accuracy in English speech recognition.
For English and several other common languages, the medium-sized model performs well.
However, for other languages, we used the large SeamlessM4T model, although it increased processing time, the improved outputs justified the extra time spent.

Once the text is processed and necessary actions are identified, if the user's selected language is not English, the output text from the LLM is translated back into the user’s language using SeamlessM4T.
This ensures that the AR system's responses are accessible and understandable, regardless of the user's language. % (\cref{fig:Text to Speech Flow}). 
For generating spoken responses in English, we utilize the open-source TTS technology from Coqui.ai.
This technology converts the English text outputs from the LLM into natural-sounding spoken words, enhancing the interaction quality and making the communication feel more engaging and natural.

\textbf{Object Extraction}

For extracting meaningful content from the transcribed or translated text, we employ the Llama2 7B LLM, which identifies objects and their attributes within the user’s commands.
To this end, we use an optimized version of Llama, llama.cpp, designed to run with minimal hardware.
This enables \textit{Matrix} framework to be used on a wide variety of hardware locally, ensuring broader accessibility and flexibility.
Additionally, for the extraction of objects and their attributes, we utilized the langchain library, which provides the capability of prompt template for prompt engineering and optimization.

\textbf{Object Recommendation}

Once the objects are identified and modeled from the transcribed or translated texts, the next step is to enhance user interaction by suggesting related objects.
This is achieved through a sophisticated use of the Llama2 LLM, which processes the extracted object data to generate recommendations.
The LLM considers the context, properties, and functions of the identified objects to suggest objects that could logically coexist or complement the primary objects in a given environment.
This process not only enriches the user experience by providing creative and contextually appropriate options but also aids in developing a more interactive and immersive AR environment.

\textbf{Search Repository}

Simultaneously with the Object Recommendation phase, the application conducts a search within a pre-established 3D object repository to find matches or similar items that are related to the user's requested object.
This repository is an essential component of our framework, as it contains a vast array of previously generated items based on past user requests as well as other pre-generated 3D models that can be readily utilized in various AR scenarios.

\begin{figure} [ht]  
\centering  % Centers the image
\includegraphics[width=0.5\linewidth]{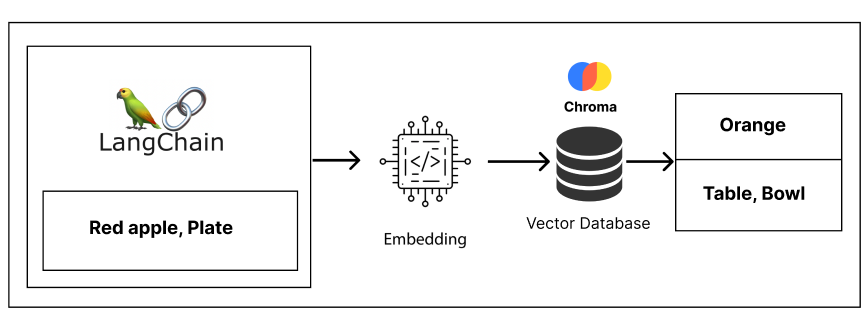}  % Adjusts the image's width to 80% of the text width
\caption{Objects repository semantic search.}  % Optional caption
\label{fig:Search Repository}  % Optional label for referencing
\end{figure}

To enhance object search based on context, we have utilized the vector database to identify similar objects, maximizing the display of related items to the user and reducing the creation of duplicate objects, thereby minimizing GPU usage.
This method ensures that when a user searches for an object, the system can efficiently scour the database for vectors that closely match the user's request in terms of context and similarity.
By leveraging these pre-embedded vectors, the system can quickly suggest objects that are not only relevant but also previously interacted with, thereby avoiding unnecessary replication and excessive computational demand.
The process is visually represented in \cref{fig:Search Repository}.

This dual approach, where the system not only generates recommendations via the LLM but also searches existing resources, ensures a comprehensive range of options for the user.
The repository is constantly updated with new models and user inputs, which enhances its utility and relevance over time.
By integrating this search process with the recommendation phase, the framework efficiently narrows down the best possible matches, significantly speeding up the object selection process for the user and enhancing the overall interaction quality within the AR environment.

This seamless integration of generating new recommendations and searching an existing repository allows for a dynamic and responsive system that can adapt to user preferences and evolving interaction contexts, providing a robust solution for real-time object integration in AR.

\textbf{Object Generation and Placement}

\begin{figure} [ht]  
\centering  % Centers the image
\includegraphics[width=0.5\linewidth]{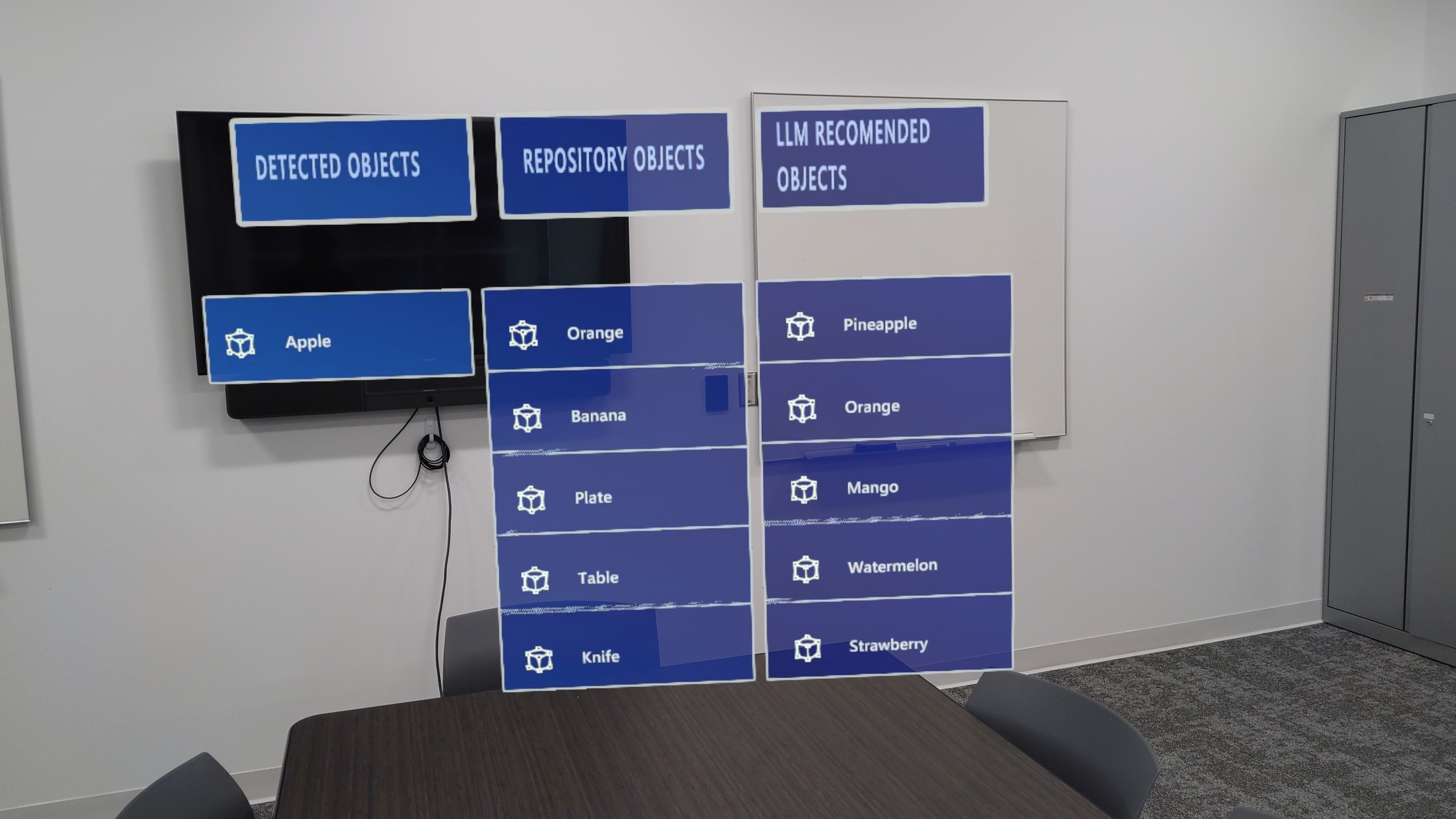}  % Adjusts the image's width to 80% of the text width
\caption{Object selection menu example in the \textit{Matrix} AR application.
The menu displays three categories: Detected Objects, Repository Objects, and LLM Recommended Objects.
The objects shown are: \textit{Detected Objects}: Apple; \textit{Repository Objects}: Orange, Banana, Plate, Table, Knife; \textit{LLM Recommended Objects}: Pineapple, Orange, Mango, Watermelon, Strawberry.}

    \label{fig:Object Selection Menu}  % Optional label for referencing
\end{figure}

The extracted objects, along with suggested related items from both the LLM and the repository, are displayed to the user through an interactive menu, as shown in \cref{fig:Object Selection Menu}.
Users receive auditory prompts informing them that they can select their desired objects from these categories to add to the AR space.
When an object is selected from the repository, it is loaded directly into the AR environment without the need for additional GPU resources, significantly reducing computational load and enhancing system responsiveness.

For selections from other menus, a request is sent to convert the chosen text into a 3D model using Shap-E.
Key parameters used for Shap-E's model generation include 64 Karras-inspired sampling steps for balanced efficiency, a sigma range of 1e-3 to 160 to control noise levels during diffusion, and  s\_churn=0 for stable denoising
Once created, the model is displayed to the user, and a voice message—in the user's selected language—confirms the successful creation of the requested object.

When a user selects an option from the menus, the chosen object is embedded using the all-MiniLM-L6-v2 model and stored in the Chroma vector database, an AI-native open-source embedding repository.
This ensures that in future interactions, if the user requests an object that is related to previously selected items, the system can efficiently suggest these related objects by querying the database.
This framework leverages the Chroma database to enhance user experience by providing quick access to related objects based on past selections, further streamlining the process and improving the interactivity of the AR environment.

\textbf{Interactive Environment Customization}

Following the placement of each object, users can interact with them through intuitive gestures.
They have the ability to move, resize, and rotate each item effortlessly, allowing them to customize their AR spaces to their liking.
This level of interaction is facilitated by the advanced gesture recognition technology of the Microsoft HoloLens 2 device, which interprets user movements with high accuracy, ensuring a seamless and engaging user experience.
 
\begin{figure} [ht]  
\centering  % Centers the image
\includegraphics[width=0.5\linewidth]{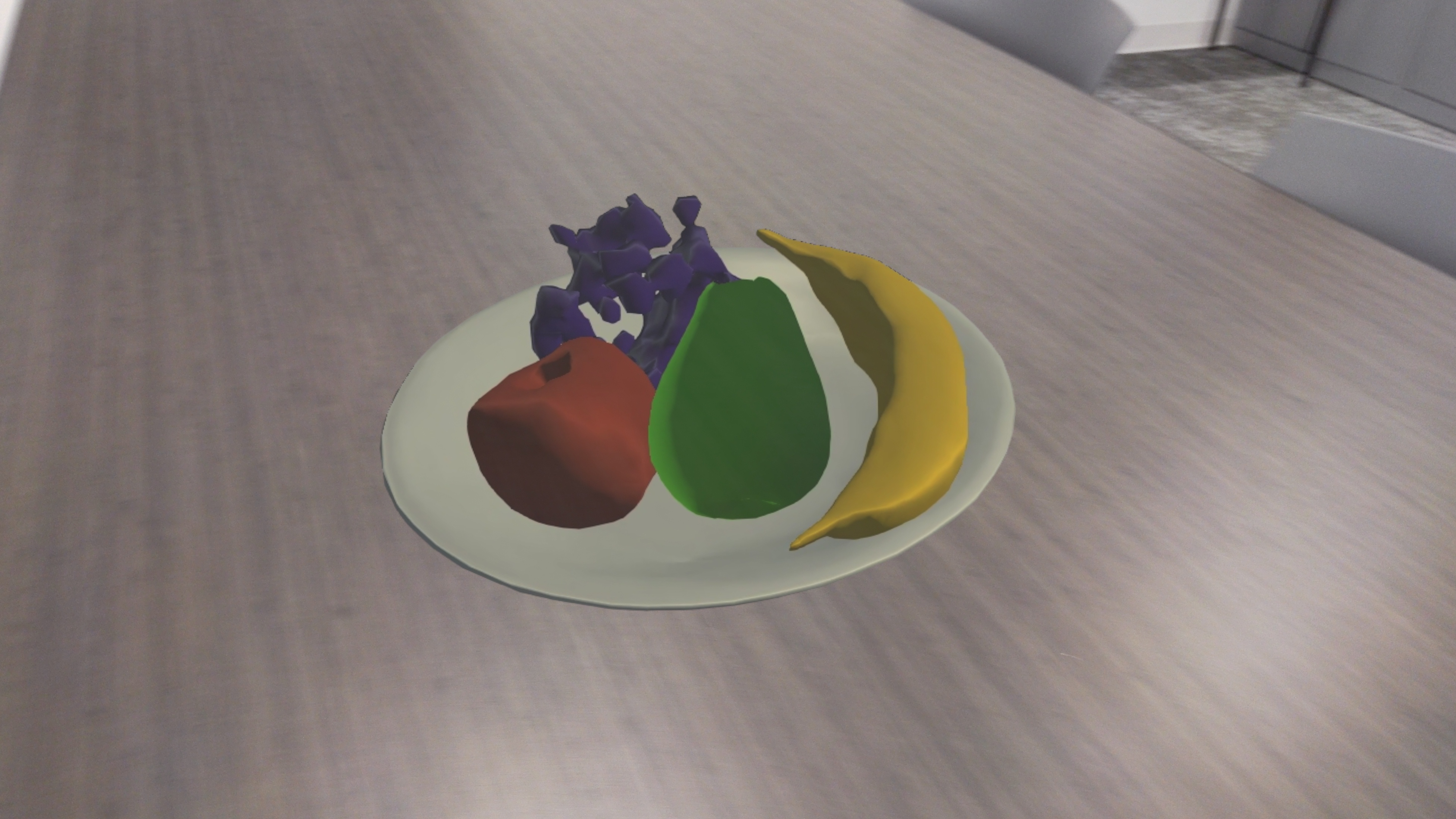}  % Adjusts the image's width to 80% of the text width
\caption{A plate containing a banana, apple, and grapes, generated in real time using \textit{Matrix}, is positioned on a table.
    % The final AR environment in the Matrix AR application.
    }  % Optional caption
\label{fig:fruit}  % Optional label for referencing
\end{figure}

To illustrate, consider the scenario of designing a fruit arrangement in AR.
Initially, a user places a white plate on the table, containing a variety of colorful fruits.
The fruits include a red apple, a green pear, a yellow banana, and some purple grapes, as shown in \cref{fig:fruit}.

Using gestures, the user can move the plate to different positions on the table or change the arrangement of the fruits.
This interaction highlights the flexibility of AR for creating and customizing virtual fruit displays with vibrant and realistic items.

\begin{figure} [ht]  
\centering  % Centers the image
\includegraphics[width=0.5\linewidth]{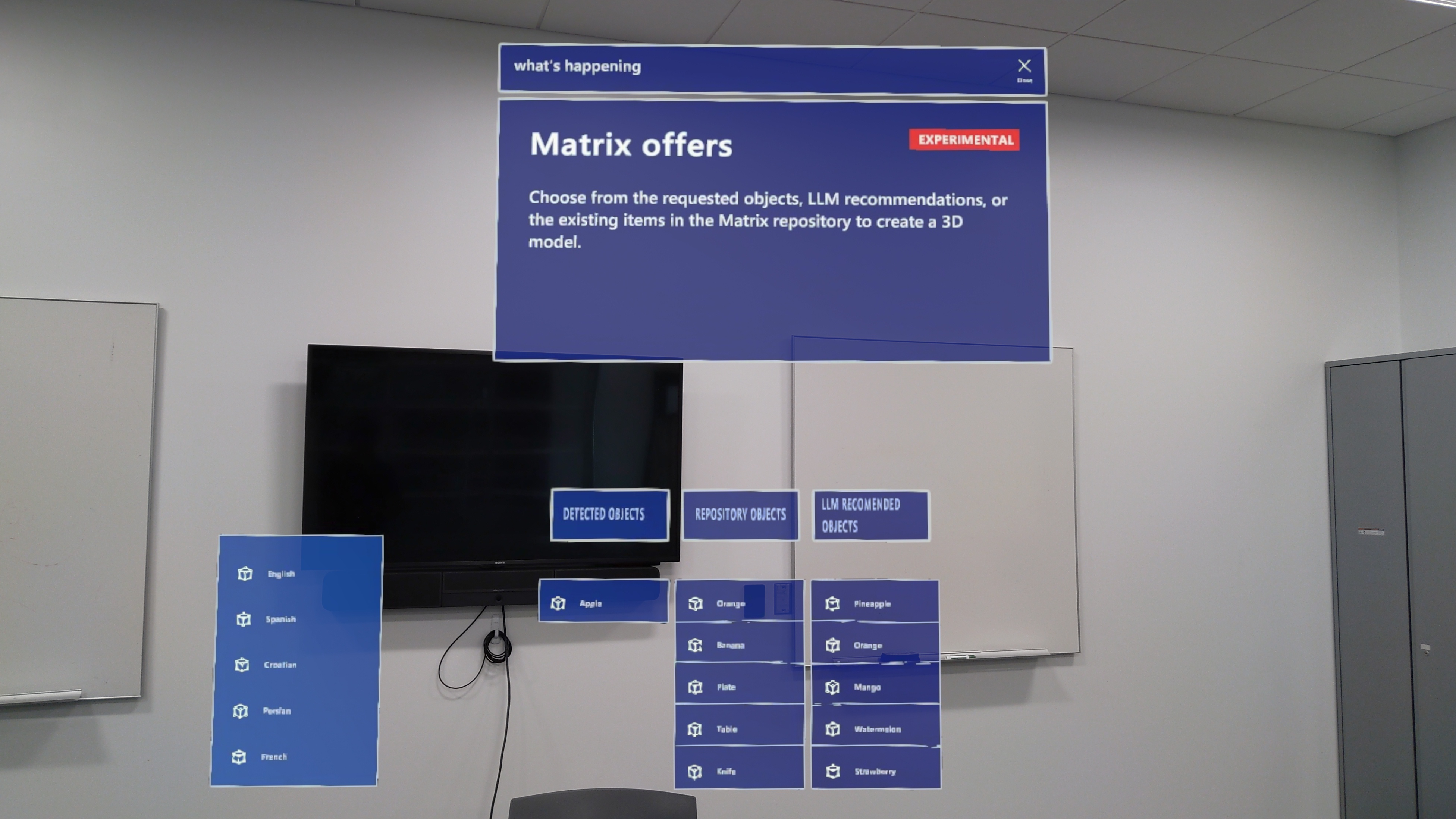}  % Adjusts the image's width to 80% of the text width
\caption{
Status board example in \textit{Matrix} AR application.
The ``what's happening'' board displays the ``Matrix offers'' message, informing users to choose from requested objects, LLM recommendations, or items from the repository to create a 3D model.
}  % Optional caption
\label{fig:status}  % Optional label for referencing
\end{figure}

Finally, to ensure optimal user experience and system transparency, we have integrated a status board at the top of the user interface, as seen in \cref{fig:status}.
This board continuously displays real-time updates about the system's status, showing messages such as `Welcome,' `Listening,' `Thinking,' `Offers,' `Baking,' and `Presenting,' depending on the current state of the system when processing-intensive tasks are being executed.

Given the time-consuming nature of tasks like 3D object generation or language model translation, this feature ensures that users remain informed about the ongoing operations, preventing confusion or uncertainty about system performance.
The status board is critical for maintaining a smooth, engaging, and transparent AR experience, allowing users to know exactly what is happening at all times. 

In the \textit{Matrix} Speech-to-3D Subsystem, we leverage advanced open-source technologies such as SeamlessM4T, Coqui.ai TTS, OpenAI’s Shap-E, Llama2, and Chromadb.
These technologies enhance the technological robustness of the AR application.
Additionally, for rendering objects in AR, we utilize the Unity Engine with the MRTK development toolkit, which is renowned for its powerful graphics capabilities and flexibility in developing interactive environments.
This integration not only makes our AR system technologically robust but also highly adaptable and user-friendly across multiple languages and interaction modalities.

% =====================chapter 3.2
\section{Context-Aware Object Recommendation}

\begin{figure} [ht]
\centering
\includegraphics[width=0.7\linewidth]{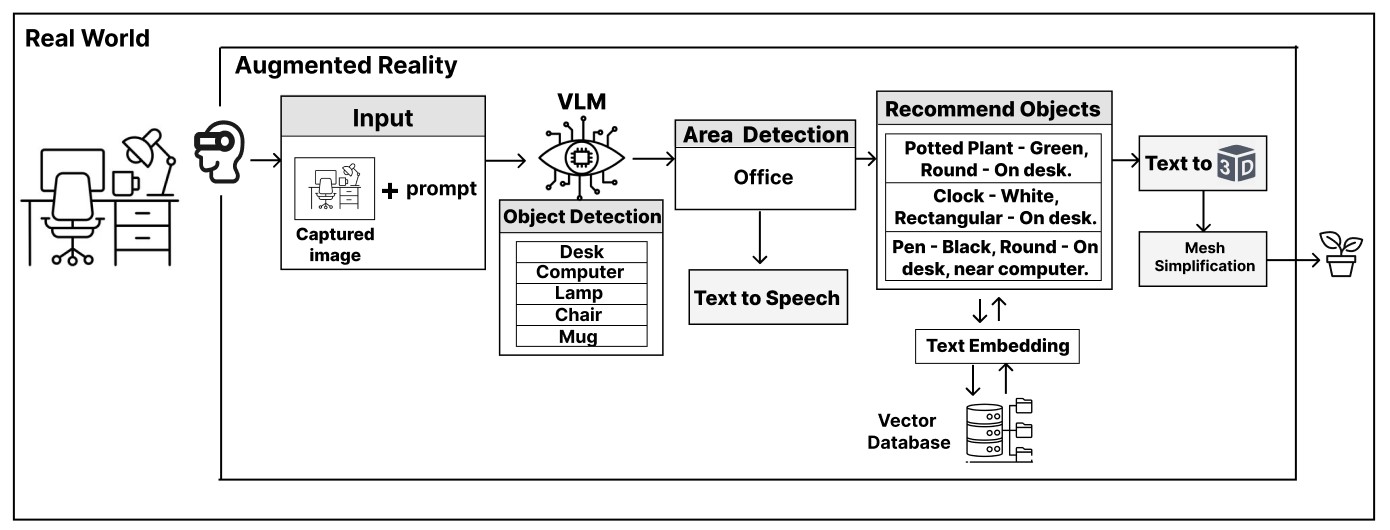} 
\caption{Workflow of integrating a VLM for context-aware object recommendation in AR.}  
\label{fig:VLM_workflow}
\end{figure}

The methodology for our framework integrates a VLM to enhance object recommendation within an AR environment\cite{behravan2024AIxVR}, as depicted in the provided workflow diagram.
This approach involves capturing the real-world scene through AR hardware, processing the captured image alongside a user-provided prompt, and utilizing the VLM to detect objects and recommend contextually relevant items.
This section provides a detailed explanation of each component and the overall workflow as shown in \autoref{fig:VLM_workflow}.

\textbf{Capture Real-World}

\begin{figure}
\centering
\includegraphics[width=0.5\linewidth]{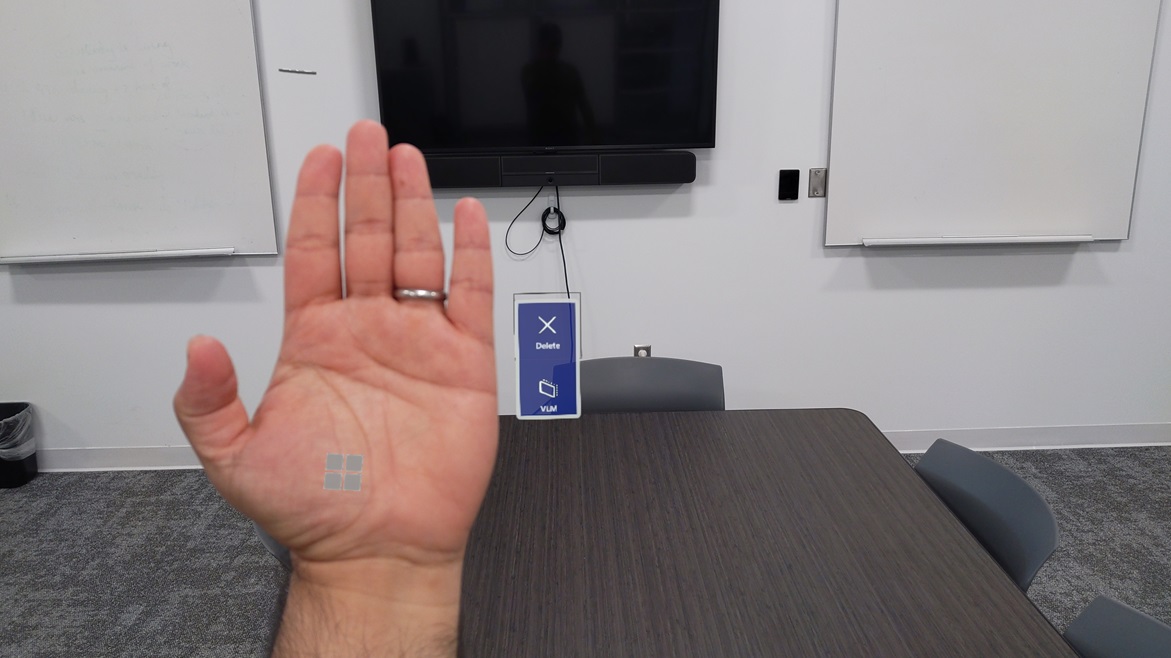}
\caption{User hand menu with VLM button for capturing images and sending them to the VLM model, along with a delete button for removing 3D-generated objects.}
\label{fig:Hand_menu}
\end{figure}

The process begins with the real-world environment, which is captured using a head-mounted AR device, such as HoloLens.
The device’s camera captures a high-resolution image of the user’s surroundings via the hand menu interface as shown in \autoref{fig:Hand_menu}.

\begin{figure}
\centering
\includegraphics[width=0.5\linewidth]{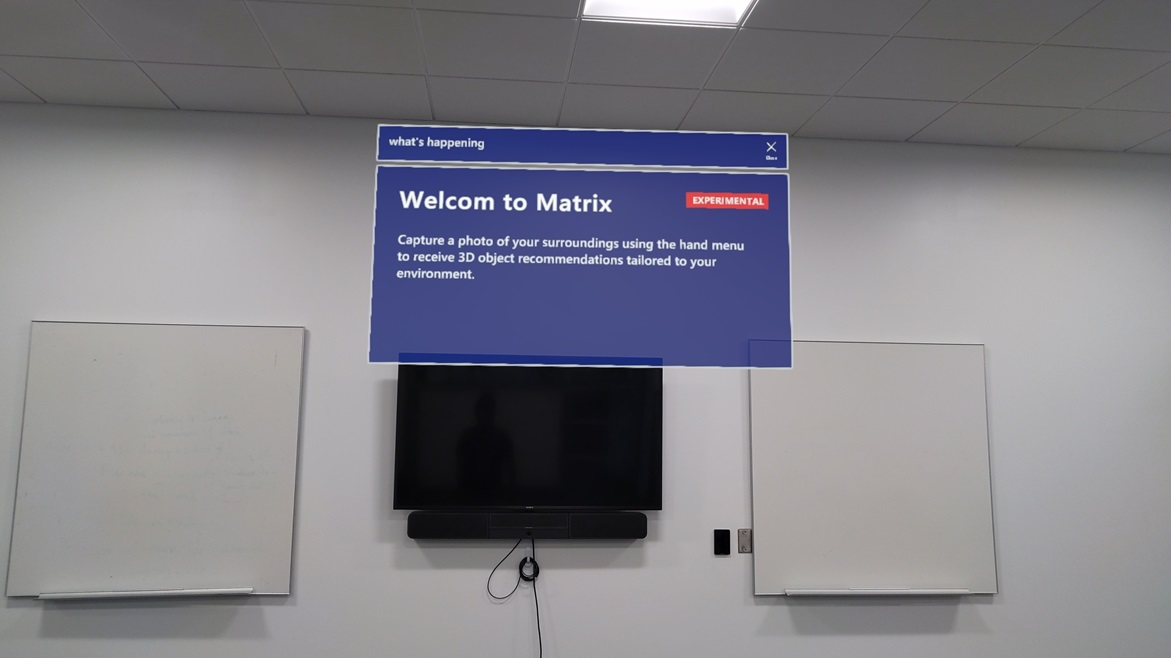}
\caption{Status board in AR environment prompting the user to capture a photo using the hand menu for personalized 3D object recommendations based on their surroundings.}
\label{fig:Status_board}
\end{figure}

This image serves as the primary input for the VLM, providing the visual context necessary for accurate object detection and recommendation.
We have also integrated a status board at the top of the user interface, as seen in \autoref{fig:Status_board}.
This board continuously displays real-time updates about the system's status, showing messages such as `Welcome,' `Observing,' `Suggestions,' `Baking,' and `Presenting,' depending on the current state of the system when processing-intensive tasks are being executed.

After clicking on VLM on the hand menu, the status board will be in observing mode.
 
% This image, as shown in \autoref{fig:CaptureImage}, serves as the primary input for the VLM, providing the visual context necessary for accurate object detection and recommendation.

% \begin{figure} [ht]
% \centering
% \includegraphics[width=0.5\linewidth]{Images/lobby2.jpg}   
% \caption{Captured image from the environment using the HoloLens webcam.}  
% \label{fig:CaptureImage}
% \end{figure}

\textbf{Input Prompt}

Simultaneously, we provide a textual prompt that describes the desired objects or contextual information.
Our prompt specifies the following:

\begin{enumerate}
\item
``write where am I (location name and a short description of the location).''
\item
``As a designer, recommend 5 simple objects (name , color, shape and suggest a location for each object within this space relative to the other objects in the picture) that would be suitable for this place but are currently not present.''
\end{enumerate}

VLMs as a multi-modal generative AI need a text to analyze the images, so the combination of the captured image and our prompt forms a comprehensive input that guides the VLM's processing.
These inputs are sent to the API we created for the VLM.

\textbf{VLM Processing and Object Detection}

The core of our framework is the VLM, specifically utilizing LLaVA. LLaVA processes the combined input to perform object detection and generate recommendations. 

\nomenclature{ViT}{Vision Transformer}

LLaVA uses a Vision Transformer (ViT) to extract visual features.
Vision Transformers have become popular for their ability to capture global context and relationships within images more effectively than traditional CNNs.
The input image is divided into fixed-size patches (e.g., 16x16 pixels). Each patch is then flattened into a vector and linearly projected into a lower-dimensional space, resulting in patch embeddings.
Since the transformer architecture does not inherently capture spatial relationships, positional embeddings are added to the patch embeddings to retain spatial information.
The patch embeddings, along with the positional embeddings, are fed into a transformer encoder.
The encoder consists of multiple layers of self-attention and feed-forward neural networks, which enable the model to capture complex relationships and dependencies between different patches of the image. The output of the transformer encoder is a set of contextualized embeddings representing the visual features of the image.
These embeddings are then used in subsequent stages of the model for tasks.
By using a Vision Transformer, LLaVA benefits from the ability to model long-range dependencies and capture more comprehensive global context within the image, which is particularly useful for tasks that require a deep understanding of the visual content.

LLaVA analyzes the captured image to identify and label objects present in the environment.
Concurrently, it interprets the user prompt to understand the context and the specific requirements of the user.
Within this VLM, the object detection module plays a crucial role by scanning the captured image to identify various objects, labeling them with their respective names.
For instance, in an office setting, LLaVA might detect and label objects such as a desk, computer, lamp, chair, and mug.
This step ensures that the VLM has a detailed understanding of the existing items within the user's environment.

\textbf{Contextual Understanding}

To ensure that the recommendations are contextually relevant, the VLM integrates the detected objects with the user's prompt.
By understanding the prompt, the VLM can filter and prioritize objects that align with the user's needs and the existing environment.
If the prompt specifies the need for workspace organization, the VLM might focus on recommending items that enhance organization and productivity.

\begin{figure} [ht]
\centering
\includegraphics[width=0.6\linewidth]{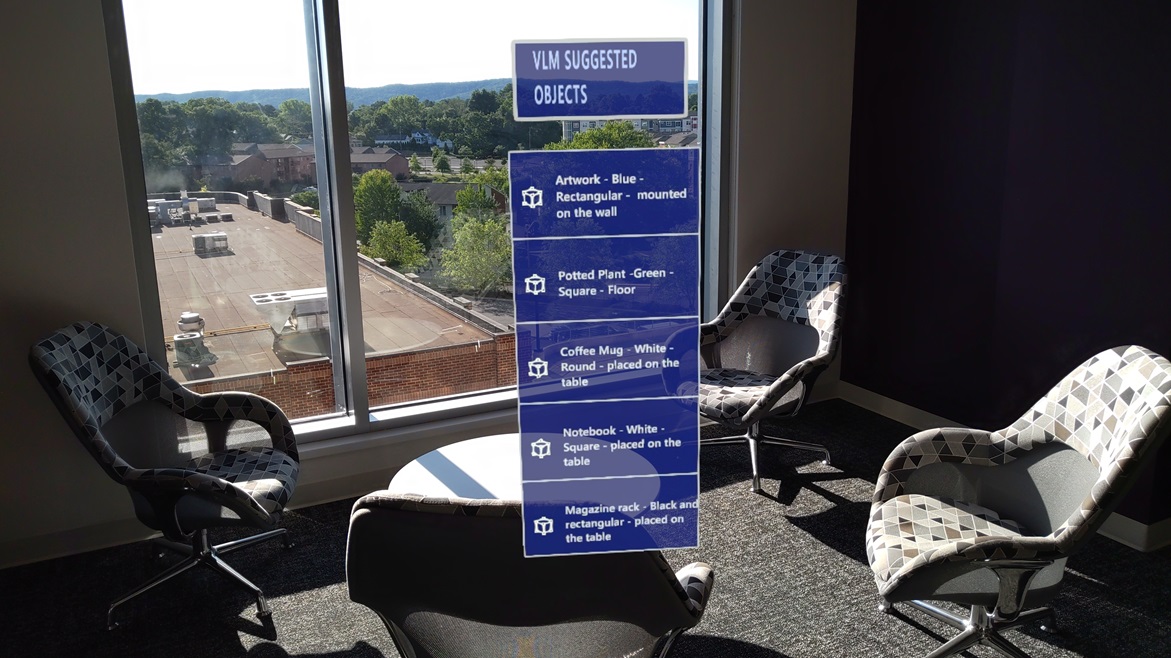}
\caption{Objects recommendation menu.}  
\label{fig:Menu}
\end{figure}

For instance, in response to the first prompt regarding \autoref{fig:Menu}, it describes the location as follows:
``You are in an office space with a purple wall, a window, and a view of the city.
The room features a round table with chairs around it. The chairs are arranged in a semi-circle, and the room appears to be a conference room or a meeting area.''

\textbf{Object Recommendation}

The final output from the VLM is a list of recommended objects.
These objects are suggested based on their logical similarity to the items detected in the environment and the context provided by the user's prompt.
The VLM analyzes the user's surroundings and provides items that not only fit within the spatial and aesthetic context but also serve to enhance the functionality and visual appeal of the environment.

For instance As shown in \autoref{fig:Menu}, the system recommend a ``Potted Plant - Green, Square - Floor.''
This recommendation is aimed at bringing a touch of nature into the workspace, offering a refreshing green element that can boost the user's mood and productivity.
The Square shape of the plant pot is chosen to complement other objects in the vicinity, creating a harmonious visual flow.

The VLM's recommendations are contextually relevant and designed to seamlessly integrate with the existing environment.
By suggesting items like the green, round potted plant, the VLM ensures that the user's space is not only practical but also aesthetically pleasing, enhancing the overall user experience.

\textbf{User Selection}

After generating the recommended objects, they are displayed to the user as a selective list in a menu.
The user can browse through the recommendations and select the most suitable item based on their preference and the context of their environment.
This interactive menu provides an intuitive interface where users can view detailed descriptions and images of each suggested item, helping them make informed decisions about which objects will best enhance their space.

The user interface is designed to be user-friendly, allowing for easy navigation through the list of recommended items.
As shown in \autoref{fig:Menu}, each item in the list includes relevant details such as color, shape, and suggested placement, ensuring that users have all the necessary information at their fingertips.
By presenting the recommendations in this manner, the system ensures that the selection process is both efficient and enjoyable, enabling users to customize their environment to better suit their needs and preferences.

\textbf{AR Environment Integration}

Our context-aware recommendation system works as a module in our framework that generates 3D models from text and voice.
Once the user selects an object from the menu, the selected object is sent as text input to a text-to-3D generative AI model.
This model processes the text description and generates a corresponding 3D model that matches the user's selection.

\begin{figure} [ht]
\centering
\includegraphics[width=0.6\linewidth]{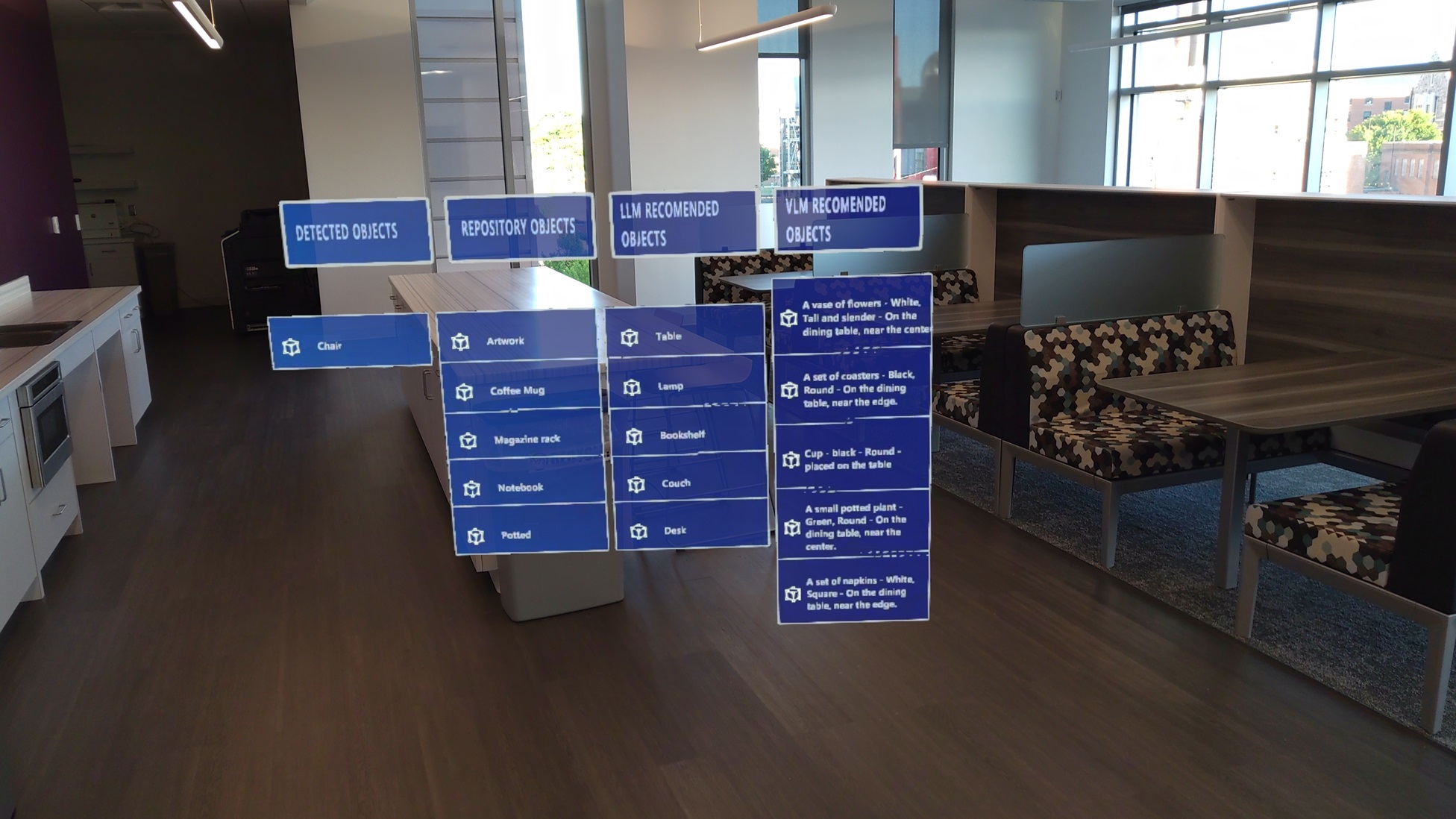} 
\caption{Multi-menu setup.}  
\label{fig:Multi Menu}
\end{figure}

The menu showing the recommended objects will appear alongside other menus generated from user requests by the LLM and repositories, as well as those requested by the user's voice commands.
This Multi-menu setup as shown in \autoref{fig:Multi Menu} allows the user to easily compare and select from both context-aware suggestions and specific requests they have made.
By providing a comprehensive and organized interface, the system ensures that users have access to a wide range of options tailored to their needs and preferences.

Additionally, the generated text by the VLM that describes the area and recommended objects in detail will be converted to speech by our framework's TTS module and played for the user in the AR environment.

The generated 3D model is then displayed within the AR environment.
The user can interact with the virtual object, including moving and rotating it to fit their desired placement and orientation within the real-world space.
This interaction enhances the immersive experience and ensures that the virtual objects are seamlessly integrated into the user's environment.
The combination of context-aware recommendations and user-requested items offers a highly personalized and flexible AR experience.

\begin{figure}[t]
\centering
\includegraphics[width=0.5\linewidth]{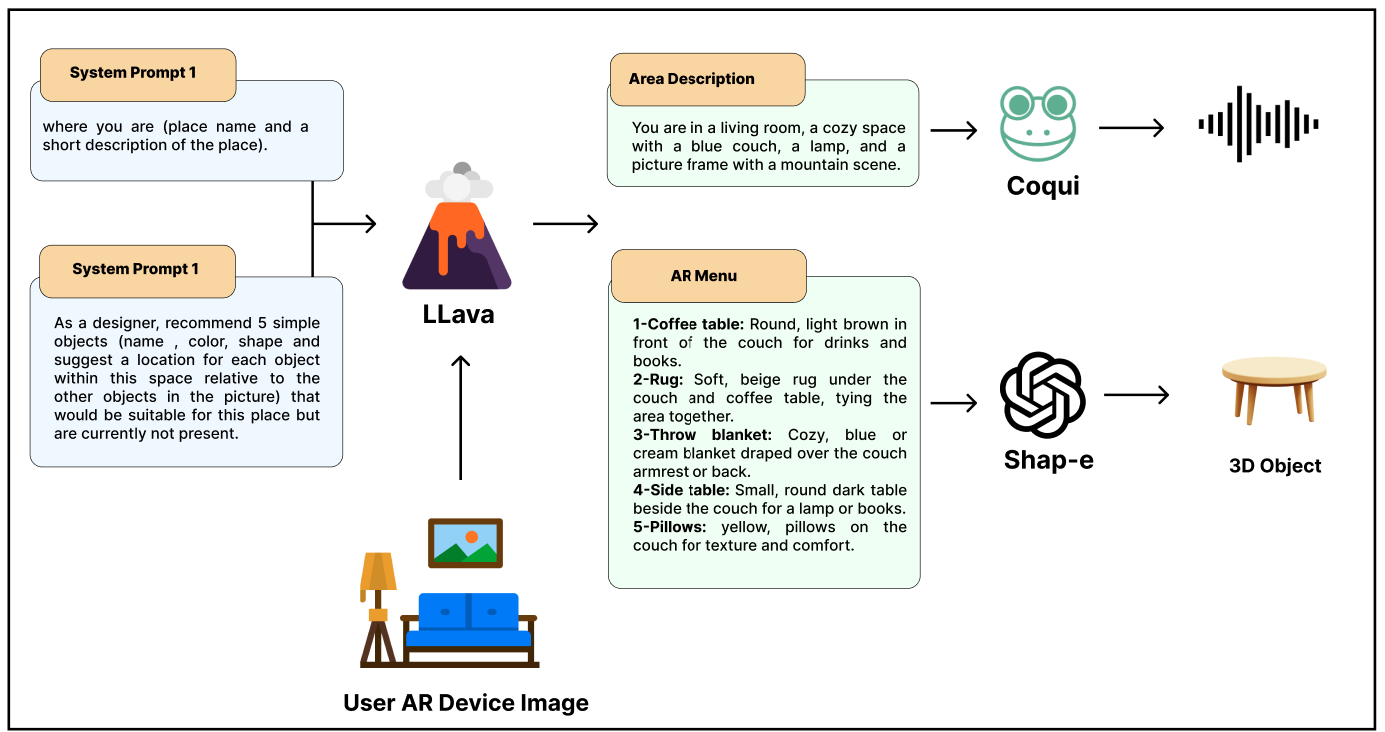}
\caption{System architecture: A user captures an image via an AR device, processed by VLM to suggest and integrate contextually relevant 3D objects into the AR environment.}
\label{fig:architecture}
\end{figure}

In conclusion, our framework leverages the strengths of VLMs to create a dynamic, context-aware AR experience (\autoref{fig:architecture}).
By combining real-world captures with user prompts and advanced AI processing, we achieve a seamless integration of virtual objects that enhance the user's environment in real time.

% ==============================chapter 3.3
\section{Image-to-3D Subsystem}
\begin{figure}
\centering
\includegraphics[width=1\linewidth]{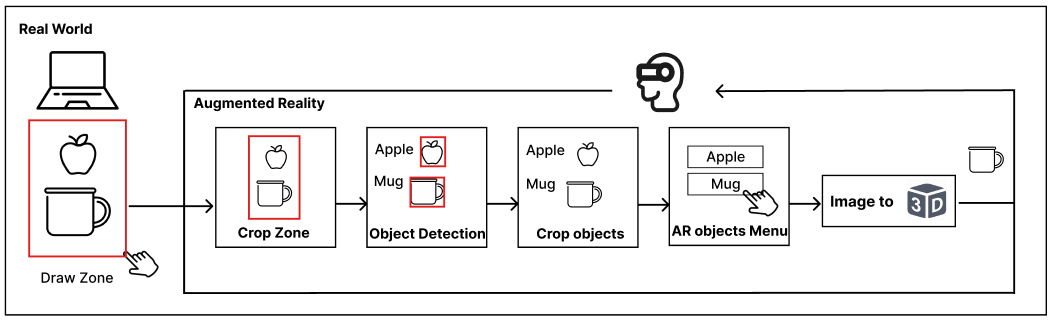} 
\caption{Workflow for the Image-to-3D Subsystem: The process begins with drawing objects in the real-world draw zone, followed by cropping and detecting these objects in the AR environment. The detected items are individually cropped and listed in an AR objects menu for user selection. The selected object is then converted from a 2D image into a 3D model for integration into the AR space.}
\label{fig:ImageWorkflow}
\end{figure}

Our primary contribution lies in the development of a process for converting 2D images into 3D models that can be seamlessly integrated into AR environments.
Our approach focuses on enhancing the capabilities of generative AI models to achieve this conversion.
We have meticulously designed a series of steps as shown in (\autoref{fig:ImageWorkflow}), each tailored to address the specific challenges and limitations identified in our previous research.
By doing so, we ensure that the generated 3D models are not only accurate and detailed but also optimized for real-time application in AR. 

% Our approach revolves around enhancing the capabilities of generative AI models to convert images into high-quality 3D models for AR environments.
% The approach includes several steps, each designed to address specific challenges identified in the previous sections.

\textbf{Zone Selection and Image Capture}

\begin{figure} [ht]
\centering
\includegraphics[width=0.6\linewidth]{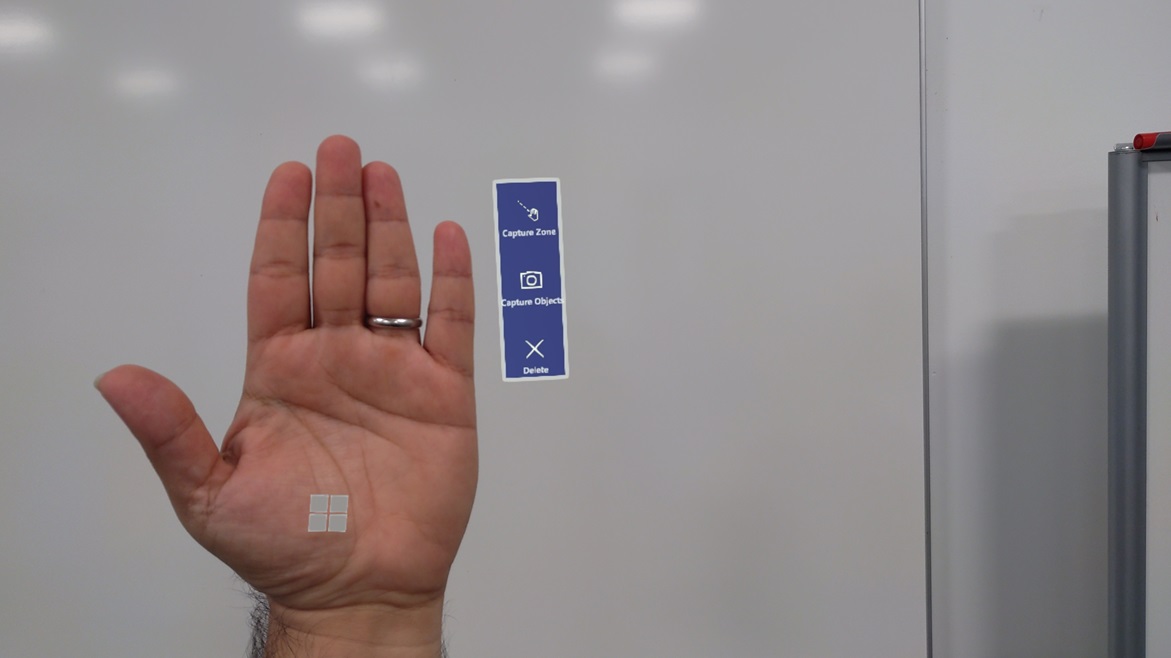}
 
\caption{AR hand menu.}  
\label{fig:Hand_Menu}
\end{figure}

\begin{figure} [ht]
\centering
\includegraphics[width=0.6\linewidth]{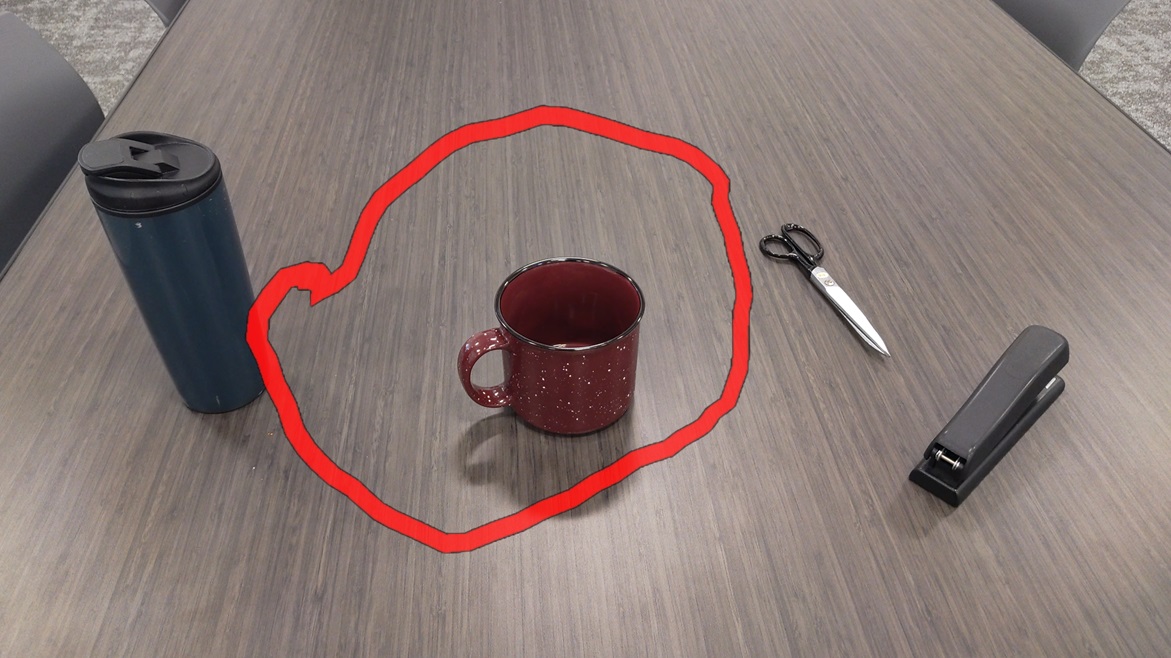}
 
\caption{Drawing selection zone (Lasso selection interaction technique).}  
\label{fig:Drawing}
\end{figure}

In the AR hand menu shown in \autoref{fig:Hand_Menu}, users can click on the ``capture zone'' button to draw a red line around the object or objects they want to generate 3D models of in the AR environment.
This lasso selection interaction technique is used in 2D and 3D settings~\cite{Yu-2016-a}.
The interactive process, illustrated in \autoref{fig:Drawing}, allows for precise zone selection, ensuring that the desired objects are accurately isolated from the background.
The process of drawing the line is stopped by pinching, and after three seconds, an image is captured, including the line drawn by the user and the objects within the user's view.

Additionally, users can click on the ``capture objects'' button in the hand menu to capture an image and request the creation of 3D models for all objects in their view without drawing a line.

\textbf{Crop Selected Zone}

\begin{figure} [ht]
\centering
\includegraphics[width=0.6\linewidth]{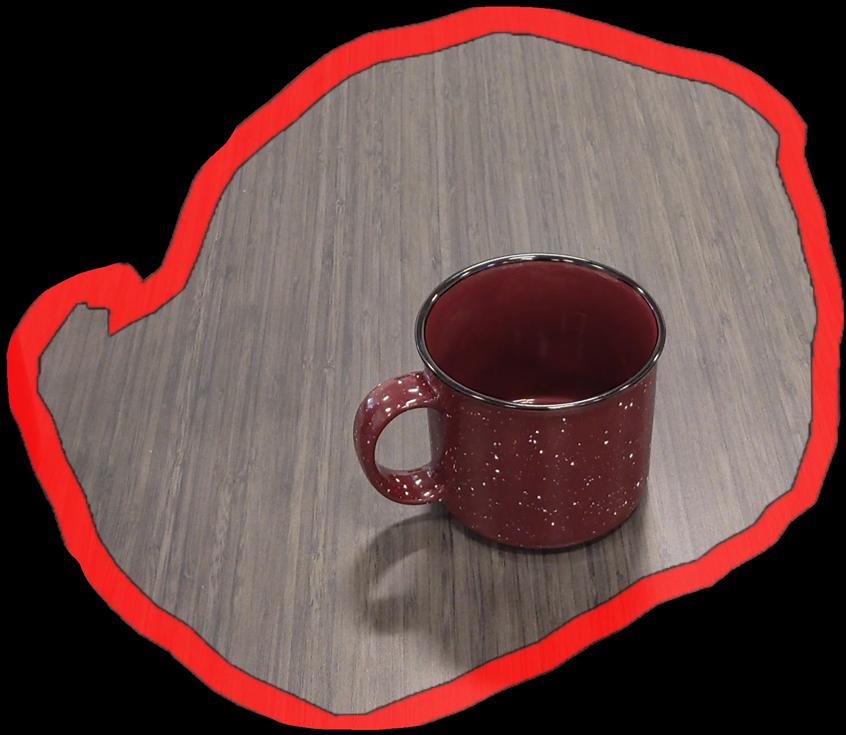}
 
\caption{Crop selected zone.}  
\label{fig:Crop}
\end{figure}

The selected zone is then cropped to focus on the area of interest, as seen in \autoref{fig:Crop}.
This cropping helps in reducing the computational load and improves the model's focus on the relevant part of the image.
This is achieved by combining the user's actions within the AR environment with the cropping process.
 
\textbf{Object Detection}
 
\begin{figure} [ht]
\centering
\includegraphics[width=0.6\linewidth]{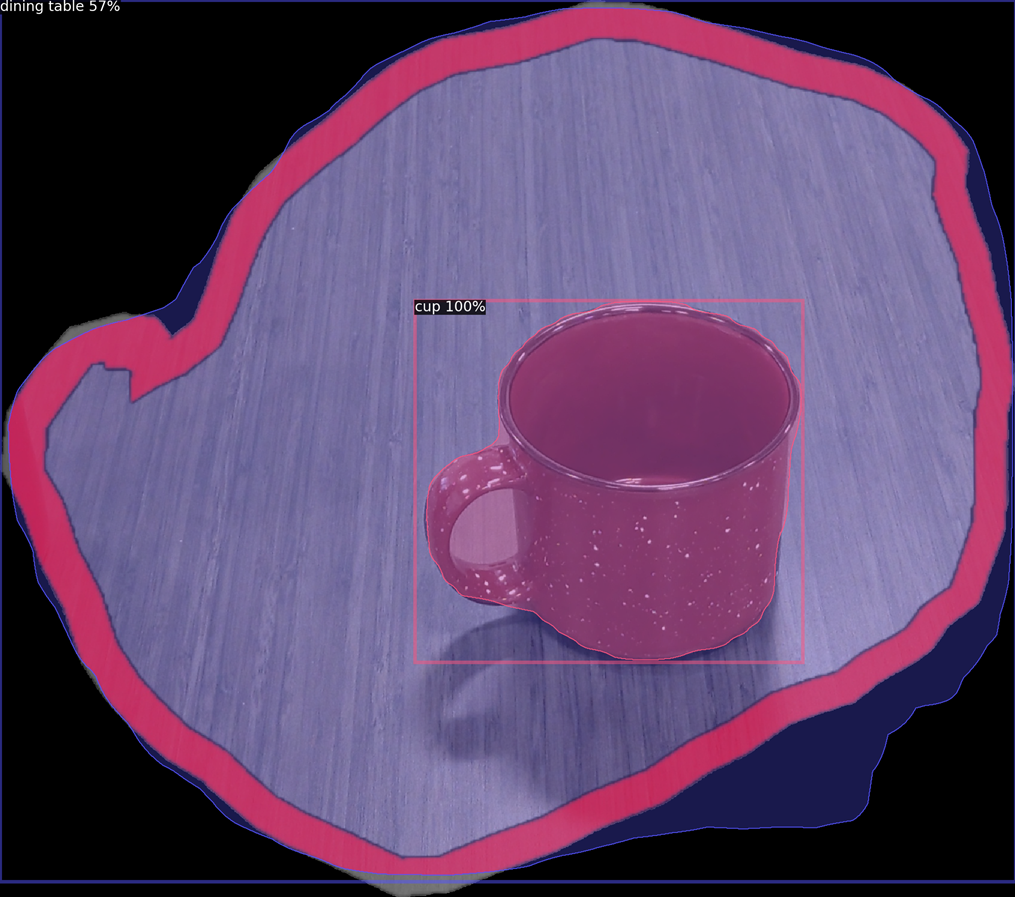} 
\caption{Object detection in zone.} 
\label{fig:Obj_detection}
\end{figure}

Advanced object detection algorithms are employed to identify and outline objects within the selected zone, as illustrated in \autoref{fig:Obj_detection}.
This step is crucial in ensuring the accuracy and robustness of the object detection process, which subsequently influences the quality of the generated 3D models.
For this purpose, we utilized the Mask R-CNN (Region-based Convolutional Neural Networks) algorithm, specifically configured for instance segmentation.
The Mask R-CNN model is well-regarded for its ability to perform pixel-level segmentation, enabling precise detection and delineation of objects within an image.

To optimize the performance of our detection model, we utilized pre-defined settings tailored for instance segmentation tasks.
These settings include parameters that control various aspects of the model's operation.
One critical parameter is the confidence threshold for making predictions, which we set at 0.5.
This threshold determines the minimum confidence score required for an object to be considered a valid detection.
By fine-tuning this parameter, we ensured that our model strikes a balance between precision and recall, reducing the likelihood of false positives while maintaining high detection accuracy.

Another critical aspect of our approach involves the use of pre-trained weights for the detection model.
These weights, derived from extensive training on large datasets such as the COCO dataset, provide a robust foundation that enhances the model's ability to recognize and segment objects accurately.
Leveraging these pre-trained weights allows our model to benefit from the knowledge encoded during training, thereby improving its performance on new, unseen images.
This approach ensures that our detection process is both efficient and effective, capable of handling the variability and complexity of real-world scenes.

The integration of the detection model with the AR system is designed to utilize available computational resources optimally.
By automatically selecting the most suitable hardware, such as GPUs when available, the system ensures efficient processing and faster detection times.
This optimization is crucial for real-time applications, where timely and accurate detection is essential for a seamless user experience.
Through this integration, our system provides robust and reliable object detection capabilities, forming the foundation for subsequent 3D model generation and AR interactions.

\begin{figure} [ht]
\centering
\includegraphics[width=0.6\linewidth]{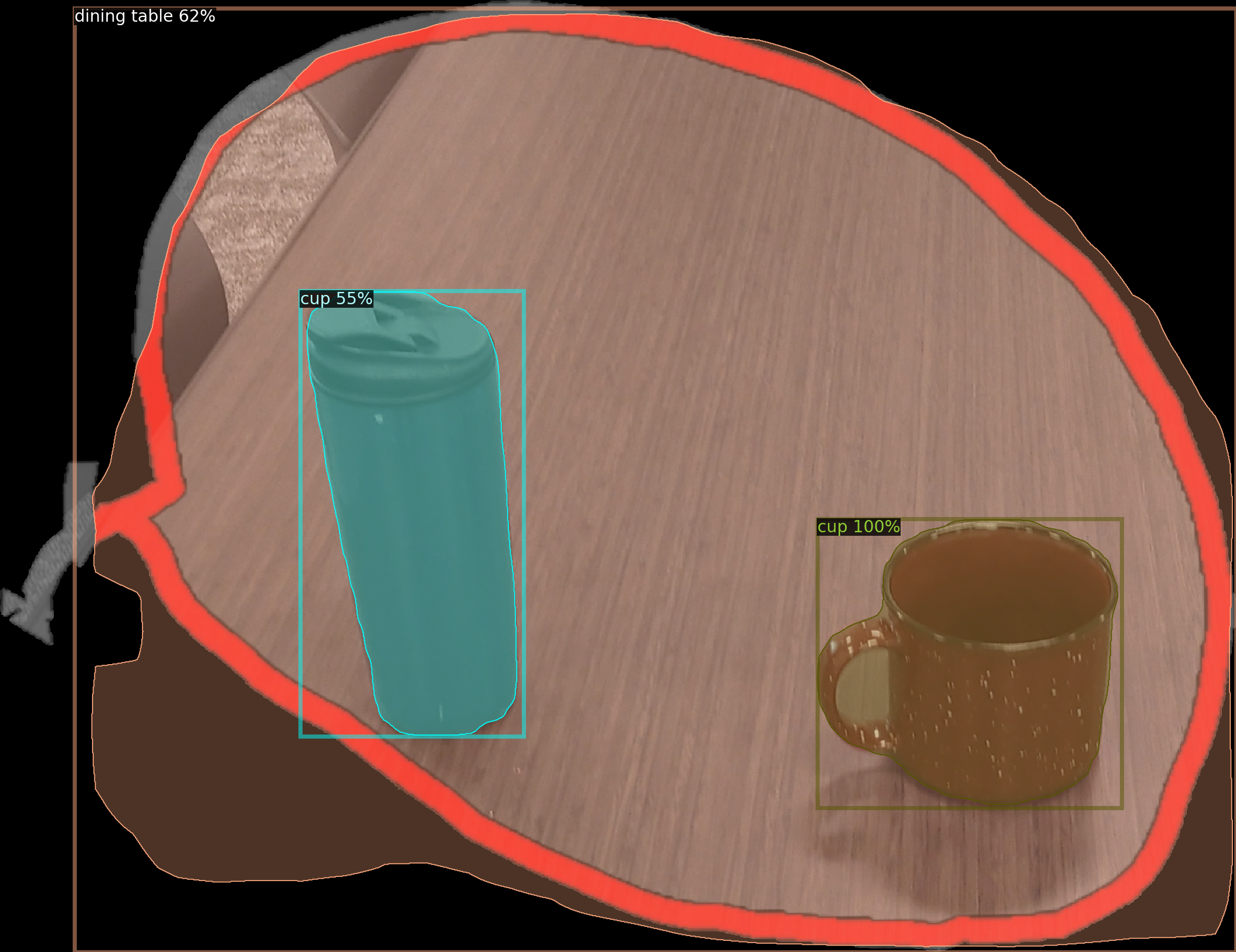} 
\caption{Multiple object detection in zone.}  
\label{fig:Multiobj_inZone}
\end{figure}

For scenes with multiple objects, the detection algorithm differentiates between various items, ensuring each object is accurately identified and processed separately, as depicted in \autoref{fig:Multiobj_inZone}.
This capability is essential for creating detailed and accurate 3D models, as it ensures that each object within the scene is individually recognized and reconstructed.
By employing advanced object detection techniques with well-configured parameters, our approach ensures high precision and reliability, facilitating the seamless integration of virtual objects into the real world and enhancing the overall AR experience.

\begin{figure} [ht]
\centering
\includegraphics[width=0.6\linewidth]{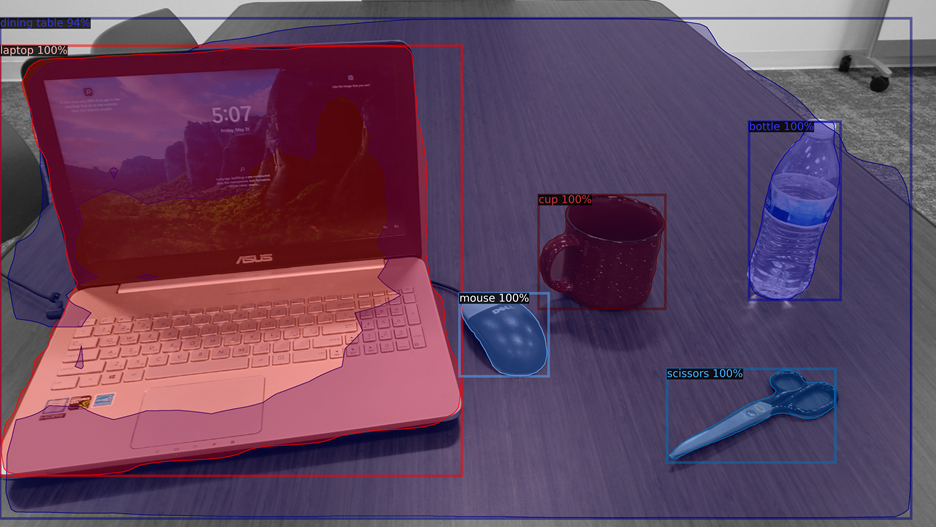} 
\caption{Capture all objects.}  
\label{fig:Multiobj_Capture All}
\end{figure}

If the user has utilized the ``capture objects'' button, similar to \autoref{fig:Multiobj_Capture All}, all objects will be detected.

\textbf{Objects Cropping}

After detecting the objects within the selected zone, each object is precisely cropped around the edges to isolate it from the background.
To achieve this, we utilize a combination of contour detection and Mask R-CNN for instance segmentation.
Initially, the image is converted to the HSV color space to detect specific color ranges, such as red, to create a mask.
Contours within the mask are identified, and the largest contour is presumed to be the boundary of the desired zone.
This contour is used to create a mask isolating the object from the background.

Using OpenCV, the bounding box around the largest contour is calculated to crop the object.
This cropped region undergoes further processing using Mask R-CNN to detect and segment individual objects within the cropped zone.
The detected objects' masks and bounding boxes are used to extract precise regions corresponding to each object.
These regions are then labeled with their respective class names, and the objects are encoded to base64 format for efficient data transfer via API.
This technical approach ensures high accuracy in object cropping, facilitating seamless integration with generative AI models for 3D model generation.

\begin{figure} [ht]
\centering
\includegraphics[width=0.6\linewidth]{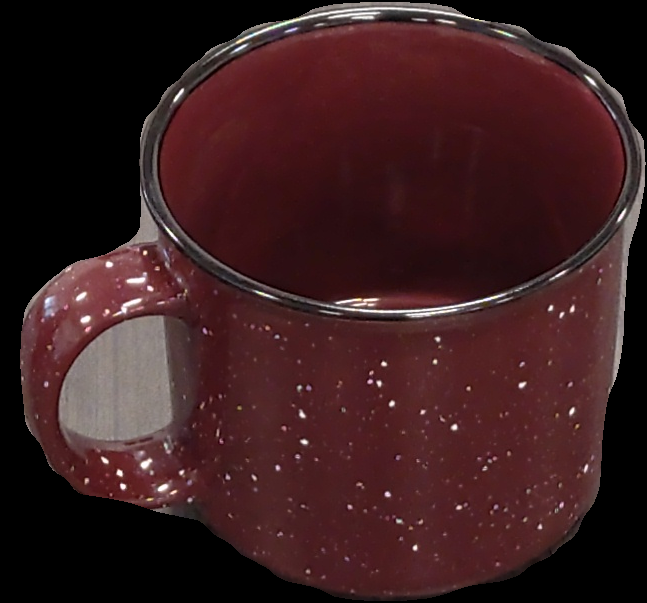} 
\caption{Object precisely cropped around the edges.}  
\label{fig:Croped_obj}
\end{figure}

As shown in \autoref{fig:Croped_obj}, this step involves detailed edge cropping to obtain a clean object image.
Once cropped, these objects are labeled with their respective names and prepared for integration.
The final output, consisting of the cropped objects along with their names, is then sent via an API.
This integration allows for seamless communication between the image processing system and the generative AI models, facilitating efficient and accurate 3D model generation.

\textbf{User Verification and Object Selection}

\begin{figure} [ht]
\centering
\includegraphics[width=0.6\linewidth]{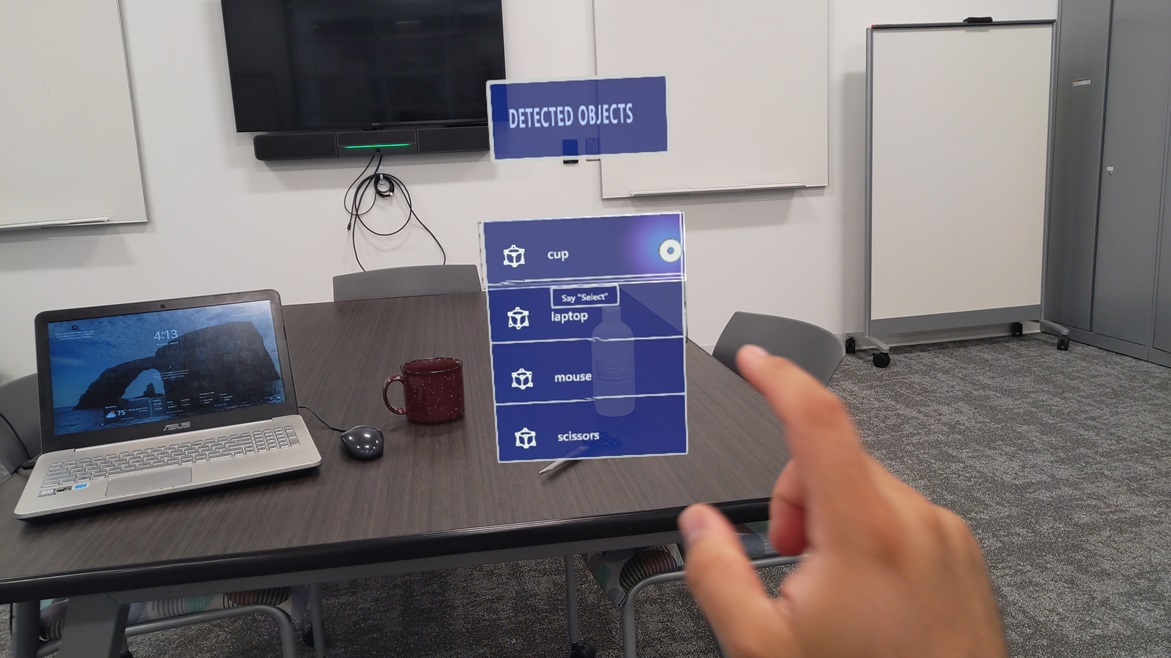} 
\caption{Detected objects menu.}  
\label{fig:objects_menu}
\end{figure}

The list of extracted objects from the previous step is presented to the user in the form of a menu (\autoref{fig:objects_menu}).
This interface allows the user to review the detected objects and prevents the creation of any inaccurately identified items.
The user can select objects based on their priorities and preferences for conversion into 3D models.
This verification process enhances the overall reliability and user control within the AR environment.

\textbf{3D Model Generation and Display}

\begin{figure} [ht]
\centering
\includegraphics[width=0.6\linewidth]{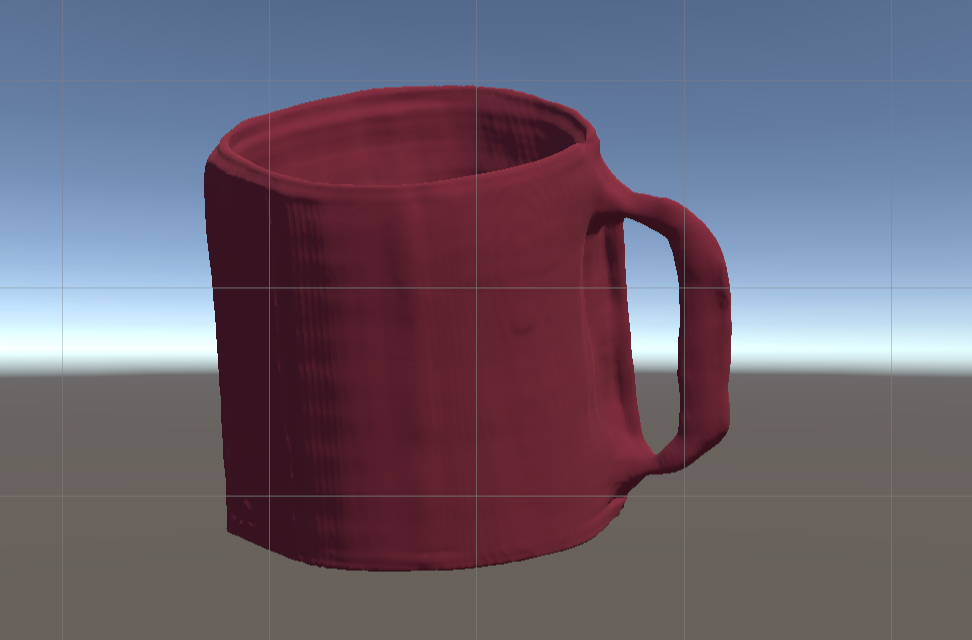} 
\caption{3D model.}  
\label{fig:output}
\end{figure}

Based on the user's selection from the menu, the extracted image is sent to the Shap-E model for 3D model generation.
The Shap-E model processes the image and creates a corresponding 3D model, which is then displayed to the user.
This step ensures that the user can visualize the generated 3D model and make any necessary adjustments.
The output from the Shap-E model is shown in \autoref{fig:output}.

% =================================User Study
      \chapter{User Study and Results} \label{ch:User Study and Results}
\section{Introduction}
This chapter describes the methodology, procedures, and results of the user study conducted to evaluate the usability and user experience of the system. Using a combination of pre- and post-study questionnaires, as well as real-time data collection, this study aimed to assess user perceptions of immersion, cognitive load, and potential cybersickness while interacting with the AR environment. By gathering these insights, we seek to understand the strengths and limitations of the system, as well as identify areas for improvement in enhancing user engagement and comfort in future iterations.

\section{Participants Demographics and Background}
A total of 35 participants were recruited to take part in this study. The participants represented diverse backgrounds with varying levels of experience with AR and VR technologies. This recruitment aimed to include both novice and experienced AR users to gauge the system's usability across a spectrum of familiarity levels. 

\begin{figure}[h!]
    \centering
    \begin{subfigure}{0.5\linewidth}
        \centering
        \includegraphics[width=\linewidth]{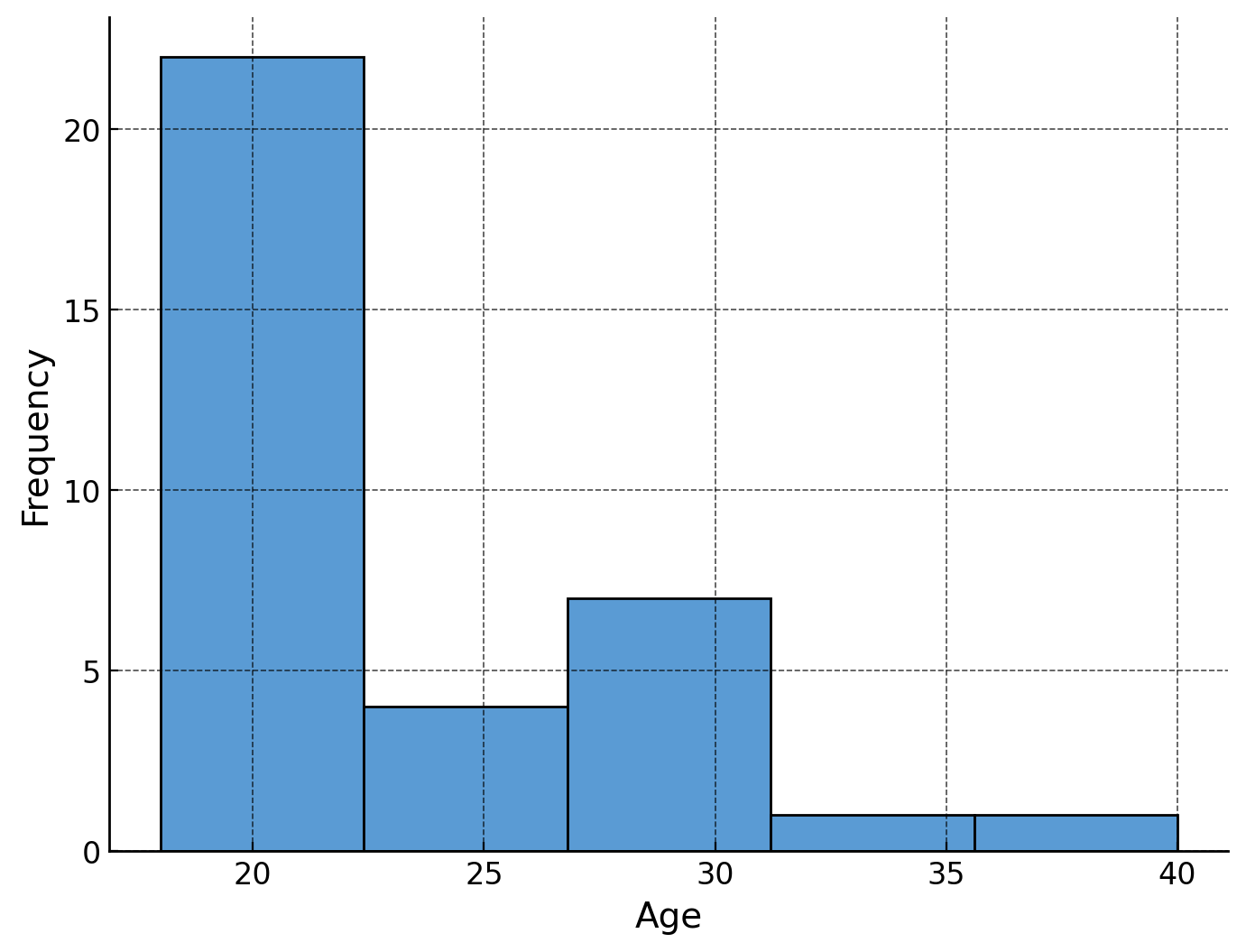}
        \caption{Histogram of user age ranges}
        \label{fig:age2}
    \end{subfigure}%
    \begin{subfigure}{0.5\linewidth}
        \centering
        \includegraphics[width=\linewidth]{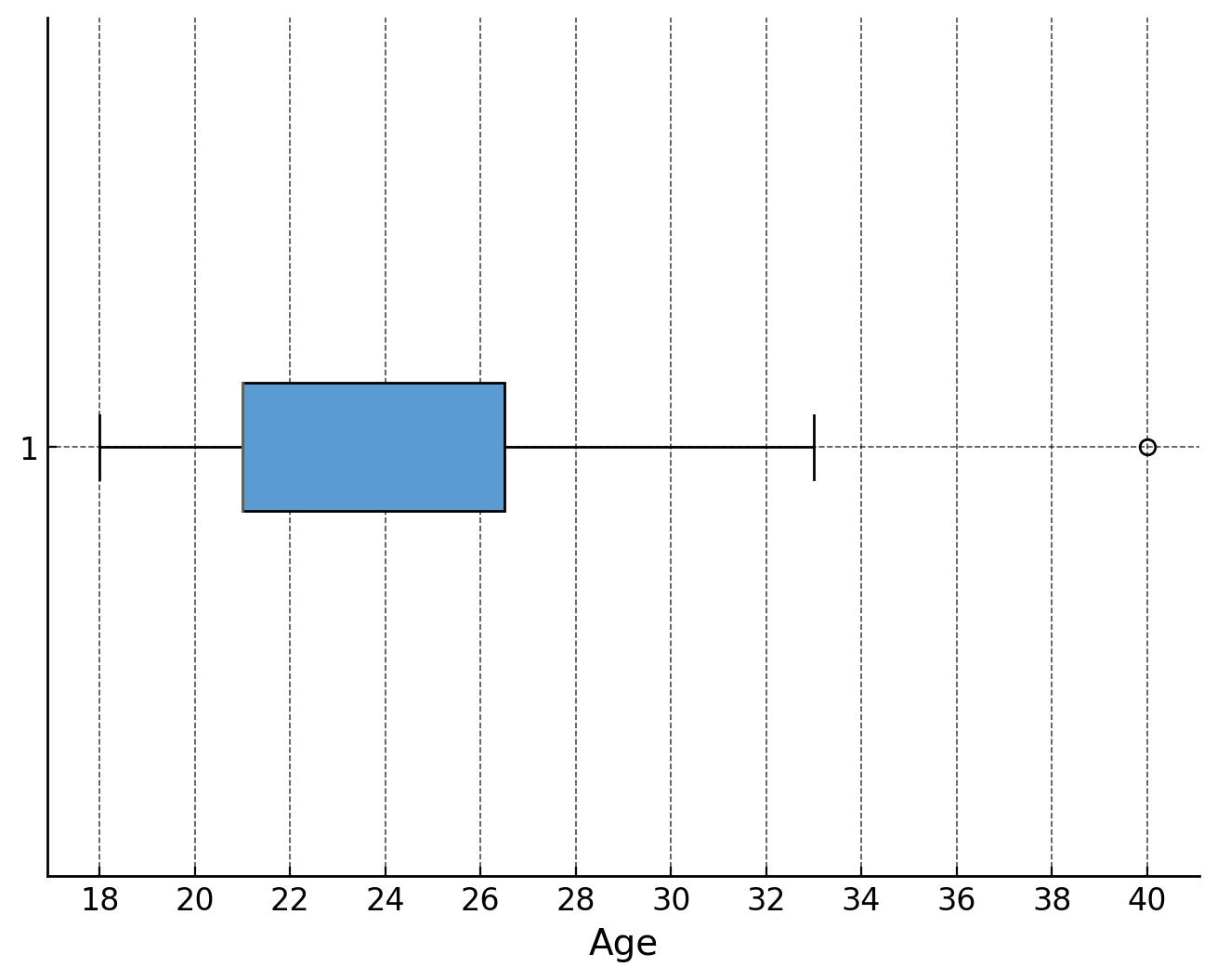}
        \caption{Box plot showing age distribution of users}
        \label{fig:age_dis}
    \end{subfigure}
    \caption{Age Distribution of Study Participants: A comparison of age ranges (left) and age distribution (right) among users.}
    \label{fig:combined_charts}
\end{figure}

\textbf{Age Distribution:} \\
The participants' ages ranged from 18 to 40, providing insights across different generational perspectives (see Figure \ref{fig:combined_charts}).

\begin{figure}[h!]
    \centering
    \includegraphics[width=0.5\linewidth]{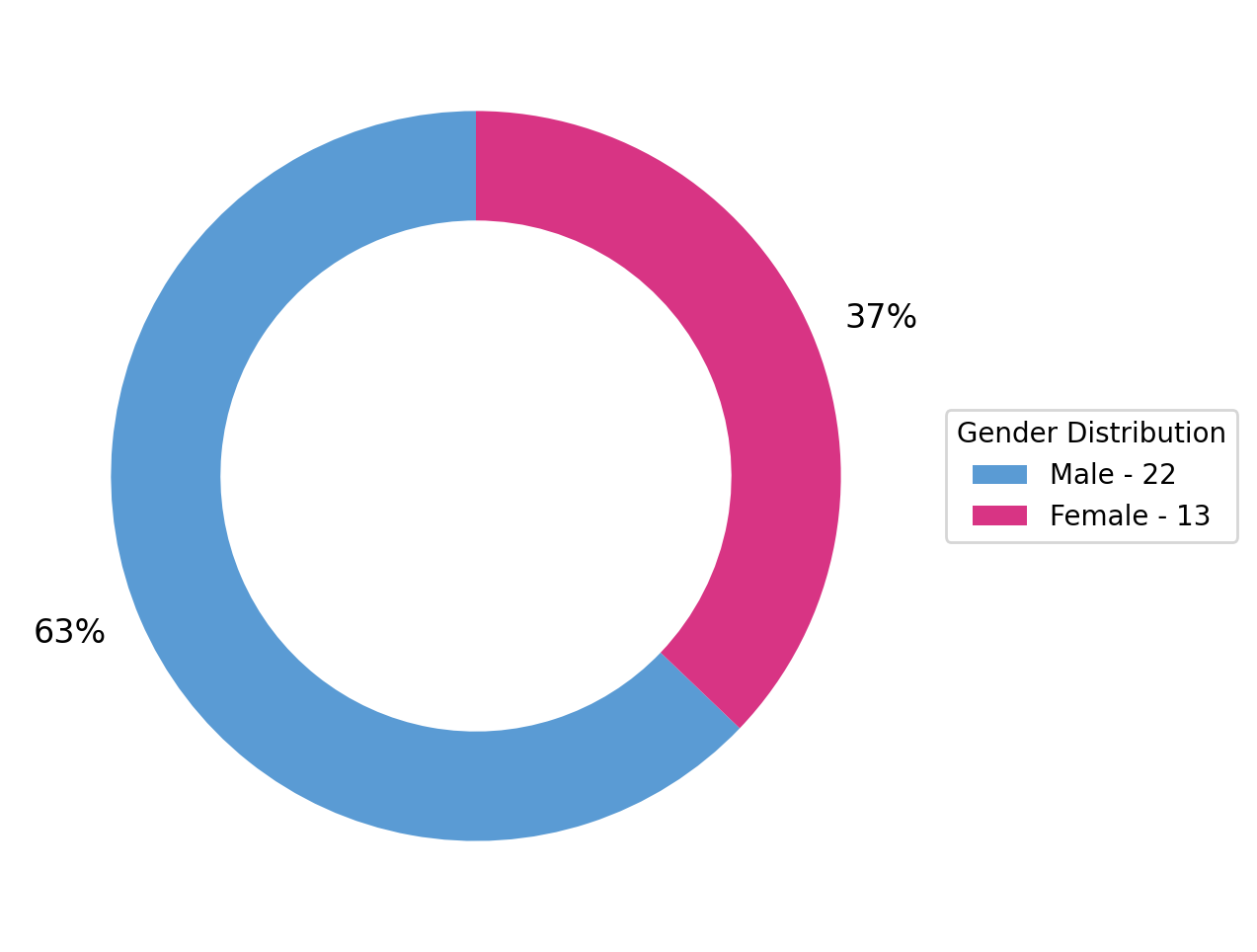}
    \caption{Gender distribution of participants.}
    \label{fig:gender_distribution}
\end{figure}

\textbf{Gender Distribution:} \\
The study's gender distribution is illustrated in Figure \ref{fig:gender_distribution}. Out of 35 participants, 63\% identified as male (22 participants), and 37\% identified as female (13 participants). Additionally, the chart includes categories for individuals who selected ``Other'' or ``Prefer not to answer,'' ensuring a more inclusive representation of gender diversity in the participant pool.

\begin{figure}[h!]
    \centering
    \includegraphics[width=0.5\linewidth]{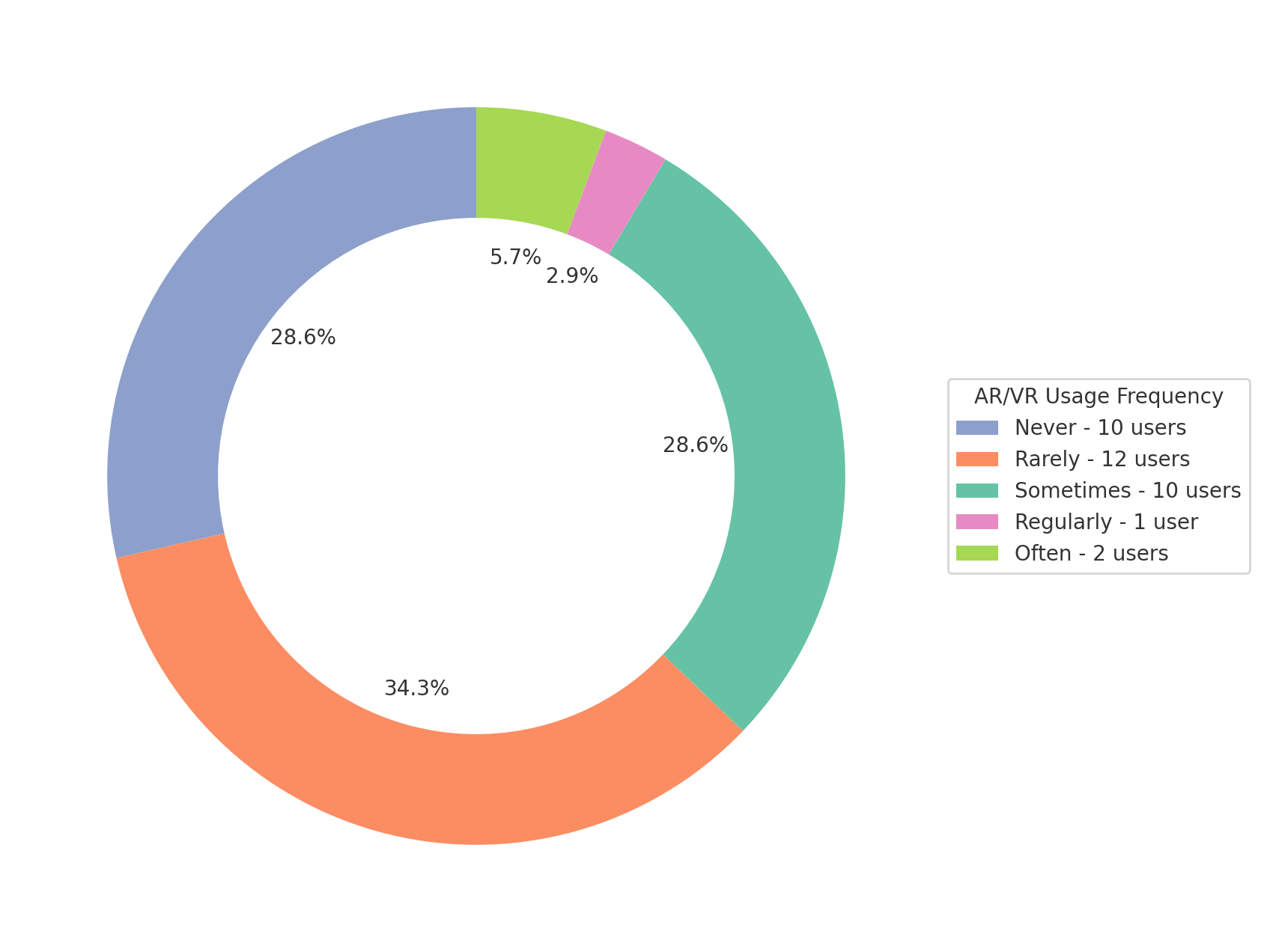}
    \caption{AR/VR usage frequency among participants.}
    \label{fig:ar_vr_usage}
\end{figure}

\textbf{Familiarity with AR/VR:} \\
Participants reported varying levels of familiarity with AR/VR environments. Specifically, 34.3\% of participants indicated they ``Rarely'' use AR/VR, 28.6\% use it ``Never,'' and another 28.6\% ``Sometimes.'' A small portion of participants, 5.7\%, reported ``Often'' using AR/VR, and 2.9\% indicated ``Regularly.'' This distribution highlights the inclusion of both novice and moderately experienced users in the study, offering insights from diverse experience levels, as shown in Figure \ref{fig:ar_vr_usage}.

\begin{figure}
    \centering
    \includegraphics[width=0.5\linewidth]{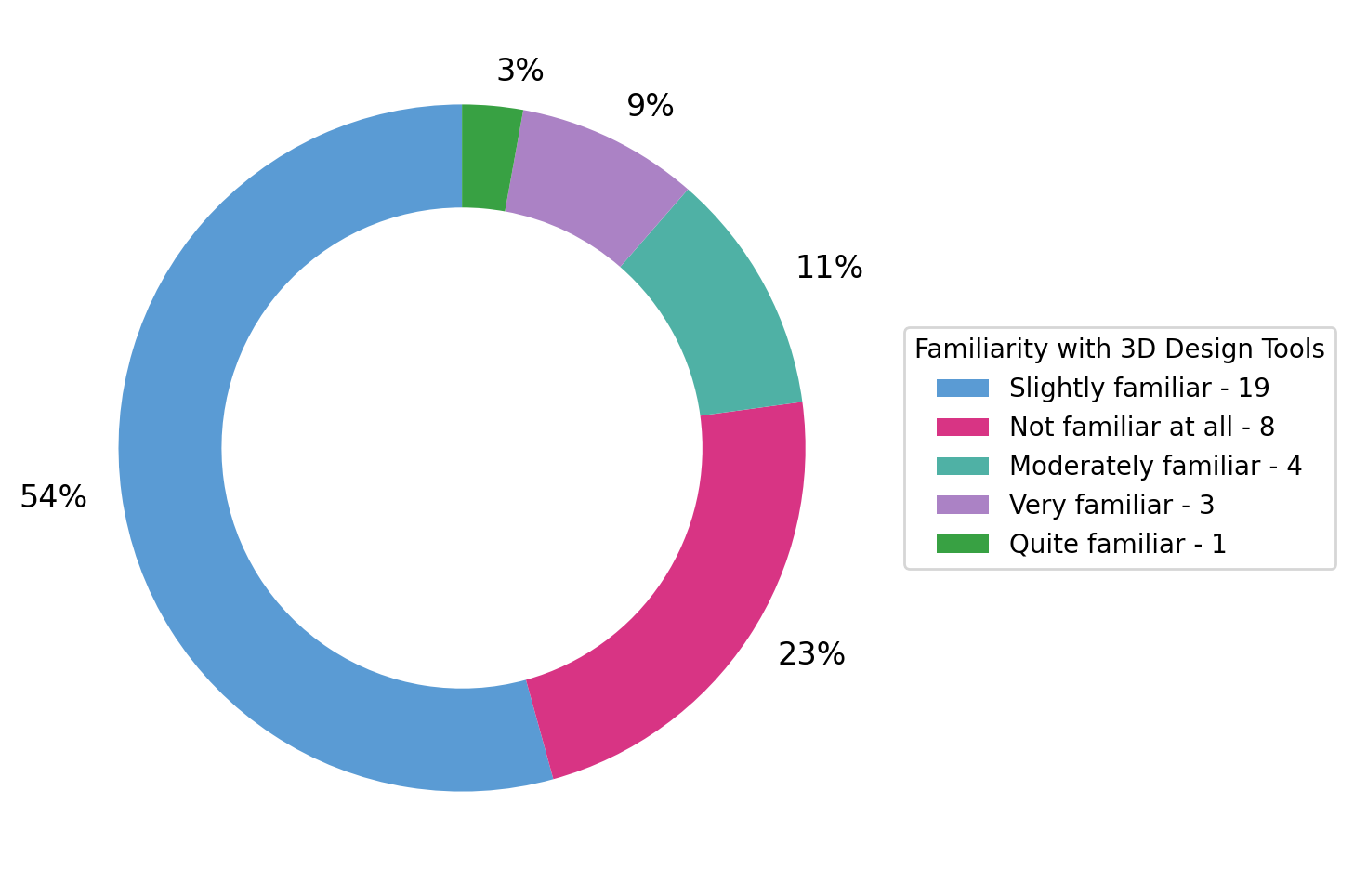}
    \caption{Familiarity with 3D design tools among participants.}
    \label{fig:3d_design_experience}
\end{figure}

\textbf{Experience with 3D Design Tools:} \\
Participants reported varying levels of familiarity with 3D design and modeling tools. Most participants (54\%) indicated they were ``Slightly familiar,'' while 23\% reported being ``Not familiar at all.'' Smaller groups identified as ``Moderately familiar'' (11\%), ``Very familiar'' (9\%), and ``Quite familiar'' (3\%). This range highlights a mix of novice users and those with some experience, providing insights into how different familiarity levels might influence interaction with 3D design tools, as illustrated in Figure \ref{fig:3d_design_experience}.

\begin{figure}[h!]
    \centering
    \includegraphics[width=0.5\linewidth]{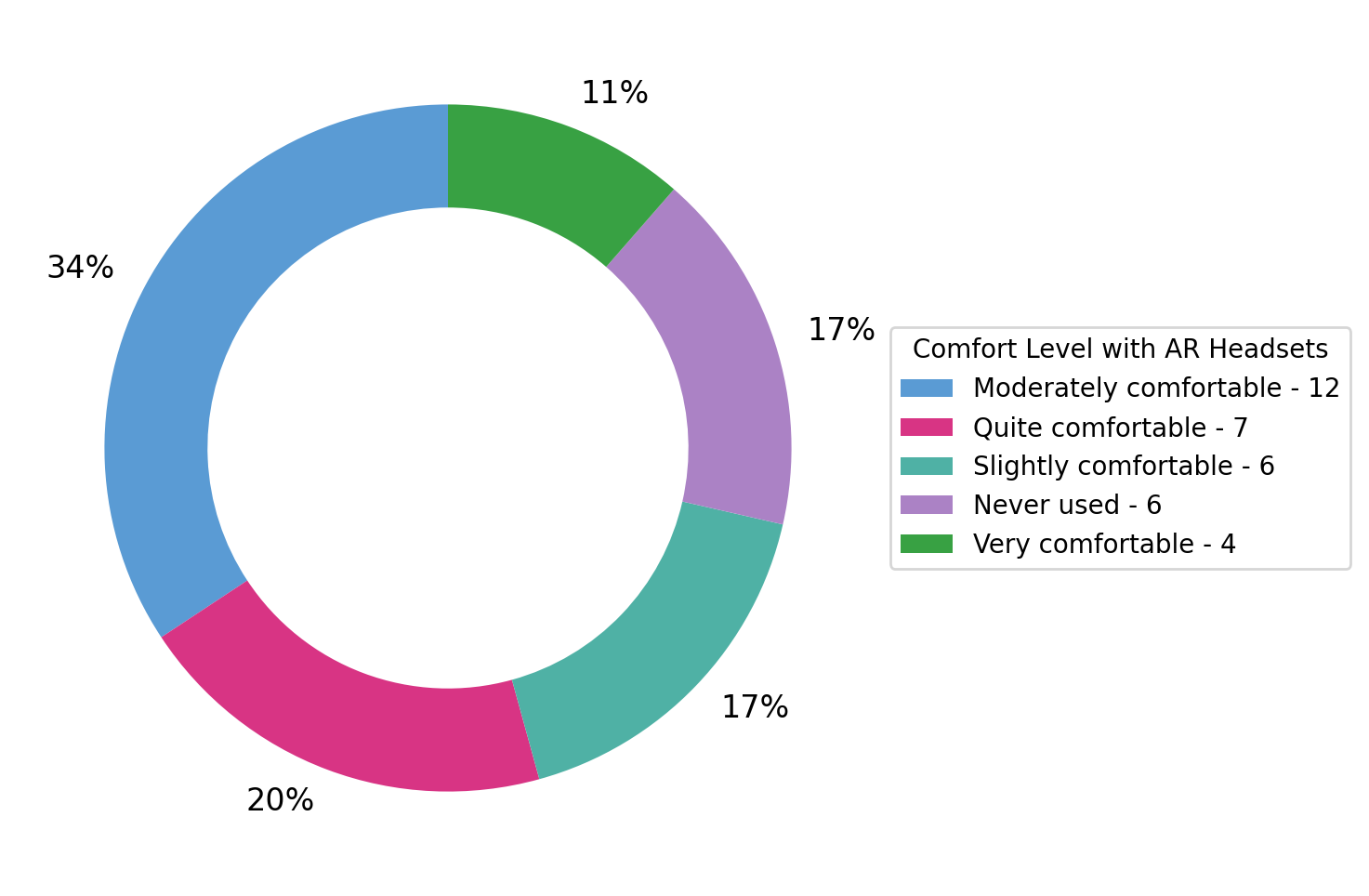}
    \caption{Comfort levels with AR headsets among participants.}
    \label{fig:ar_headset_comfort}
\end{figure}

\textbf{Comfort Level with AR Headsets:} \\
Participants expressed a range of comfort levels with AR headsets, spanning from ``Very comfortable'' to ``Never used.'' The largest group, 34\%, reported being ``Moderately comfortable,'' while 20\% indicated they were `Quite comfortable.'' Smaller groups were ``Slightly comfortable'' (17\%), ``Never used'' (17\%), and ``Very comfortable'' (11\%). This distribution provides insights into user experience with AR headsets across different comfort levels, offering a nuanced understanding of the system's usability, as shown in Figure \ref{fig:ar_headset_comfort}.

% \subsection{Apparatus and Environment}
% The study was conducted in a controlled environment where participants used a Microsoft HoloLens 2 AR headset. The device enabled participants to interact with 3D objects through voice commands, real-time object recommendations, and physical gestures. The setup was disinfected before each session to ensure hygiene and comfort, and participants were informed about possible mild cybersickness symptoms.

\section{Study Procedure}
The user study followed a structured procedure, which included a pre-study questionnaire, a training phase, a task phase involving three distinct tasks, and a post-study questionnaire. Each session lasted approximately 40 minutes.

\textbf{Pre-Study Questionnaire}
Prior to beginning the experiment, participants completed a preliminary questionnaire to capture their prior experience with AR, familiarity with voice commands, and comfort levels with using AR headsets. This data served as a baseline to compare with post-study responses, providing insight into how user experience might shift after interacting with the AR environment.

\textbf{Training Phase}
In this phase, participants were given an orientation on using the HoloLens 2 headset and a demonstration of voice commands to generate and manipulate 3D objects. The use of the system was shown in the form of three pre-recorded videos.

\textbf{Task Phase}
During the main task phase, participants were asked to complete three specific tasks designed to assess the system’s performance, usability, and user experience metrics:

\begin{itemize}
    \item \textbf{Task 1: Speech-to-3D Object Generation} \\
    Participants issued voice commands to generate various 3D objects in the AR environment. For example, a prompt might include, “Generate a red chair” or “Create a table with four legs.” The system processed these commands in real-time, interpreting the inputs and rendering the requested objects.

    \item \textbf{Task 2: Context-Aware Object Recommendations} \\
    Participants used the HoloLens 2 webcam to capture an image of their surrounding environment. Based on this image, the VLM provided object recommendations suited to the detected context. Participants reviewed these suggestions and selected objects to render in the AR space.

    \item \textbf{Task 3: Image-to-3D Model Generation} \\
    Participants selected objects within the AR environment using a virtual pen tool to trace the outline of specific items. After completing the outline, the system captured and processed the traced image to generate a corresponding 3D model that participants could view and interact with in real-time.
\end{itemize}

\nomenclature{SUS}{System Usability Scale}
\nomenclature{PQ}{Presence Questionnaire}
\nomenclature{NASA-TLX}{NASA Task Load Index }

\textbf{Post-Study Questionnaire}
After completing the tasks, participants answered a series of questions to evaluate their overall experience with the AR system. The questionnaire included standard usability scales, such as the System Usability Scale (SUS) \cite{brooke1996quick} and Presence Questionnaire (PQ) \cite{6788002}, to assess how immersive the experience felt. Additionally, NASA Task Load Index (NASA-TLX) \cite{HART1988139} was used to measure task load.

% \section{Data Collection and Analysis}

% \subsection{Usability and Presence}
% The SUS provided an aggregate usability score for the AR system, while the PQ captured participants' feelings of immersion. Higher scores on the PQ indicated a stronger sense of "presence" or feeling of being immersed in the virtual environment, which was essential to measure given the focus on enhancing user interaction within the AR space.

% \subsection{Cognitive Task Load}
% The NASA-TLX was used to assess cognitive load across multiple subscales, including mental demand, physical demand, and frustration level. The scores were analyzed to understand the mental effort required by participants to navigate the system and complete each task. Lower task load scores indicated a smoother and more intuitive user experience, while higher scores highlighted potential areas for reducing complexity.

\section{Results}

\subsection{Evaluation of System Usability Scale}

The SUS was used to evaluate user perceptions of the system's usability. The SUS provides an overall score based on users' responses to ten statements  \ref{table:susTB}, alternating between positively and negatively worded items. Responses were collected from participants, with scores for each question adjusted to align with the SUS scoring guidelines. Specifically, scores for positively worded statements were adjusted by subtracting 1, while scores for negatively worded statements were scored as \(5 - \text{score}\). Each participant’s total score was then multiplied by 2.5 to convert it to a scale of 0-100.

The average SUS score across all participants was \textbf{69.64}, which is slightly above the established average threshold of 68 for usability. This score indicates that the system is perceived as usable but suggests areas for further enhancement. Specifically, the score reflects:

\begin{itemize}
    \item \textbf{Frequent Use and Ease of Learning}: Responses to statements like ``I would like to use this system frequently'' and ``I would imagine that most people would learn to use this system very quickly'' suggest that users generally view the system favorably regarding routine use and ease of learning. This is reflected in relatively higher scores for these items, supporting the system’s accessibility and appeal for regular use.

    \item \textbf{Complexity and Need for Support}: Lower scores on items related to complexity, such as ``I found the system unnecessarily complex,'' and the perception of needing technical support indicate that some users may find certain aspects of the system challenging to navigate. This suggests that while the system is overall usable, there may be specific functions or interfaces that could benefit from simplification or additional support features.

    \item \textbf{Confidence in Use}: Users generally felt confident using the system, as indicated by their responses to ``I felt very confident using the system.'' This positive feedback highlights the system's success in creating an environment where users feel competent and secure in their interactions.
\end{itemize}

To further analyze the results, participants were divided into two groups based on their frequency of use: \textbf{Rarely/Never} and \textbf{Regularly/Sometimes/Often}. An ANOVA test was conducted to determine whether there was a statistically significant difference in SUS scores between these two groups. 

\subsubsection{ANOVA Analysis}

The results of the ANOVA test indicated a significant difference in SUS scores between the two groups (\textit{F} = 18.212, \textit{p} < 0.001). The mean SUS score for the \textbf{Rarely/Never} group was \textbf{64.38}, while the mean SUS score for the \textbf{Regularly/Sometimes/Often} group was significantly higher at \textbf{80.71}, as shown in Table \ref{table:anova_sus}.

\begin{table}[h!]
\centering
\caption{Detailed ANOVA calculations for SUS (System Usability Scale).}
\label{table:anova_sus}
\begin{tabular}{|l|c|}
\hline
\textbf{Statistic} & \textbf{Value} \\ \hline
Group 1 (Rarely/Never) Mean Score & 64.38 \\ \hline
Group 2 (Regularly/Sometimes/Often) Mean Score & 80.71 \\ \hline
Group 1 (Rarely/Never) Variance & 50.32 \\ \hline
Group 2 (Regularly/Sometimes/Often) Variance & 42.17 \\ \hline
Number of Participants in Group 1 & 20 \\ \hline
Number of Participants in Group 2 & 15 \\ \hline
F-Statistic & 18.21 \\ \hline
P-Value & 0.000032 \\ \hline
\end{tabular}
\end{table}

\begin{itemize}
    \item \textbf{Rarely/Never Group}: Participants in this group rated the system lower on average, suggesting usability challenges that may discourage frequent use.
    \item \textbf{Regularly/Sometimes/Often Group}: This group had higher average scores, indicating a more favorable perception of the system's usability among participants who engage with it more frequently.
\end{itemize}

\begin{figure}[h!]
    \centering
    \includegraphics[width=0.7\textwidth]{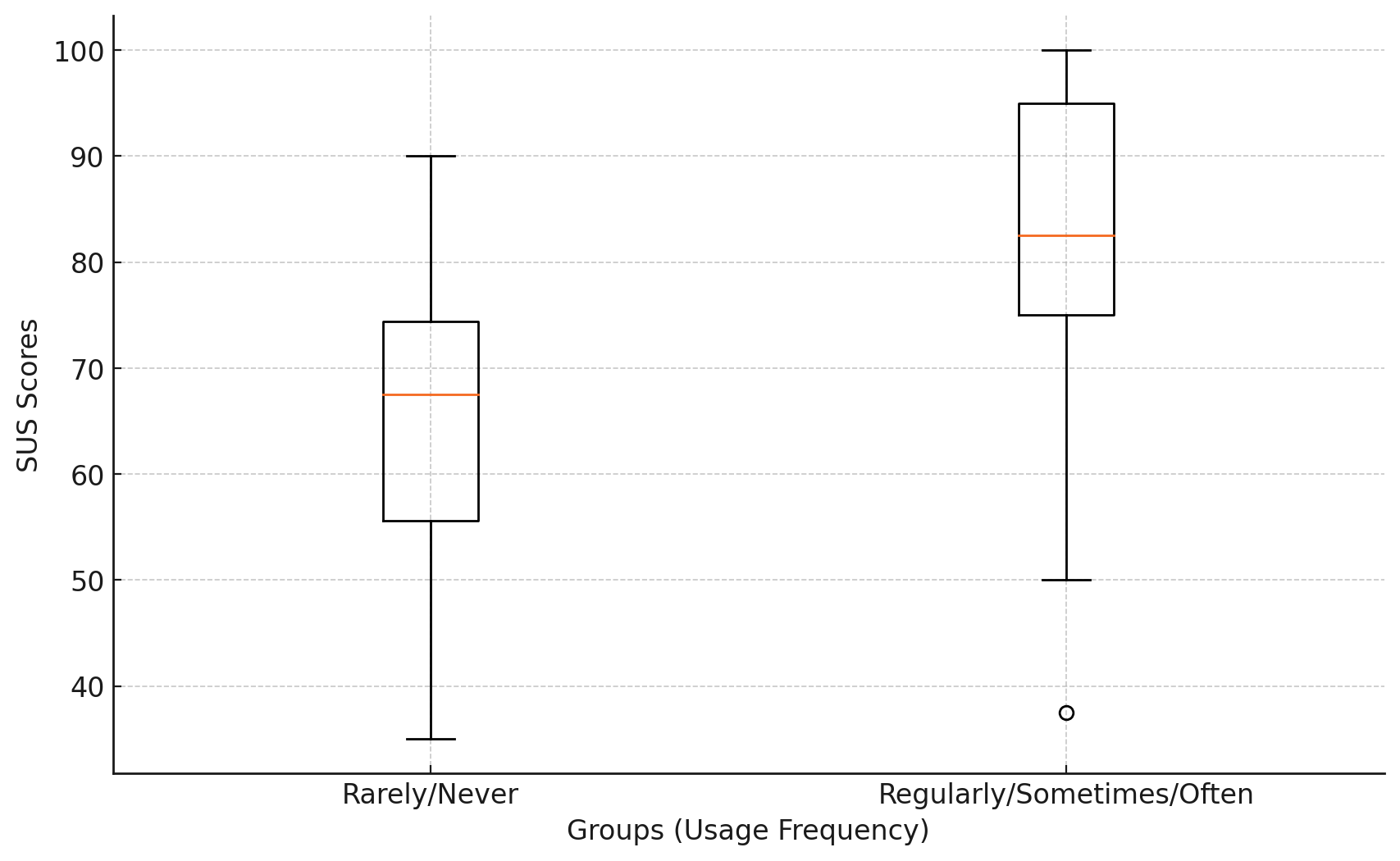}
    \caption{SUS Score Distribution by Group (Rarely/Never vs. Regularly/Sometimes/Often)}
    \label{fig:boxplot_sus}
\end{figure}

The boxplot in Figure~\ref{fig:boxplot_sus} visually demonstrates the differences in score distributions between the two groups.  

The statistically significant difference in scores supports the hypothesis that frequency of use influences users' perceptions of system usability. Specifically, participants categorized as \textbf{Regularly/Sometimes/Often} tend to rate the system more favorably, achieving higher SUS scores compared to those categorized as \textbf{Rarely/Never}. These findings suggest that engagement frequency may play a crucial role in shaping user experiences and perceptions of usability.

In summary, the SUS score of \textbf{69.64} suggests that the system is largely usable with strengths in ease of use and user confidence. However, addressing areas related to complexity and support may further enhance user satisfaction and effectiveness. Additionally, the ANOVA analysis highlights the importance of engagement levels in influencing perceptions of usability. Future iterations of the system could benefit from design adjustments that streamline complex features, reduce the need for user support, and encourage more frequent use, ultimately aiming to improve the overall SUS score and user experience.

\begin{table}[h!]
\centering
\caption{Simplified SUS scores by statement.}
\begin{tabular}{|l|c|}
\hline
\textbf{Statement} & \textbf{Mean Score (1-5)} \\ \hline
I would like to use this system frequently & 3.91 \\ \hline
I found the system unnecessarily complex & 1.80 \\ \hline
I thought the system was easy to use & 3.71 \\ \hline
I think that I would need technical support to use this system & 2.37 \\ \hline
I found the various functions in this system were well integrated & 4.14 \\ \hline
I thought there was too much inconsistency in this system & 1.71 \\ \hline
I would imagine that most people would learn to use this quickly & 4.14 \\ \hline
I found the system very cumbersome to use & 2.00 \\ \hline
I felt very confident using the system & 4.00 \\ \hline
I needed to learn a lot before I could get going with this system & 2.00 \\ \hline
\end{tabular}
\label{table:susTB}
\end{table}

\subsection{Evaluation of Task Load Using NASA-TLX}

To assess the cognitive and physical workload imposed on participants during the study, the NASA-TLX was used. This tool captures subjective workload perceptions across six dimensions: Mental Demand, Physical Demand, Temporal Demand, Performance, Effort, and Frustration. The analysis provides an overall workload score, along with diagnostic subscores for each category.

\textbf{Overall Workload Score}  
The average overall workload score across participants was calculated as \textbf{28.42}. This score suggests a moderate level of perceived workload among participants, indicating that the task complexity and demands were manageable but involved notable cognitive and physical efforts.

\subsubsection{Analysis of Variance (ANOVA)}

To further understand the variations in workload perception, participants were categorized into two groups based on their frequency of use:
\begin{itemize}
    \item \textbf{Rarely/Never (Group 1)}: Participants who indicated ``Rarely'' or ``Never'' in their usage frequency.
    \item \textbf{Regularly/Sometimes/Often (Group 2)}: Participants who indicated ``Regularly,'' ``Sometimes,'' or ``Often.''
\end{itemize}

An ANOVA test was conducted to determine whether there was a statistically significant difference in overall workload scores between the two groups. The results revealed a statistically significant difference (\textit{F} = 7.89, \textit{p} = 0.007). The mean workload score for Group 1 (\textbf{Rarely/Never}) was \textbf{28.65}, while the mean workload score for Group 2 (\textbf{Regularly/Sometimes/Often}) was lower at \textbf{23.69}, as shown in Table \ref{table:anova_nasatlx}.

\begin{table}[h!]
\centering
\caption{Detailed ANOVA calculations for NASA-TLX (Task Load Index).}
\label{table:anova_nasatlx}
\begin{tabular}{|l|c|}
\hline
\textbf{Statistic} & \textbf{Value} \\ \hline
Group 1 (Rarely/Never) Mean Score & 28.65 \\ \hline
Group 2 (Regularly/Sometimes/Often) Mean Score & 23.69 \\ \hline
Group 1 (Rarely/Never) Variance & 18.72 \\ \hline
Group 2 (Regularly/Sometimes/Often) Variance & 15.85 \\ \hline
Number of Participants in Group 1 & 20 \\ \hline
Number of Participants in Group 2 & 15 \\ \hline
F-Statistic & 7.89 \\ \hline
P-Value & 0.007 \\ \hline
\end{tabular}
\end{table}

\begin{figure}[h!]
    \centering
    \includegraphics[width=0.7\textwidth]{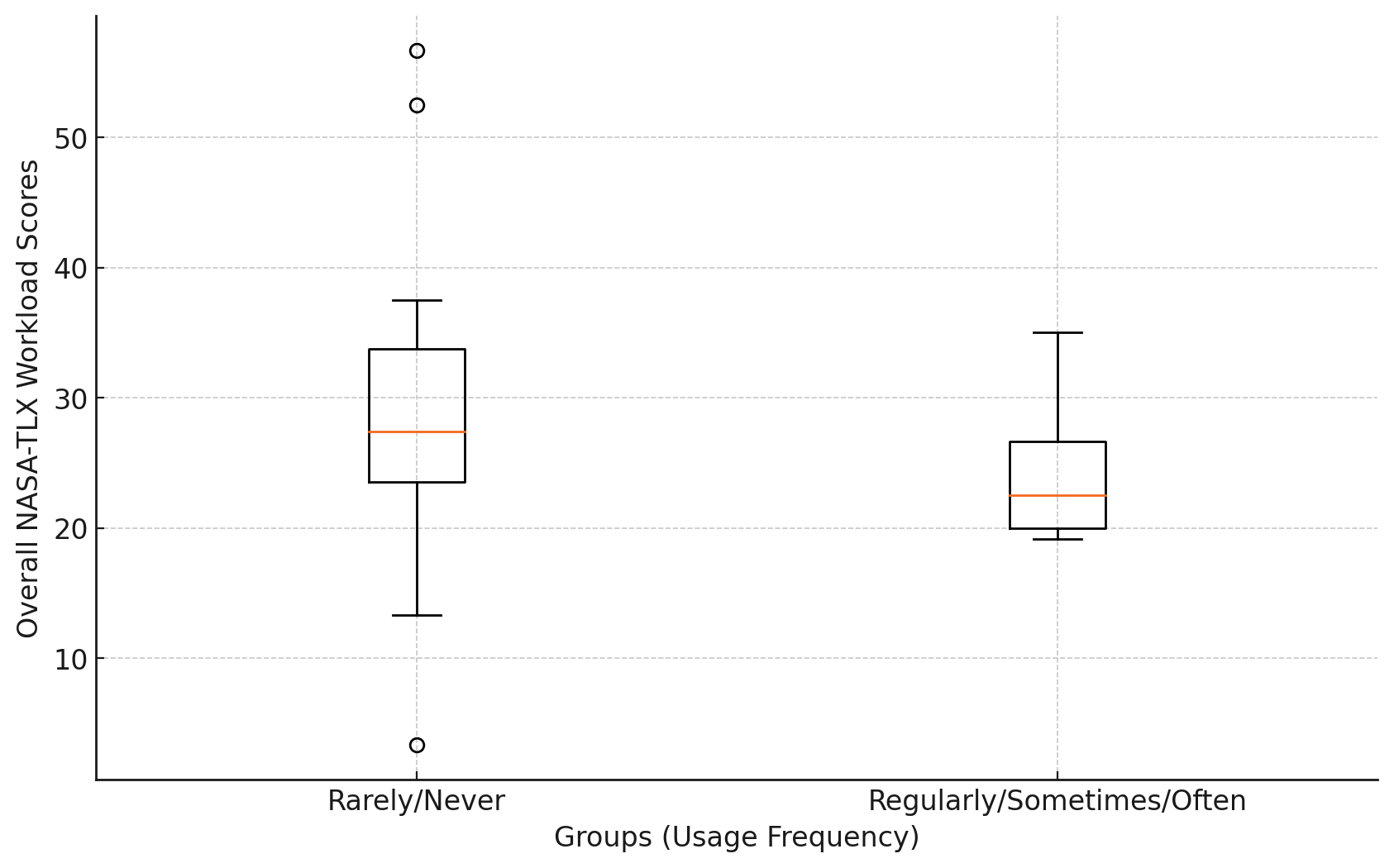}
    \caption{NASA-TLX Workload Score Distribution by Group (Rarely/Never vs. Regularly/Sometimes/Often)}
    \label{fig:nasa_boxplot}
\end{figure}

The boxplot in Figure~\ref{fig:nasa_boxplot} visually compares the overall workload score distributions between the two groups. It illustrates that Group 1 exhibited a wider range of workload scores, with a slightly higher central tendency, suggesting that participants in this group experienced a greater workload.

\subsubsection{Diagnostic Subscores Analysis}

The individual scores for each of the six NASA-TLX dimensions provide detailed insights into specific areas of task demand:
\begin{itemize}
    \item \textbf{Mental Demand}: The average score of \textbf{24.14} indicates that participants found the tasks somewhat mentally demanding, reflecting the cognitive processes involved, including memory recall, problem-solving, and decision-making.
    
    \item \textbf{Physical Demand}: With a mean score of \textbf{16.60}, physical demand was relatively low, suggesting that tasks were not physically strenuous. This aligns with the study's focus on cognitive rather than physical activities.
    
    \item \textbf{Temporal Demand}: The Temporal Demand dimension received a lower mean score of \textbf{8.97}, implying that time constraints were less of a significant factor for participants' perceptions of workload than initially anticipated.
    
    \item \textbf{Performance}: The mean score for Performance was \textbf{78.20}, indicating a high level of participant satisfaction with their own performance. This suggests that participants generally felt successful in achieving the task objectives.
    
    \item \textbf{Effort}: The Effort score, averaging \textbf{22.54}, highlights the subjective exertion required by participants to complete the tasks. Although not overly strenuous, the effort required was notable, particularly in tasks demanding sustained attention and cognitive processing.
    
    \item \textbf{Frustration}: The Frustration score of \textbf{17.20} reflects a relatively low level of frustration, indicating that most participants did not feel significantly frustrated during the tasks. This suggests that the tasks were appropriately designed to avoid excessive stress or confusion.
\end{itemize}

The statistically significant difference in workload scores between Group 1 (\textbf{Rarely/Never}) and Group 2 (\textbf{Regularly/Sometimes/Often}) suggests that usage frequency influences workload perceptions. Participants in Group 1 reported higher workload scores, indicating that less frequent engagement with the system may lead to higher perceived workload, potentially due to unfamiliarity with the system or tasks.

The diagnostic subscores provide additional insights. The high Performance and relatively low Frustration scores imply that while tasks required mental focus and attention, they were achievable, and participants generally felt competent in their completion. However, the moderate Effort scores suggest that completing tasks demanded a certain level of cognitive exertion.

The moderate overall workload score, along with the diagnostic insights, indicates that the system design successfully supports task completion without overburdening users. However, addressing the higher workload perceived by the \textbf{Rarely/Never} group may involve targeted interventions, such as enhanced onboarding or task simplification for infrequent users. Additionally, further refinement of tasks that demand cognitive exertion may help to reduce overall workload, particularly for less experienced users.

These findings provide valuable insights into how users interact with the system and inform future design iterations aimed at optimizing user workload and overall experience.

\section{Presence Questionnaire Analysis}

The Presence Questionnaire is a tool designed to assess the level of immersion and sense of presence that participants feel when interacting with AR environments. It captures various aspects of user experience, such as the perceived integration of virtual objects into the real world, engagement levels, realism, and awareness of the real environment while engaging with virtual content. By evaluating these dimensions, the questionnaire provides insights into the effectiveness of AR systems in creating a believable and immersive experience.

\subsection{Overall Presence Score}

The overall presence score was calculated as the average of all questionnaire items. The mean overall presence score across participants was \textbf{3.13}, suggesting a moderate level of perceived presence within the AR environment. This score indicates that while participants generally felt some level of immersion, there is room for improvement in enhancing the realism and integration of virtual objects into the real world.

\subsubsection{Analysis of Variance (ANOVA)}

To further analyze the data, participants were categorized into two groups based on their frequency of use:
\begin{itemize}
    \item \textbf{Rarely/Never (Group 1)}: Participants who indicated ``Rarely'' or ``Never'' in their usage frequency.
    \item \textbf{Regularly/Sometimes/Often (Group 2)}: Participants who indicated ``Regularly,'' ``Sometimes,'' or ``Often.''
\end{itemize}

An ANOVA test was conducted to determine whether there was a statistically significant difference in overall presence scores between the two groups. The results revealed a statistically significant difference (\textit{F} = 5.67, \textit{p} = 0.02). The mean overall presence score for Group 1 (\textbf{Rarely/Never}) was \textbf{3.00}, while the mean score for Group 2 (\textbf{Regularly/Sometimes/Often}) was slightly higher at \textbf{3.31}, as shown in Table \ref{table:anova_presence}.

\begin{table}[h!]
\centering
\caption{Detailed ANOVA calculations for Presence Questionnaire.}
\label{table:anova_presence}
\begin{tabular}{|l|c|}
\hline
\textbf{Statistic} & \textbf{Value} \\ \hline
Group 1 (Rarely/Never) Mean Score & 3.00 \\ \hline
Group 2 (Regularly/Sometimes/Often) Mean Score & 3.31 \\ \hline
Group 1 (Rarely/Never) Variance & 0.13 \\ \hline
Group 2 (Regularly/Sometimes/Often) Variance & 0.14 \\ \hline
Number of Participants in Group 1 & 20 \\ \hline
Number of Participants in Group 2 & 15 \\ \hline
F-Statistic & 5.67 \\ \hline
P-Value & 0.02 \\ \hline
\end{tabular}
\end{table}

\begin{figure}
     \centering
     \includegraphics[width=0.5\linewidth]{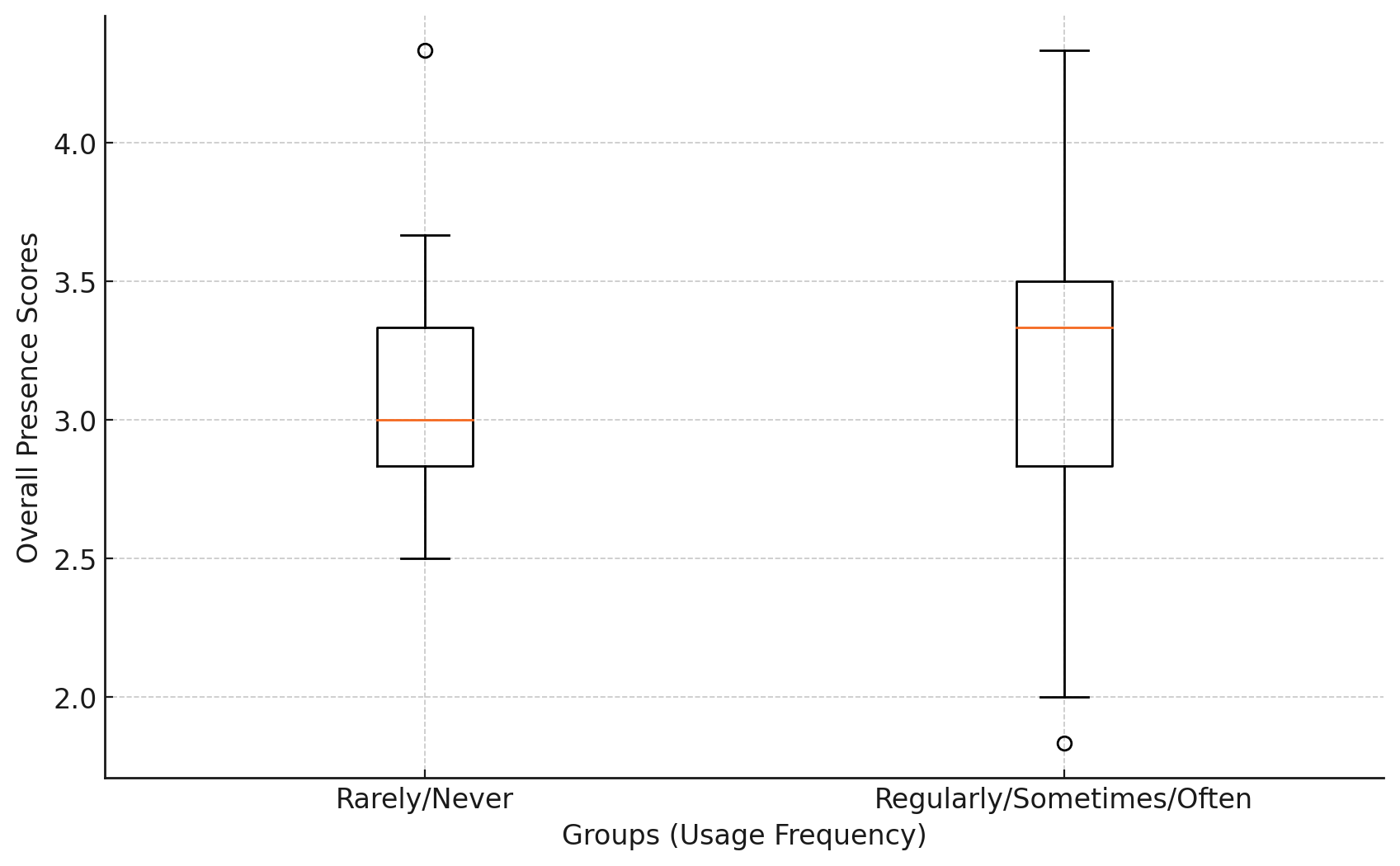}
     \caption{Presence Questionnaire Overall Score Distribution by Group (Rarely/Never vs. Regularly/Sometimes/Often)}
     \label{fig:presence_boxplot}
 \end{figure}

The boxplot in Figure~\ref{fig:presence_boxplot} visually compares the overall presence score distributions between the two groups. It illustrates that Group 2 exhibited a higher central tendency in presence scores, suggesting that participants who interact with the AR environment more frequently perceive a stronger sense of immersion and presence.

The table below summarizes the average scores for each questionnaire item, providing a snapshot of participant responses and highlighting areas where the AR experience could potentially be improved:

\begin{table}[h!]
    \centering
\caption{Presence Questionnaire average scores by item.}
\label{table:presence_scores}
    \begin{tabular}{| p{12cm} | c |}
        \hline
        \textbf{Questionnaire Item} & \textbf{Average Score} \\
        \hline
        Sense of ``Being There'' with Virtual Objects & 3.5 \\
        \hline
        Integration of Virtual Objects into Real World & 3.2 \\
        \hline
        Felt Like Viewing Overlays Instead of Interacting & 2.8 \\
        \hline
        Did Not Feel Fully Engaged with Virtual Elements & 3.0 \\
        \hline
        Sense of Interacting with Virtual Objects as Part of Real Space & 3.5 \\
        \hline
        Feeling Present with Virtual Objects in Real Surroundings & 3.4 \\
        \hline
        Sometimes Unaware of Real Environment While Engaging with Virtual Elements & 2.7 \\
        \hline
        Remained Aware of Real Surroundings While Interacting with AR Content & 3.2 \\
        \hline
        Captivated by Virtual Objects, Losing Focus on Surroundings & 2.9 \\
        \hline
        Perception of Realism in Virtual Objects within Physical Environment & 3.3 \\
        \hline
        Consistency of AR Experience with Real-World Perceptions & 3.1 \\
        \hline
        Virtual Objects Matching User's Expectations & 3.4 \\
        \hline
    \end{tabular}
\end{table}

The statistically significant difference in presence scores between Group 1 (\textbf{Rarely/Never}) and Group 2 (\textbf{Regularly/Sometimes/Often}) indicates that frequency of interaction with the AR system impacts users' sense of presence. Participants in Group 2 reported slightly higher presence scores, suggesting that regular or frequent interaction enhances the immersive quality of the AR experience. 

The questionnaire item analysis reveals that participants generally experienced a moderate sense of ``Being There'' with virtual objects, as well as moderate levels of integration and interaction with virtual elements. However, lower scores in areas such as losing awareness of the real environment and perceiving virtual objects as highly real indicate opportunities for improvement. Enhancing graphical fidelity, spatial integration, and interaction design could strengthen the sense of immersion and presence.

The Presence Questionnaire results highlight the system's strengths in creating a moderately immersive experience while identifying areas for further enhancement. By focusing on improving the realism, consistency, and integration of virtual objects, the system can better align with user expectations and create a more compelling AR environment. These insights can guide future development efforts aimed at optimizing presence and immersion within AR experiences.

\section{Qualitative User Study Results}

\subsection{Key Aspects Found Useful or Enjoyable by Users}
\begin{figure} [ht]
    \centering
    \includegraphics[width=0.5\linewidth]{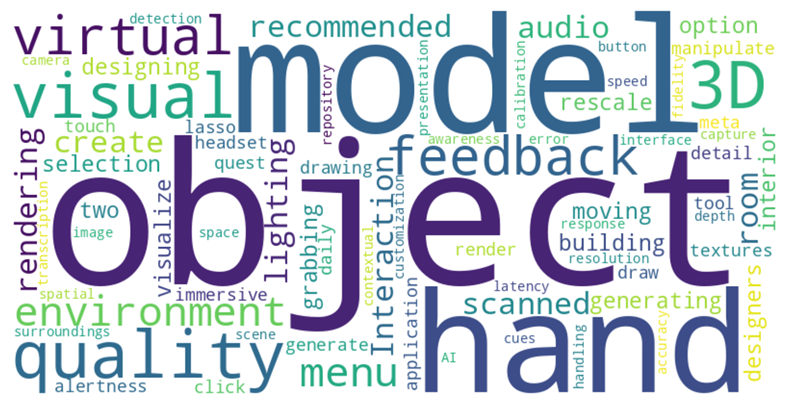}
    \caption{Word cloud illustrating frequently mentioned features and sentiments related to the AR system's usability and immersive qualities.}
    \label{fig:word_cloud}
\end{figure}

Participants highlighted several features that enhanced their experience and facilitated intuitive interaction with the AR system. Specifically, users enjoyed the ease of interaction using simple voice commands and touch gestures. Many participants found the \textit{object manipulation capabilities} (e.g., moving, resizing, rotating objects) particularly satisfying. Additionally, the system's ability to \textit{scan environments and provide contextual object recommendations} was appreciated, with several users recognizing its potential for applications in interior design, prototyping, and other visualization tasks (see Figure~\ref{fig:word_cloud}).

The real-time generation and placement of objects based on user commands were cited as engaging and beneficial, especially in fields that require the visualization of virtual objects within physical spaces, such as architecture and retail. Furthermore, the interaction with the system's voice assistant was commended for making the experience hands-free and easy to use, with minimal learning curve required.

\subsection{Challenges and Difficulties Experienced by Users}
Despite the system’s strengths, participants encountered several challenges that impacted their overall experience. \textit{Latency in generating 3D models} was a common concern, as some users felt that the generation process was slower than expected, disrupting the flow of interaction. Additionally, some users noted that the \textit{rendering quality and realism of 3D objects} needed improvement, mentioning that textures and colors of virtual objects did not always align with their real-world expectations.

Participants also encountered difficulties with the \textit{interface for selecting options and interacting with virtual menus}, especially in terms of accurately clicking or selecting objects. The \textit{depth perception and alignment} of virtual objects within the real-world environment was also challenging for some users, occasionally leading to a reduced immersive experience. For new users or those unfamiliar with AR technology, \textit{navigation and object scaling controls} felt unintuitive and difficult to handle. A few participants expressed confusion regarding the system’s status, particularly when waiting for commands to process, which led to unclear moments of system feedback.

\subsection{Suggested Improvements}
To enhance usability and overall user experience, participants provided several suggestions:
\begin{itemize}
    \item \textbf{Improving Visual and Audio Feedback}: Adding visual indicators and audio cues to signal when the system is processing commands or generating models would improve user awareness and reduce moments of confusion.
    \item \textbf{Enhancing 3D Model Quality}: Participants suggested improving textures and increasing the resolution of 3D models to make virtual objects appear more realistic, enhancing immersion.
    \item \textbf{Reducing Latency}: Optimizing the system to decrease response and loading times would allow for smoother, more immediate interactions.
    \item \textbf{Simplifying Interface Navigation}: Adjustments such as larger selection buttons or voice commands to hide or reveal interface elements could streamline user control and make the system more accessible.
    \item \textbf{Improving Depth and Placement Adjustments}: Enhancing the system’s ability to accurately align virtual objects with real-world elements was recommended. Implementing physics-based adjustments would make object placement more natural within the environment.
\end{itemize}

\subsection{Potential Applications and Contexts for Use}
Most participants agreed that the system holds significant potential in \textit{professional contexts}, particularly in fields such as \textit{interior design, architecture, and engineering}. The ability to visualize designs or layouts within a real-world setting was identified as valuable for industry professionals. Additionally, several participants highlighted the system's usefulness for \textit{educational and training purposes}, where immersive visualization can aid in learning within hands-on environments.

Some participants noted its potential in \textit{retail applications}, where customers could view products in their own spaces before making purchase decisions. However, a few users indicated that while the system was beneficial in professional settings, it may be less practical for casual or personal use in its current form, citing the technological learning curve and occasional latency as barriers.

\subsection{Additional Features Suggested by Users}
Participants provided suggestions for additional features that could improve functionality:
\begin{itemize}
    \item \textbf{Model Preview Option}: Allowing users to preview generated objects before fully rendering them within the AR environment could improve interaction flow.
    \item \textbf{Object Categorization and Library of Pre-Loaded Models}: A categorized library of pre-generated models for quicker access would reduce reliance on real-time generation and minimize latency.
    \item \textbf{Enhanced Editing Tools}: Adding more editing options for objects post-creation, such as adjusting size, orientation, and textures, would enable users to customize their AR environment more effectively.
    \item \textbf{Improved Voice Command Capabilities}: Extending voice commands to support multi-step tasks and more detailed object descriptions could reduce the need for extensive menu navigation.
\end{itemize}

These qualitative insights from the user study provide a balanced view of the AR system's strengths in promoting intuitive, interactive 3D model generation and its limitations, which could benefit from further refinement. Implementing these user recommendations may significantly enhance the system’s usability, making it more accessible and effective for a wide range of applications.

% ===========================Evaluation and results

      \chapter{Performance Evaluation and Results} \label{ch:Performance Evaluation and Results}
      
\section{Speech-to-3D Subsystem Performance Evaluation}

To comprehensively assess the effectiveness and performance of \textit{Matrix}  framework for real-time 3D object generation in AR environments, we defined key performance metrics that spanned system performance.
These metrics allowed us to evaluate the technical efficiency of the framework.
To validate these metrics, we conducted a user study with 35 participants, each interacting with the system to generate, modify, and manipulate 3D objects in AR. 

The evaluation was split into multiple phases, focusing on task performance, GPU utilization, system resource consumption, and model accuracy. By analyzing these aspects in detail, we aimed to understand how efficiently the system operates.

\subsection{Tasks}

Participants were asked to complete a set of tasks that highlighted various features of \textit{Matrix} framework.
The tasks were designed to test ease of interaction, system responsiveness, and overall user experience.
The tasks included:

\begin{description}
\item[Task 1:]
Generate a simple 3D object using a voice command.
This task was meant to test the core functionality of the speech-to-3D conversion process, including how well the system transcribes the user's speech and generates an object based on that input.
\item[Task 2:]
Retrieve pre-generated objects from the repository.
This task was designed to evaluate how efficiently the system retrieves objects from the repository, and whether this retrieval is faster than generating objects from scratch.
\item[Task 3:]
Customize object placement and size in the AR environment.
This task required participants to manipulate the generated objects in real-time, testing the interaction capabilities of the system, including how intuitive and responsive the object manipulation felt to the users.
\end{description}

\subsection{Task Performance Evaluation}

We analyzed the performance of the tasks using four key metrics: task completion time, success rate, error rate, and system responsiveness.
Each of these metrics provided insights into different aspects of the system's usability.

\begin{description}
\item[Task Completion Time:]
This metric measured the average time it took participants to complete each task, including speech recognition, object generation, and rendering in AR.
Shorter times indicate higher system efficiency.
AR-based systems have been shown to significantly reduce task completion times by providing real-time guidance~\cite{RajSub2023}. 
\item[Task Success Rate:]
This metric evaluated how often participants successfully completed tasks on their first attempt.
The small failure rate could be attributed to speech transcription errors or difficulty in object manipulation, which we aim to improve in future iterations.
Similar usability studies have demonstrated that AR environments often face these challenges, impacting task completion rates due to multimodal input challenges like speech errors and object recognition~\cite{ErrorRate2013}.
\item[System Error Rate:]
This metric tracked the number of errors encountered per task, such as incorrect object generation or speech-to-text errors.
With a low error rate per task, the system demonstrates a relatively low occurrence of issues during interaction, further showcasing its robustness~\cite{ErrorRate2013}.
\item[System Responsiveness:]
This refers to the time between the user's speech input and the rendering of the 3D object in the AR environment.
A faster response time is critical in ensuring smooth user interactions, especially when multiple tasks are performed consecutively.
This is in line with the requirements for efficient human-computer interaction in AR systems, where rapid feedback is essential for maintaining task flow and user engagement~\cite{Bi2023MISARAM}.
\item[Mesh File Size:]
This metric measures the file size of the generated 3D object, both for the default Shap-E model output and the reduced mesh size.
Reducing the number of vertices and faces can significantly decrease file size, which is crucial for improving system performance in AR environments with limited resources, as it reduces load times.
Compression techniques have been shown to effectively minimize file sizes for storage and transmission without compromising the quality of the reconstructed mesh~\cite{Siddeq2016}.
\end{description}

\begin{figure}
\centering
\includegraphics[width=0.5\linewidth]{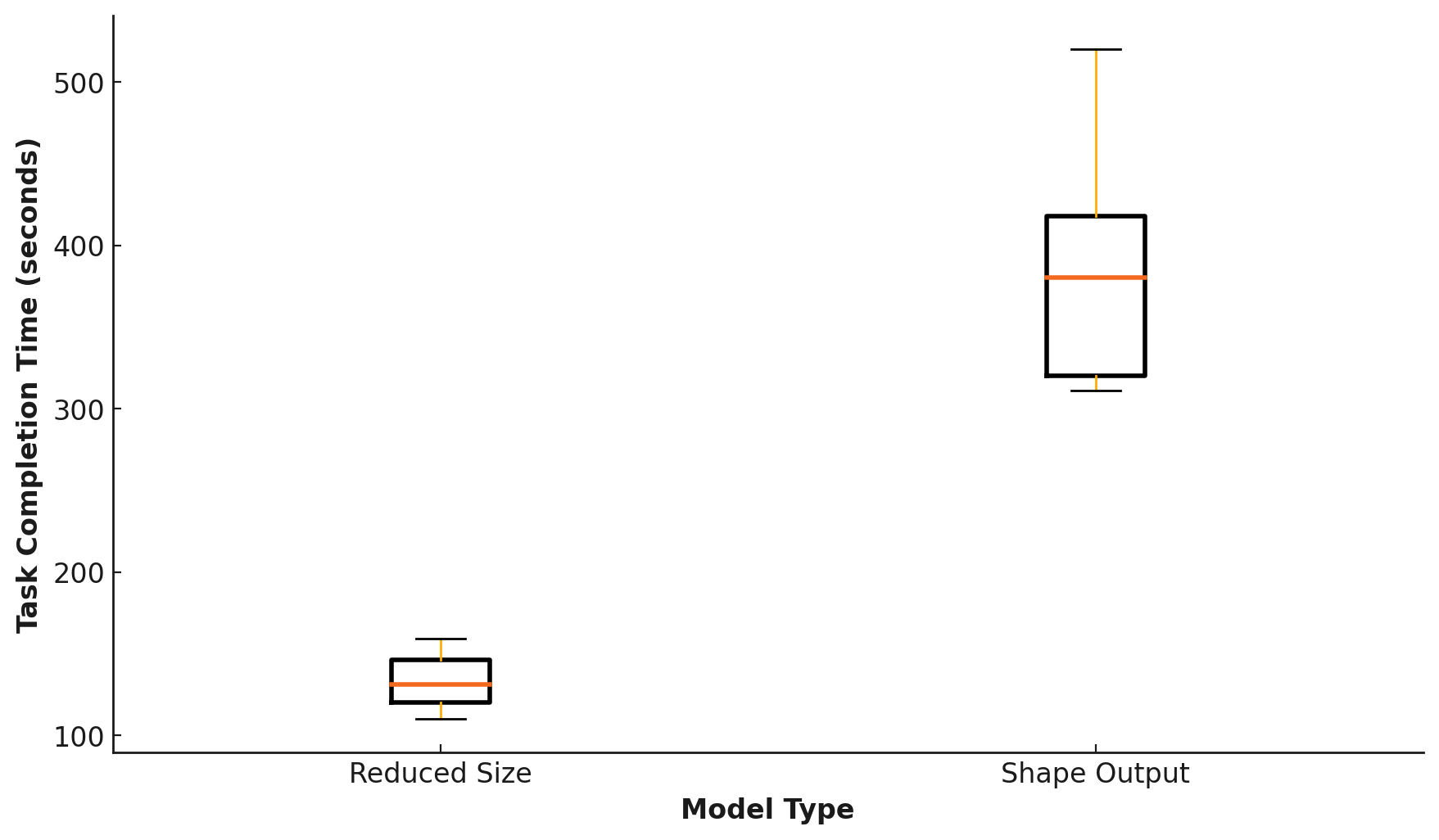}
\caption{Task completion time for the Reduced Mesh Size model compared to the Shape Output model.}
\label{fig:Task_completion}
\end{figure}

The task performance evaluation demonstrates the system's overall efficiency and usability, as summarized in \cref{table:task_performance_reduced_mesh}.
As shown in \cref{fig:Task_completion}, the Task Completion Time is significantly reduced when using the Reduced Mesh Size model compared to the default Shape Output model.
The median task completion time for the Shape Output model was approximately 386 seconds, whereas the Reduced Size model achieved a much lower median time of around 122.4 seconds.
This drastic reduction in task completion time highlights the importance of optimizing file size for faster processing in AR environments and demonstrates the system's enhanced performance, making it more suitable for real-time AR applications that require rapid processing.

\begin{table} [ht]
\centering
\caption{Task performance with default Shap-E model output and reduced mesh size.}
\label{table:task_performance_reduced_mesh}
\begin{tabular}{lcc} 
\toprule
\textbf{Metric} & \textbf{Shap-E Output} & \textbf{Reduced Size} \\ 
\midrule
\small Task Completion Time (seconds) & \small 386.2  & \small 122.4  \\ 
\small Task Success Rate (\%) & \small 76\% & \small 88\% \\ 
\small System Error Rate (errors/task) & \small 0.15  & \small 0.14  \\ 
\small System Responsiveness (seconds) & \small 282.73  & \small 74.32  \\ 
\small Mesh File Size (MB) & \small 2.5  & \small 0.16  \\ 
\bottomrule
\end{tabular}
\end{table}

With a success rate of 88\% for the reduced mesh size option, the system shows promising improvements in user interactions and task accuracy.
The error rate remains consistently low across both models, indicating robustness in object generation and speech-to-text conversion.
Additionally, system responsiveness is considerably enhanced with mesh simplification, reducing response times and ensuring smoother real-time interactions.
By reducing the mesh file size to just 0.16 MB, the system not only improves load times but also becomes more resource-efficient, ensuring smoother user experiences in environments with limited computational power.

\subsection{GPU and System Performance}

To better understand how \textit{Matrix} framework uses system resources, we evaluated the GPU and system performance across multiple metrics.
These evaluations are particularly important because real-time 3D object generation in AR environments can be computationally expensive, particularly when it involves complex models.
The following metrics were used to assess performance:

\begin{description}
\item[Average GPU Utilization:]
This metric measures the percentage of GPU resources used during object generation and rendering.
Reducing GPU load in resource-constrained environments is essential for maintaining system responsiveness.
Optimizing GPU usage improves performance and ensures smooth operation during high-computation tasks like rendering and object generation~\cite{Jin2023S3IG}.
\item[GPU Memory Consumption:]
This tracks the amount of GPU memory consumed during the object generation process, particularly when handling multiple or complex 3D objects.
Excessive memory consumption can slow down the system and lead to performance bottlenecks, especially when rendering large models in real-time.
Efficient memory usage is crucial for maintaining optimal performance, as high GPU consumption is often a key challenge in 3D object detection and generation tasks~\cite{liu2019pointvoxel}.
\item[Time to Object Generation:]
This metric measures the total time taken by the Shap-E model to generate a 3D object from the input text.
Reducing this generation time is crucial for improving system efficiency, as faster processing enhances user experience and reduces task completion times in AR environments~\cite{9352121}.
\end{description}

We also conducted an experiment to evaluate how using a pre-generated object repository impacts system performance.
Two scenarios were tested:

\begin{description}
\item[Without Repository (Full Generation):]
All 3D objects were generated from scratch.
\item[With Repository (Partial Retrieval:]
The system retrieved objects from a pre-existing repository, avoiding the need for full generation.
\end{description}

The impact of using the repository is summarized in \cref{table:gpu_performance}, where we observe a significant reduction in both GPU utilization and memory consumption when objects were retrieved from the repository.
For example, average GPU utilization dropped from 54\% to 31\%, and GPU memory consumption decreased from 6.1 GB to 2.3 GB.
These results highlight the importance of optimizing system resources by utilizing pre-generated objects, which also reduced the time to object generation from 29.60 seconds to 16.32 seconds.

\begin{table} [ht]
\centering
\caption{GPU and system performance (with and without repository).}
\label{table:gpu_performance}
\begin{tabular}{lcc} 
\toprule
\textbf{Metric} & \textbf{No Repo} & \textbf{With Repo} \\ 
\midrule
Average GPU Utilization (\%) & 54\%  & 31\%  \\ 
GPU Memory Consumption (GB) & 6.1  & 2.3  \\ 
Time to Object Generation (seconds) & 29.60  & 16.32  \\ 
\bottomrule
\end{tabular}
\end{table}

\subsection{Model Accuracy}

Finally, we measured the accuracy of the AI models used in \textit{Matrix} framework, focusing on speech recognition and object generation. Accurate model performance is essential for delivering a smooth and error-free user experience.
\cref{table:model_accuracy} summarizes these accuracy metrics.

\begin{description}
\item[Speech-to-Text Accuracy:]
This metric evaluates the rate at which the system correctly converts spoken commands into text. 
\item[Object Recognition Precision:]
This measures how accurately the system identifies and suggests relevant objects based on the transcribed text. 
\item[Object Rendering Fidelity:]
This metric compares the accuracy of the generated 3D objects with the user’s original spoken command.
\end{description}

\begin{table} [ht]
\centering
\caption{Summary of AI model accuracy metrics for speech recognition, object recognition with LLM and object generation.}
\label{table:model_accuracy}
\begin{tabular}{lcc} 
\toprule
\textbf{Metric} & \textbf{Value} \\ 
\midrule
Speech-to-Text Accuracy (\%) & 91\% \\ 
Object Recognition Precision (\%) & 85\% \\ 
Object Rendering Fidelity (\%) & 87\% \\ 
\bottomrule
\end{tabular}
\end{table}

The system demonstrates strong performance in speech recognition with a 91\% accuracy rate, indicating reliable understanding of user commands, though there is room for improvement in handling diverse accents and noisy environments.
With an object recognition precision of 85\%, \textit{Matrix} framework accurately identifies and suggests relevant objects based on user input, showing a high level of contextual understanding.
A rendering fidelity of 87\% reflects the system’s overall effectiveness in generating 3D objects that align with user expectations.
However, there is potential for further enhancement, especially in complex objects.

\subsection{Comparison with Similar Works}

Our AI-powered \textit{Matrix} framework for real-time 3D object generation is compared with Dream Mesh: A Speech-to-3D Model Generative Pipeline in Mixed Reality~\cite{Weng2024}.
Both approaches generate 3D models from speech, but key differences exist in areas such as generation time, consistency, hardware requirements, and additional interactive features.

\begin{description}
\item[Model Generation Time:]
Dream Mesh uses the DreamFusion model, which takes 30--40 minutes to generate 3D models due to 10,000 iterations~\cite{Weng2024}.
In contrast, \textit{Matrix}  framework, using Shap-E, reduces generation time to under 50 second by optimizing for AR environments with mesh simplification.
\item[Model Consistency:]
Dream Mesh faces challenges with inconsistent results when generating the same object from identical prompts~\cite{Weng2024}.
Our system improves on this by implementing a Semantic search through a vector database and retrieval, allowing reuse of previously generated objects, ensuring better consistency.
\item[Hardware Requirements:]
The two systems also differ in hardware demands.
Dream Mesh relies on a high-performance desktop setup with Nvidia 3090Ti GPU with 24 GB of VRAM, offering more power with increased VRAM, CUDA cores, Tensor cores, and memory bandwidth.
\textit{Matrix}  framework, however, is optimized for more resource-constrained environments, functioning efficiently on mid-range GPUs like the Nvidia Tesla T4 with 16GB of VRAM, providing broader device support.
\item[Additional Features:]
\textit{Matrix}  framework offers several interactive features not present in Dream Mesh.
These include a LLM-based object recommender that suggests relevant objects based on user input, multilingual support for generating 3D objects from speech in multiple languages, and system speech feedback, which provides real-time verbal responses to users to enhance the interactivity and user experience in AR environments.
These features make \textit{Matrix}  framework more adaptable and user-friendly, especially in diverse, multilingual settings.
\end{description}

\begin{table} [ht]
\centering
\caption{Comparison of \textit{Matrix} vs. Dream Mesh.}
\label{table:comparison_framework_dreammesh}
\renewcommand{\arraystretch}{1.5} % Adjust this value for more or less spacing
\begin{tabular}{p{2.5cm}p{2.5cm}p{2cm}} 
\toprule
\textbf{Feature} & \textbf{\textit{Matrix}} & \textbf{Dream Mesh} \\ 
\midrule
Model Generation Time & \textless{} 50 Seconds & 30-40 minutes \\ 
Model Consistency & Consistent via reuse of objects & Varies with same prompt \\ 
GPU Requirements & Nvidia Tesla T4, 16GB VRAM & Nvidia 3090Ti, 24GB VRAM  \\ 
% Device Support &  HoloLens 2 & Meta Quest Pro\\ 
\bottomrule
\end{tabular}
\end{table}

This comparison highlights the improved efficiency and hardware flexibility of our system, making it more suitable for a wider range of AR applications.
 
% ==========================VLM Eval

\section{Context-Aware Object Recommendation Evaluation}

We present a comprehensive evaluation of the proposed AI system for dynamic real-time context-aware content creation in AR.
The evaluation focuses on key metrics such as accuracy, relevance of recommendations, responsiveness, computational efficiency, error rates, diversity of suggestions, reliability, and adaptability to different environments. 

\subsection{Accuracy of Object and Environment Detection and Recognition}

We evaluated the VLM for object and environment detection using a subset of the COCO 2017 dataset~\cite{lin2014microsoft} and real-world AR testing. We selected a balanced subset of 150 images from the COCO 2017 train and validation sets, which represent diverse indoor and outdoor environments.
These images were annotated with ground truth labels provided in the COCO 2017 annotations.
Additionally, we conducted tests using the HoloLens in 12 different locations to assess the VLM's real-time performance in dynamic settings.

To assess the VLM's contextual understanding and recommendation capabilities, we utilized prompts that asked the model to generate concise descriptions of the location, including both the name and a short description of the environment based on the visual inputs it received.
This allowed us to test the VLM's ability to interpret the environment and provide useful textual outputs for AR applications.

The evaluation focused on key metrics such as precision, recall, and F1-score, as summarized in \autoref{vlm_metrics}.

\begin{table} 
\centering
\caption{VLM performance metrics in AR environments.}
\label{tab:vlm_metrics}
\resizebox{\columnwidth}{!}{%
\begin{tabular}{lccc}
\toprule
\label{vlm_metrics}
\textbf{Environment} & \textbf{Precision (\%)} & \textbf{Recall (\%)} & \textbf{F1-Score (\%)} \\
\midrule
Indoor & 92.5 & 88.0 & 90.2 \\
Outdoor & 90.0 & 85.0 & 87.4 \\
\midrule
\textbf{Average} & \textbf{91.3} & \textbf{86.5} & \textbf{88.8} \\
\bottomrule
\end{tabular}%
}
\end{table}

The VLM demonstrated strong performance in both indoor and outdoor environments. High precision and recall values indicate effective object detection and recognition, while the F1-score confirms the model's reliability.

By integrating these contextual description prompts into the evaluation, we highlighted the VLM's capability to understand and interpret both visual and textual information, providing context-aware recommendations that enhance user experience in AR.

\subsection{System performance metrics}

The evaluation focuses on key system metrics such as Average Response Time, GPU Utilization, and Memory Consumption during the object recommendation and creation processes.

The Average Response Time is a critical metric for real-time applications.
It measures the time from when the user captures an image to the display of the recommended and generated objects in the AR environment.
The system achieved an average response time of 82 seconds.
This indicates that our system can provide reasonable object recommendations and generate corresponding 3D models swiftly, ensuring a seamless user experience, which is crucial for immersive AR interactions.
In contrast, Dream Mesh, which uses the DreamFusion model, takes 30--40 minutes just to generate a single 3D object~\cite{Weng2024}.

Efficient utilization of computational resources is essential to maintain performance without overloading the hardware.
The GPU Utilization was monitored during the object recommendation and 3D model generation processes.
The system exhibited an average GPU utilization of 68\%, with peaks reaching 85\% during intensive text-to-3D model generation phases.
This demonstrates that the system effectively leverages GPU capabilities to handle computationally demanding tasks without causing bottlenecks or significant slowdowns.
\autoref{tab:metrics} summarizes the key performance metrics of the system.

\begin{table} 
\centering
\caption{Summary of key performance metrics.}
\label{tab:metrics}
\begin{tabular}{lcc}
\toprule
\textbf{Metric} & \textbf{Average} & \textbf{Peak} \\
\midrule
Average Response Time & 82 seconds & 89 seconds\\
GPU Utilization & 68\% & 85\% \\
\bottomrule
\end{tabular}
\end{table}

The integrated evaluation of these metrics confirms that our system efficiently handles object recommendation and creation in real-time AR applications.
The acceptable response times and resource utilization ensure that users can interact with the system seamlessly, without noticeable delays or performance issues.

\subsection{Evaluation of Recommendation Accuracy and Quality}
  
To assess the effectiveness of our proposed system for dynamic, context-aware content creation in AR, we conducted a pilot study, focusing on two primary aspects: contextual appropriateness of recommendations and the diversity and novelty of recommendations.

We conducted this pilot study with nine participants who interacted with our AR system in various real-world environments such as living rooms, offices, and outdoor spaces.
Each participant used the system to receive object recommendations tailored to their immediate surroundings and preferences.
The recommendations were generated based on the following prompt: ``\textit{As a designer, recommend 5 simple objects (name, color, shape, and suggest a location for each object within this space relative to the other objects in the picture) that would be suitable for this place but are currently not present.}''

Participants rated the contextual fit of the recommendations, assessing how well the suggested objects matched their environment.
Additionally, we tracked the occurrence of duplicate object recommendations, which refer to objects that already exist in the user's environment but are also recommended by the system's VLM.
The average contextual fit rating was 4.1 out of 5, and the average rate of duplicate object recommendations was 13\%, indicating the system occasionally recommended objects that were already present in the user’s environment.
\autoref{tab:contextual_fit} summarizes the key performance metrics related to the contextual fit and duplicate object recommendations.

\begin{table}
\centering
\caption{Summary of Contextual Fit and Duplicate Recommendations metrics.}
\label{tab:contextual_fit}
\begin{tabular}{l c}
\toprule
\textbf{Metric} & \textbf{Value} \\
\midrule
Contextual Fit Rating (out of 5) & 4.1 \\
Duplicate Object Recommendation Rate (\%) & 13 \\
\bottomrule
\end{tabular}
\end{table}

In addition to evaluating contextual fit, we analyzed the variety and novelty of recommendations provided to each participant over multiple sessions.
The Diversity Index was calculated using Shannon Entropy to quantify the diversity of recommended objects. 

Diversity of Recommendations measures how varied the objects recommended by the system are.
If the system repeatedly recommends the same objects, the entropy (diversity) will be low.
On the other hand, if the system suggests a wide variety of different objects, the entropy will be higher.
This ensures that the system provides unique and fresh recommendations across different sessions, maintaining user engagement.

Shannon Entropy ($H$) is calculated using \autoref{eq:entropy}~\cite{zhao_fairness_2024}:
\begin{equation}
H(X) = - \sum_{i=1}^{n} p(x_i) \log p(x_i)
\label{eq:entropy}
\end{equation}

\noindent where $X$ is the set of possible objects recommended by the system;
$p(x_i)$ is the probability of each object $x_i$ being recommended; and $\log$ is the logarithm (base 2 or natural logarithm, depending on the context).

This formula calculates the amount of ``uncertainty'' or ``diversity'' in the system’s recommendations.
A high entropy value indicates that the system frequently recommends different objects, whereas a low value indicates repetitive suggestions.

Along with the Diversity Index, Novelty Scores were assigned based on the frequency of unique recommendations that the user had not previously received. These metrics ensured that the system avoided repetitive suggestions, thereby enhancing the user experience.

The average Diversity Index across sessions was 0.72, indicating a moderate level of variety in the recommendations.
The Novelty Score averaged 0.59, demonstrating that the system consistently provided new, non-repetitive suggestions to users.
\autoref{tab:contextual_fit} summarizes Diversity and Novelty metrics scores.

\begin{table} 
\centering
\caption{Summary of Diversity and Novelty metrics.}
\label{tab:diversity_metrics}
\begin{tabular}{lcc}
\toprule
\textbf{Metric} & \textbf{Average Score}   \\
\midrule
Diversity Index & 0.72   \\
Novelty Score & 0.59  \\
\bottomrule
\end{tabular}
\end{table}

The results of this pilot study demonstrate that our system effectively generates contextually appropriate object recommendations in AR environments.
The recommendations, which were based on the prompt asking designers to suggest five simple objects (including their name, color, shape, and ideal placement within the space), were rated highly for contextual fit.
This highlights the system's ability to understand spatial layouts and recommend objects that not only complement the existing environment but also fill gaps in aesthetic or functional design.

The high contextual fit scores suggest that integrating a VLM with text-to-3D generative AI models successfully tailors suggestions to users' surroundings.
The diversity and novelty metrics further indicate that the system avoids repetition and maintains user engagement by introducing varied and unique content~\cite{castells_novelty_2013,zhao_fairness_2024}.
This is crucial for enhancing the immersive experience in AR and ensuring long-term user satisfaction.

% =================== Evaluation image to 3d

\section{Image-to-3D Subsystem Performance Evaluation}
To assess the performance of the Image-to-3D subsystem within the Matrix framework for real-time 3D object generation in AR environments, we evaluated two main phases: \textit{image processing for object detection} and \textit{image-to-3D conversion}. These phases were analyzed for their time efficiency and GPU usage, which are critical metrics in determining the system’s suitability for real-time AR applications.

\subsection{Key Metrics}

\begin{itemize}
    \item \textbf{Image Processing for Object Detection Time}: This metric measures the time required to detect objects within a captured image, identify and list their names, and crop the objects in the image as necessary before proceeding with 3D conversion. Efficient object detection is crucial to minimizing delays in real-time AR interactions.

    \item \textbf{Image-to-3D Conversion Time}: This metric measures the time taken to convert a detected object from a 2D image into a 3D model that can be rendered in an AR environment.

 \item \textbf{Vertices Reduction and Rendering Time}: This metric captures the time required to simplify the 3D model by reducing the number of vertices for efficient storage, download, and rendering. It includes:
    \begin{enumerate}
        \item \textbf{Model Simplification Time}: The time to process and reduce the 3D model's complexity while maintaining visual fidelity.
        \item \textbf{Load and Render Time}: The duration to download the reduced 3D model and render it in the AR environment. Optimizing this metric is critical for reducing latency and ensuring smooth user interactions, particularly in resource-constrained or mobile AR scenarios.
    \end{enumerate}

    \item \textbf{GPU Utilization}: GPU resource usage during the object detection and 3D conversion phases. Reduced GPU load is crucial in AR applications with limited hardware resources.
    
\end{itemize}

\subsection{Performance Summary}

Table \ref{tab:performance_metrics} summarizes the performance metrics for the Image-to-3D subsystem, with separate metrics for image processing (object detection) and image-to-3D conversion.

\begin{table}[h!]
\centering
\caption{Performance metrics for Image-to-3D subsystem.}
\label{tab:performance_metrics}
\begin{tabular}{@{}lcc@{}}
\toprule
\textbf{Metric} & \textbf{Measured Value} \\ \midrule
Image Processing for Object Detection Time (s) & 5.2 \\
Image-to-3D Conversion Time (s)                & 43.2  \\
Model Simplification Time (s)                  & 9.1   \\
Load and Render Time (s)                       & 10.3  \\
Average GPU Utilization (\%)                   & 61\% \\
GPU Memory Consumption (GB)                    & 6.8 \\ \bottomrule
\end{tabular}
\end{table}

The evaluation of the Image-to-3D subsystem highlighted the benefits of optimized object detection and mesh size reduction during image-to-3D conversion. These improvements contribute to a more resource-efficient framework, ensuring smoother user interactions in AR environments with limited computational resources.

% ==================================Discussion 
    \chapter{Discussion} \label{ch:discussion}
Integrating generative AI for real-time 3D model generation in AR revolutionizes various industries by making AR more accessible and engaging. Our approach enables users to create and interact with 3D content without specialized skills, fostering innovation and creativity. This chapter discusses diverse applications, enhanced user experiences, system requirements, user involvement, and future research directions to improve the technology's capabilities and user-friendliness.

\section{System Requirements}

To ensure optimal performance for real-time 3D model generation in AR environments using our developed system, the recommended hardware specifications for each subsystem are as follows:

Graphics Processing Unit (GPU): Compatible with CUDA Version 12.3 and equipped with 16 GB of memory.
Central Processing Unit (CPU): \texttt{x86\_64} architecture, 8 cores, including a 512 KiB L1 Data Cache (8 instances) and a 4 MiB L2 Cache (8 instances).
Memory (RAM): A minimum of 32 GB.
Operating System: Compatible with Linux distributions for API development and Windows 10 or higher for AR environment development.
AR Development Software: Unity Version 2022.3.
AR Headset: HoloLens 2 for sufficient display resolution, built-in sensors for environment mapping and object recognition, and support for interactive hand-tracking.
These system requirements ensure that the generative AI models perform efficiently, providing real-time 3D model generation and seamless integration into AR environments. This setup facilitates high-quality, responsive experiences necessary for innovative applications in gaming, education, retail, and more.

\section{Applications Across Industries}

The potential applications of our approach span multiple industries, including gaming, education, retail, interior design, and beyond.

\textbf{Education}

Interactive 3D models can enhance learning by providing students with a more engaging way to explore complex subjects. For instance, students can visualize historical artifacts, anatomical structures, or scientific phenomena in 3D, making abstract concepts more tangible and understandable. Additionally, this approach can be further enriched by integrating it with other educational research areas, such as the study conducted on educating high school students and aspiring entrepreneurs on Controlled Environmental Agriculture principles through AR \cite{Bigonah2024}, making the learning experience even more captivating and effective. 

\textbf{Retail}

Virtual product demonstrations can improve customer experience and drive sales by allowing customers to visualize products in their own space before making a purchase. This can reduce return rates and increase customer satisfaction by providing a more accurate representation of products.

\textbf{Gaming}

Players can create and customize in-game assets, adding a new level of personalization and engagement to their gaming experience. The ability to generate unique 3D content in real-time opens up new possibilities for user-generated content, fostering a vibrant and creative community.  

\textbf{Interior Design}

The ability to convert real-world objects into 3D models and visualize them in an AR environment is a game-changer for interior design. Designers can create, modify, and preview furniture and decor items in real-time, providing their clients with a realistic representation of their envisioned space.

\section{Enhancing User Experience Through Generative AI}
The integration of generative AI into AR environments is transforming the way users interact with digital and physical spaces. By focusing on user experience, democratizing technology, and involving users directly in the process, our approach fosters a richer and more inclusive AR landscape.

\textbf{Impact on User Experience} The integration of generative AI for real-time 3D model generation significantly enhances user experiences in AR environments. By allowing users to create detailed 3D objects from real-world images or objects, our approach offers a more immersive and interactive experience. Users can interact with the AR environment more intuitively, leading to a richer and more dynamic experience. The ability to visualize and manipulate 3D objects in real-time makes AR applications more compelling and useful, providing a seamless blend of the physical and digital worlds.

\textbf{Democratization of 3D Model Creation} One of the significant contributions of our research is the democratization of 3D model creation. By making this technology accessible to a broader audience, we enable users from various backgrounds to generate and interact with 3D content. This democratization fosters creativity and innovation, allowing individuals and small businesses to leverage AR without requiring extensive technical expertise or resources. The accessibility of 3D model generation empowers users to explore new ideas and applications, breaking down barriers that previously limited the use of AR technologies to those with specialized skills. This broadens the scope of AR applications and encourages widespread adoption across different sectors.

\textbf{User Involvement in the Process} A key aspect of our approach is the involvement of users in the 3D model generation process. By incorporating interactive zone selection and object verification steps, users have greater control over the objects being converted. This user-centric approach not only enhances the accuracy of the generated models but also ensures that the final output meets the user’s specific needs and preferences. By actively engaging users in the process, we improve the relevance and quality of the 3D models, leading to a more satisfying and effective AR experience.

These enhancements will not only extend the functionality of our framework but also deepen the integration of AI technologies in everyday applications, pushing the boundaries of what is possible in XR and beyond. By continuing to innovate and expand the capabilities of our framework, we aim to set new standards for interaction and customization in XR environments.

 \section{Challenges} \label{se:Challenges}
 In the pursuit of developing robust and user-friendly AR applications, several key challenges persist in the domains of speech-to-3D translation, image-to-3D conversion and context aware recommendations. These challenges stem from the limitations of current technologies and the complexities inherent in real-world applications. Previous works have made significant strides in leveraging LLMs and AI-driven models for these tasks, yet gaps remain. 
 Addressing these challenges is crucial for advancing the practical implementation and enhancing the user experience in AR systems.

\subsection{Efficient GPU Usage and object Consistency through a Pre-Generated Objects Repository}

The high costs and extensive computational demands of GPUs present a major challenge in deploying sophisticated AR systems.
To address this, we provide a strategically developed menu that displays objects related to the user's request which are already available in a repository.
This repository contains a collection of 3D models that were previously generated based on earlier user requests, as well as other pre-generated models that could be relevant.
By allowing users to select from existing models rather than generating new ones in real-time, we not only decrease the GPU workload but also tackle the challenge of obtaining consistent results with diffusion-based models, where users might experience varying outcomes even with the same prompt.

This approach enables users to reuse previously generated objects, enhancing predictability and reliability.
As a result, the rendering process speeds up, energy consumption is reduced, and overall operational costs of running GPU-intensive tasks are slashed.
The use of a pre-generated repository thus serves as a multi-faceted solution—improving system efficiency while ensuring consistency and providing a rich, responsive user experience in the AR environment.

\subsection{Generative AI Model Output Size and Latency on AR Devices}

 One of the significant challenges with generative AI models is the large output size of the generated 3D objects, which can result in latency issues on AR devices.
 For example, even relatively simple objects can have file sizes as large as 2.5MB, which leads to extended download and loading times often exceeding 5 minutes on devices such as Microsoft HoloLens.
 This delay significantly hampers the real-time user experience that is critical for AR applications.

To optimize 3D model output sizes and enhance computational efficiency, we tested various target numbers of vertices—500, 800, 1000, 1500, and 2000—on a set of 20 objects.
The optimal trade-off between maintaining visual quality and reducing file size was consistently achieved at around 1,000 vertices, which typically resulted in approximately 2,000 faces.
By using this target, we were able to significantly reduce the size of the generated 3D models while maintaining a high level of visual fidelity.

The quadric edge collapse algorithm was employed to iteratively reduce the complexity of the mesh.
This algorithm works by collapsing edges, simplifying the structure of the mesh, and recalculating the number of vertices and faces after each iteration until the target vertex count is reached.
This process ensures that the reduction in complexity is controlled and incremental, allowing us to retain the essential geometric properties of the objects while reducing their overall size.

In comparison, the default Shape-E model 3D objects' output averages around 13,944 vertices and 27,884 faces.
While these models provide high detail, they are too large and computationally intensive for real-time applications.
Reducing the number of vertices to around 1,000 allowed to maintain sufficient detail for AR environments, while also drastically reducing model size and improving performance.
This method significantly enhances the responsiveness and overall user experience on AR devices such as Microsoft HoloLens.

\subsection{Limitations of Open-Source LLMs and Text-to-3D Models in Multilingual Applications}

One significant challenge in leveraging LLMs for XR systems lies in their support for multiple languages. Predominantly, open-source LLMs are trained on datasets composed mainly of English language text. This training bias restricts their direct applicability in multilingual contexts, where user interactions might occur in a myriad of languages. To address this limitation, we integrate AI-driven translation models that serve as intermediaries. These models translate non-English inputs into English, allowing the power of LLMs to be harnessed across diverse linguistic landscapes. However, this approach introduces an additional layer of complexity and potential for translation inaccuracies, which can impact the overall system performance and user experience.

Additionally, Text-to-3D conversion technologies often face similar challenges. The quality and accuracy of 3D models generated from textual descriptions can vary significantly when the input text is not in English. Most current systems and datasets used to train these models are optimized for English descriptions, which can lead to less accurate or contextually inappropriate 3D models when using non-English inputs. This limitation not only affects usability but also the inclusivity of AR applications in global contexts, necessitating further research and development to enhance the multilingual capabilities of these technologies.

To mitigate these limitations, we have implemented a system where all incoming speech is translated into English before being processed by the LLMs and Text-to-3D models. This translation step ensures that the input to these models remains consistent and optimized for the highest accuracy and contextual appropriateness in generated content. The reliance on translation models and the inherent limitations in training data highlight critical gaps in the capabilities of open-source LLMs and Text-to-3D models—their limited multilingual support directly affects their versatility and effectiveness in global applications.

\subsection{Speech-to-Text Transcription Errors and Their Effects on AR Solutions}

Another prevalent issue in AI-driven systems is the accuracy of STT services. In real-world applications, transcription errors are inevitable due to factors such as background noise, speech nuances, and accents. These errors can significantly degrade the user experience by misinterpreting user commands or queries. To mitigate the impact of these errors in an XR system, we designed a user interface menu that displays detected objects. This menu allows users to visually confirm and select the correct objects before they are rendered in the XR environment. By incorporating this feature, we not only enhance user interaction but also significantly reduce unnecessary GPU usage. Processing only the confirmed objects minimizes the computational load, thereby optimizing the system’s performance and reducing operational costs.

\subsection{Contextual Awareness and Environment Adaptation Challenges in LLMs for AR}
 
One major challenge is the LLM’s limited awareness of the user’s physical environment. While the model excels at understanding textual prompts and generating related suggestions, it lacks the ability to incorporate real-time environmental data into its decision-making process. This results in suggestions that might be appropriate in a general sense but are not tailored to the specific characteristics and constraints of the user’s surroundings.

Moreover, the creation of 3D objects that match the style and color of the environment requires a nuanced understanding of the surrounding space. This includes recognizing the textures, lighting conditions, and existing design elements within the environment. The LLM, primarily trained on text data, does not possess the capability to analyze and interpret these visual elements accurately. 

Another significant challenge is ensuring the consistency of the generated objects with the user’s aesthetic preferences and the overall design language of the space. In AR, it is crucial that the virtual elements blend seamlessly with the real-world environment to create a cohesive and immersive experience. This requires sophisticated algorithms that can adjust the style, color, and material properties of the 3D objects based on real-time visual feedback from the environment.

Integrating environmental context into the generative process is not straightforward. It demands a multi-modal approach that combines visual data from the user’s environment with the LLM’s language understanding capabilities. This integration poses technical challenges, including the need for real-time data processing and the development of models that can simultaneously handle visual and textual inputs.

Additionally, there are challenges related to the scalability of the system. As the complexity and diversity of AR applications increase, the generative models must be able to handle a wide range of scenarios and environments. This requires extensive training data and robust models capable of generalizing across different contexts and user needs. Privacy and security concerns also arise when integrating real-time environmental data into the generative process. Ensuring that the system handles user data responsibly and complies with privacy regulations is essential. This involves implementing secure data handling practices and providing users with transparency and control over their data.

In conclusion, while the use of LLMs in suggesting objects in AR shows promise, it is clear that a more holistic approach is needed to address the challenges of context-aware content creation. By integrating visual data, optimizing real-time performance, and ensuring a seamless user experience, we can enhance the relevance and quality of generative multi-modal AI in dynamic AR environments. For addressing this challenge, we utilized VLM on a local server, which not only enabled the processing of images and provided context-aware suggestions based on the user’s environment but also ensured the security of the data by preventing the storage of user environment images on external servers.

\subsection{Challenges in Image-to-3D Conversion for AR}

One of the primary challenges in converting images to 3D models within AR environments stems from the limitations of current AI models.
These models are typically designed to convert simple images, often with plain backgrounds, into 3D representations.
However, in AR, images are captured using a camera integrated into the headset, and these images often include complex backgrounds and multiple objects.
This complexity makes it difficult for the AI models to accurately identify and convert individual objects into 3D models, as the presence of multiple items and intricate backgrounds confuses the conversion algorithms.

\begin{figure} [ht]
\centering
\includegraphics[width=0.5\linewidth]{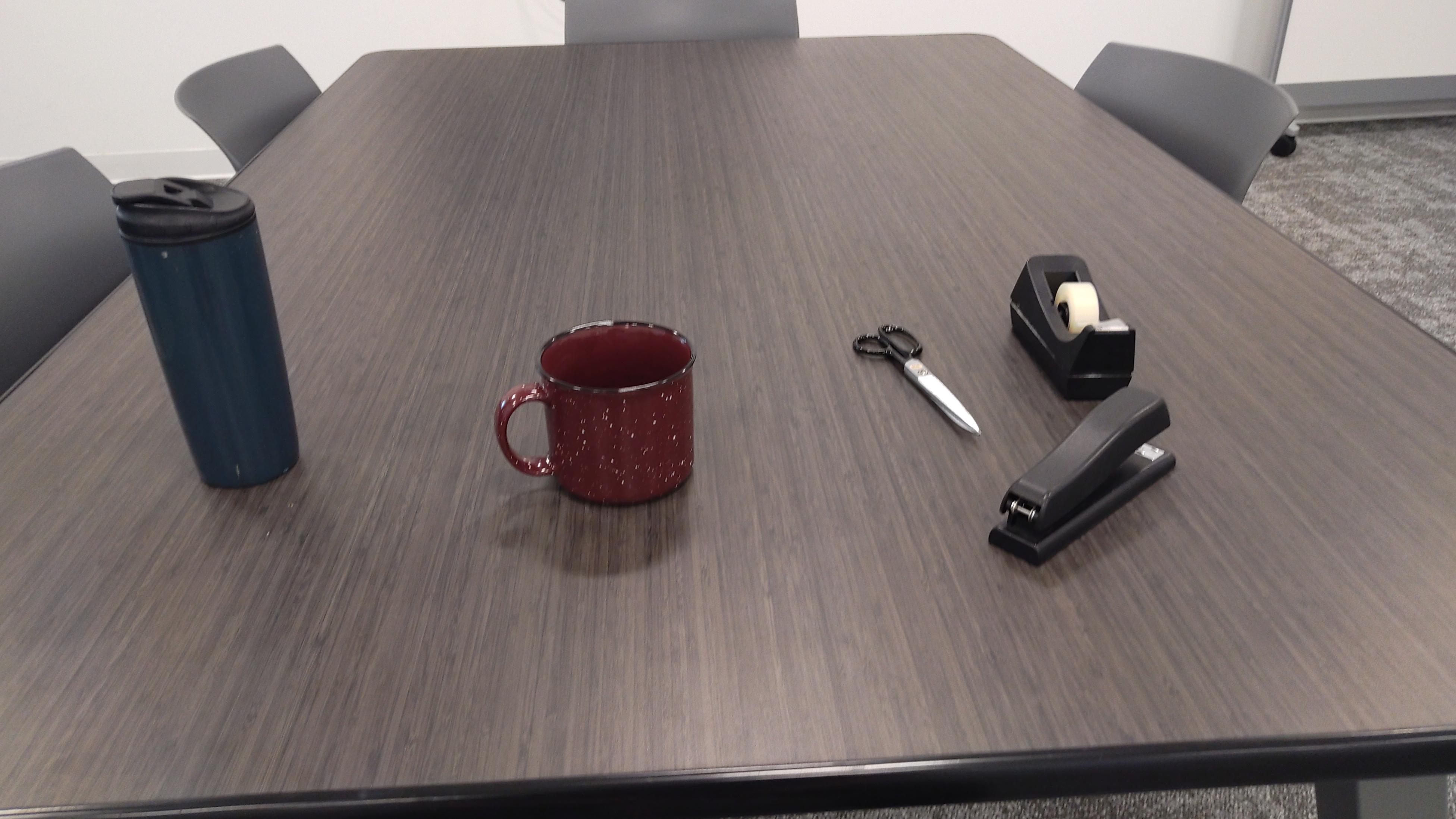} 
\caption{Augmented reality headset view of multiple objects.}  
\label{fig:Multi_ar_headset}
\end{figure}

\begin{figure} [ht]
\centering
\includegraphics[width=0.5\linewidth]{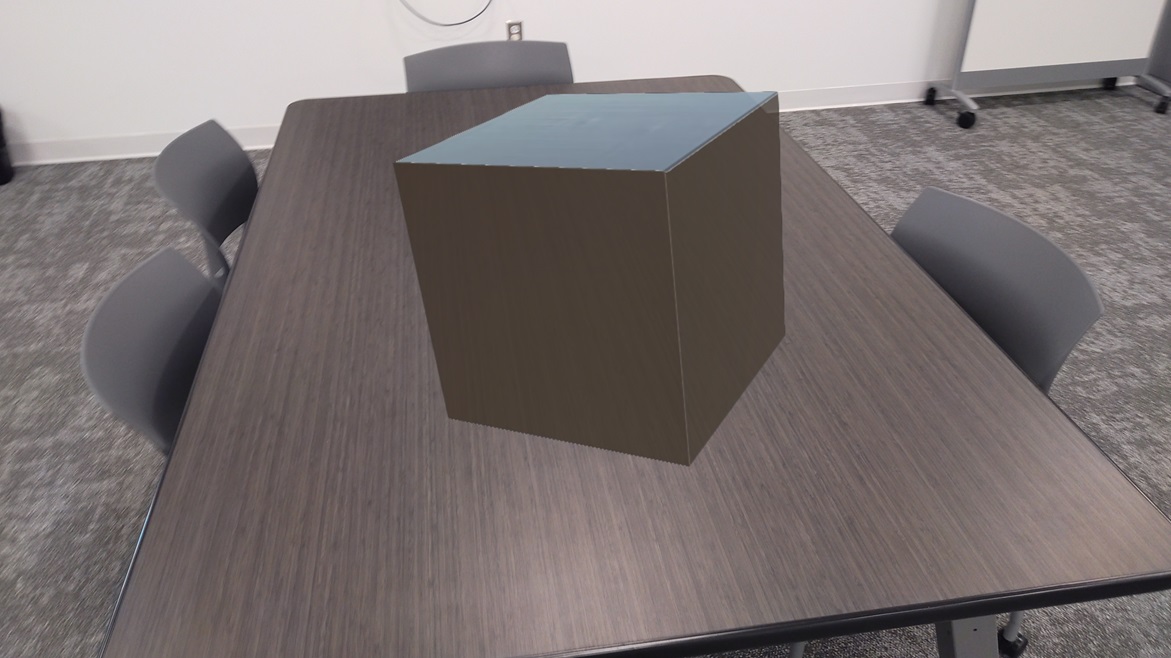} 
\caption{Shap-E 3D output for multiple objects.}  
\label{fig:Shape_Multi }
\end{figure}

As shown in \autoref{fig:Multi_ar_headset}, there are multiple objects present in the environment, and the image is captured from the view of the AR headset.
However, in  \autoref{fig:Shape_Multi }, the output from the Shap-E model for this scene is just a single cube, indicating the model's inability to handle environments with multiple objects effectively.

Another significant challenge arises when trying to convert images of objects displayed on monitors, such as items from online stores, into 3D models.
When the user attempts to capture an image of a product on their screen, the resulting picture inevitably includes the monitor and surrounding elements.
Current image-to-3D conversion models are not equipped to isolate the object of interest from its background effectively.
Consequently, the output often includes the monitor itself as part of the 3D model, which is not the intended outcome.

\begin{figure} [ht]
\centering
\includegraphics[width=0.5\linewidth]{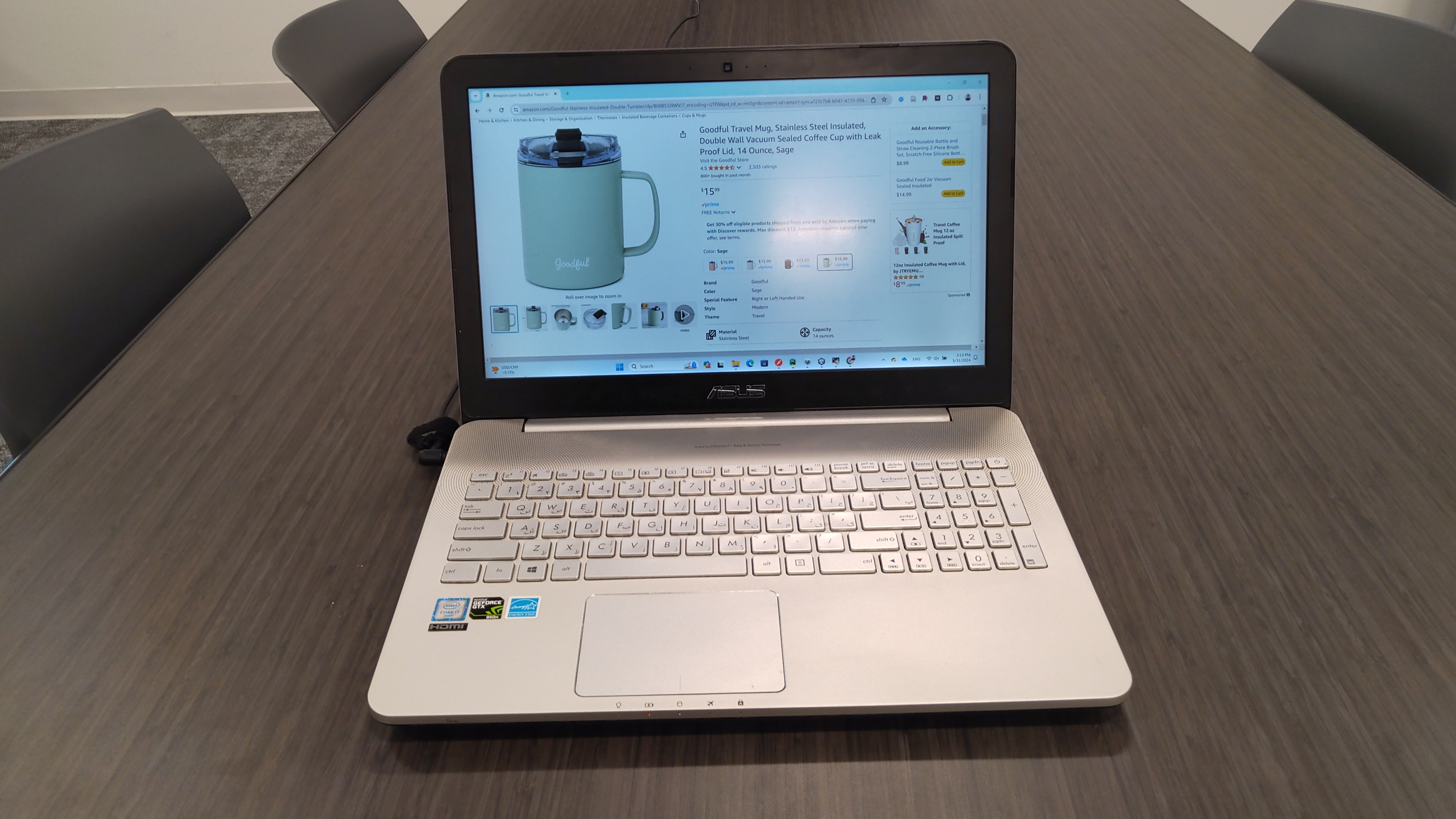} 
\caption{Augmented reality headset view of a mug in an online store.}  
\label{fig:mug_ar_headset}
\end{figure}

\begin{figure} [ht]
\centering
\includegraphics[width=0.5\linewidth]{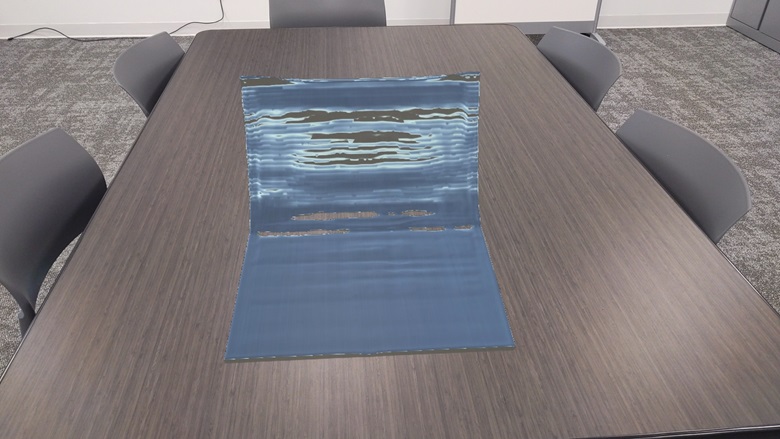} 
\caption{Shap-E 3D output of a mug displayed on a monitor.}  
\label{fig:Shape_monitor}
\end{figure}

As shown in \autoref{fig:mug_ar_headset}, a mug is displayed in an online store view captured by the AR headset.
\autoref{fig:Shape_monitor} shows the output from OpenAI's Shap-E model based on this image of the mug.
Interestingly, if the user intended to create the mug seen on the monitor, the model's output is actually a laptop.

In addition to these challenges, there is also the issue of varying lighting conditions and angles at which images are captured.
In real-world settings, lighting can significantly affect the appearance of objects, creating shadows, reflections, and other visual artifacts that can mislead the AI models.
These models often struggle to differentiate between actual object features and these artifacts, leading to inaccuracies in the generated 3D models.
Furthermore, the angle at which an image is taken can distort the perceived shape of the object, complicating the conversion process and resulting in 3D models that do not accurately reflect the original items.
 
Lastly, the integration of these 3D models into the AR environment itself poses a challenge.
Even if the conversion process is successful, the models must be seamlessly integrated into the user's real-time view.
This integration requires precise alignment and scaling to ensure that the virtual objects appear as part of the real world.
Any errors in this process can disrupt the user's experience and reduce the overall effectiveness of the AR application.
Therefore, advancements in both image-to-3D conversion and AR integration are crucial to overcoming these challenges and achieving seamless real-time content creation in AR environments.

\begin{figure*} [ht]
  \centering
  \begin{minipage}{0.24\textwidth}
    \includegraphics[width=\linewidth]{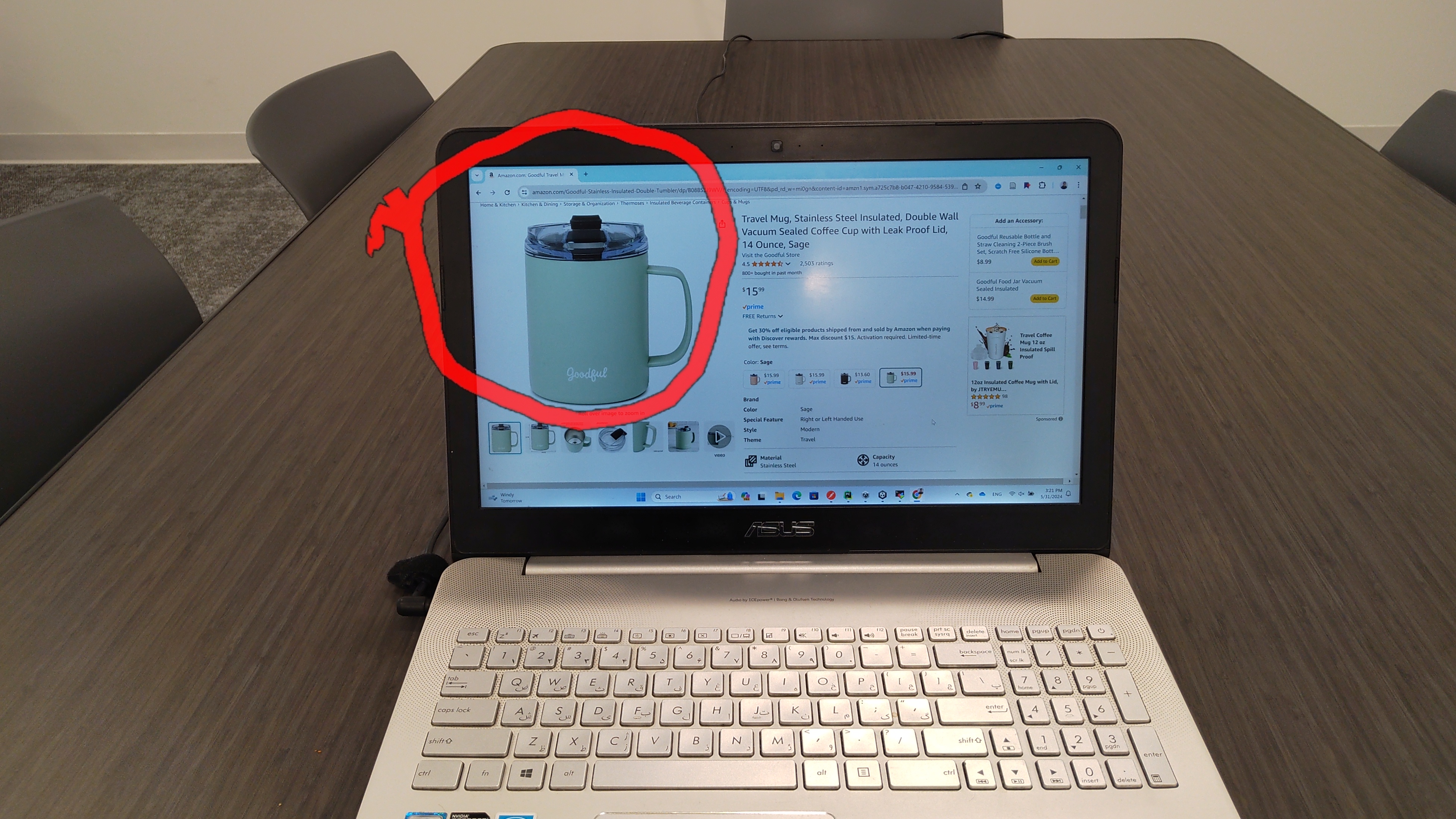}
    % \caption{Image 1} 
    \label{fig:image1}
  \end{minipage}
  % \hfill
  \begin{minipage}{0.24\textwidth}
    \includegraphics[width=\linewidth]{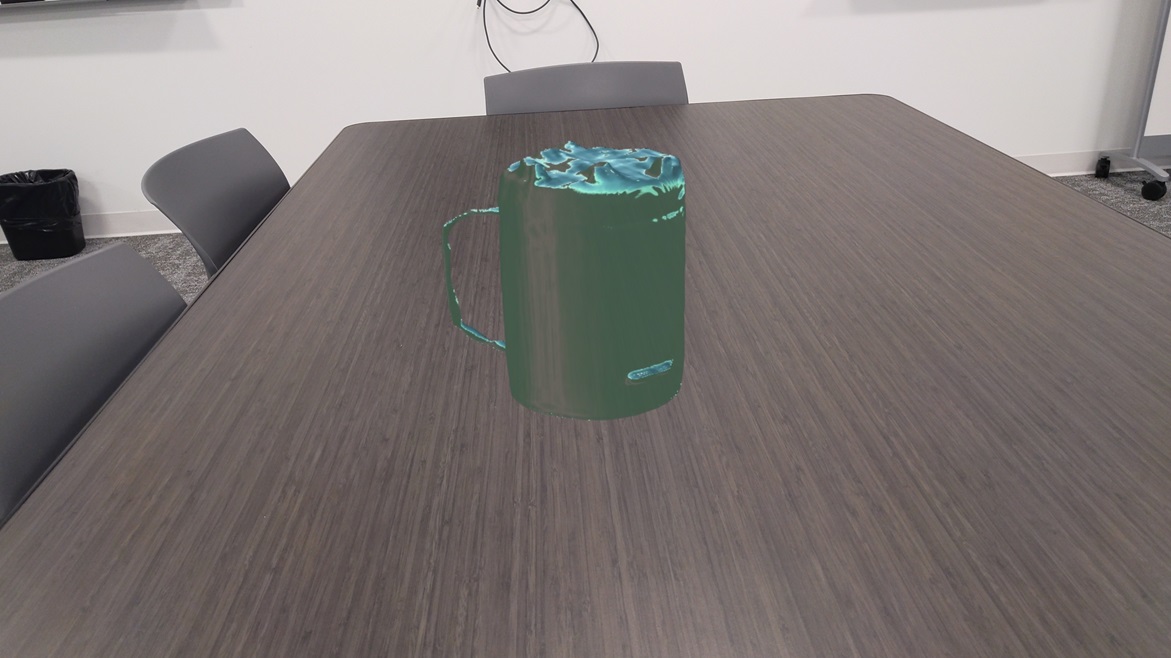}
    % \caption{Image 2} 
    \label{fig:image2}
  \end{minipage}
  % \hfill
  \begin{minipage}{0.24\textwidth}
    \includegraphics[width=\linewidth]{Images//1Obj/1.jpg}
    % \caption{Image 3} 
    \label{fig:image3}
  \end{minipage}
  % \hfill
  \begin{minipage}{0.24\textwidth}
    \includegraphics[width=\linewidth]{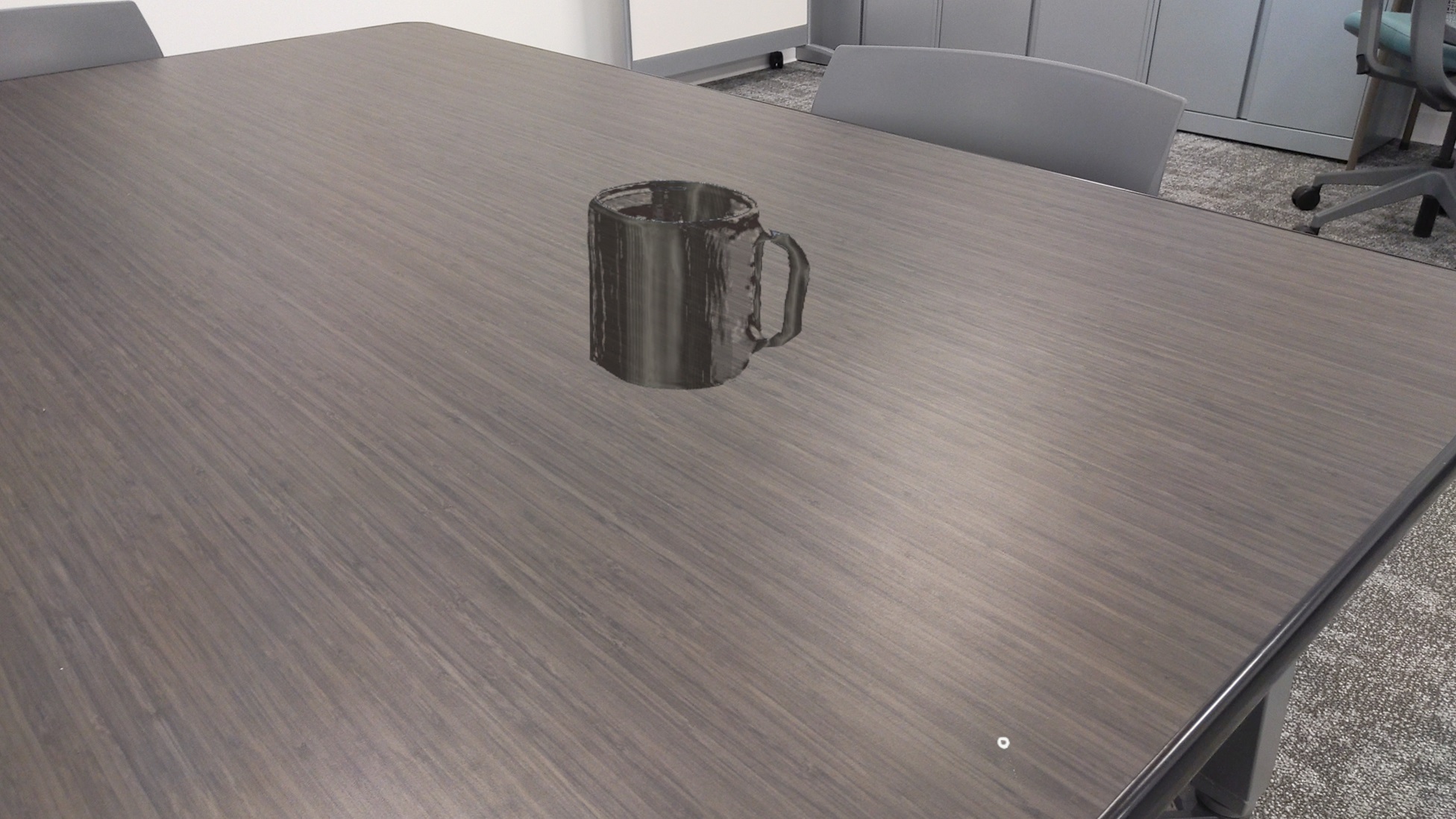}
    % \caption{Image 4} 
    \label{fig:image4}
  \end{minipage}
  \caption{From a 2D image to an AR environment.
\textbf{Left half:} From a 2D image on a computer screen to a 3D object in an AR environment.
\textbf{Right half:} From a 2D image of a real world object to a 3D object in an AR environment.
}
\label{fig:ar_environment}
\end{figure*}

To address this issue, we improved the process by including steps for identifying objects for image processing and utilizing lasso tools, as shown in \ref{fig:ar_environment}. This enhancement helps isolate and accurately convert the intended objects, reducing errors and improving the quality of 3D outputs.

These challenges underscore the necessity for continued innovation and refinement in AI and AR technologies. By addressing these obstacles, we can enhance the functionality and accessibility of AR systems, paving the way for more immersive and practical applications.
   
% =====================Chapter 5 
   
    \chapter{Conclusion and Future work} \label{ch:Conclusion and Future work}
This thesis presents significant advancements in the integration of generative AI into AR systems, showcasing the potential for enhanced user experiences and technical efficiency. Through the development of the \textit{Matrix} framework, we have demonstrated substantial improvements in areas such as real-time 3D object generation, system responsiveness, and resource management. By optimizing 3D object generation, reducing file sizes, and ensuring smooth system performance on resource-constrained devices, \textit{Matrix} addresses critical challenges in AR environments, making interactive experiences more accessible and engaging.

The incorporation of multilingual speech recognition, text-to-3D conversion, and image-to-3D capabilities has expanded the system's global reach, allowing for a broader user base and more versatile content generation. Integrating image-to-3D conversion enables users to create 3D models from photographs and other visual sources, bridging the gap between various media types and enhancing the process of digital content creation. Additionally, \textit{Matrix}'s modular design enables the flexible substitution of AI models, ensuring adaptability for various applications.

The integration of context-aware object recommendations, optimal placement guidance, and aesthetic matching contributes to a highly immersive AR experience. The system's use of open-source AI models on local infrastructures enhances data security and reduces operational costs, further supporting accessible, advanced AR experiences.

Our research emphasizes practical applications in fields such as gaming, education, and retail, where real-time, detailed 3D representations and interactive customization can greatly enhance user engagement and interactivity. The successful implementation of the Shap-E model for precise and customizable 3D object creation exemplifies how generative AI can transform AR, bridging the gap between physical and digital environments.

Looking ahead, there are numerous directions for future work to expand the capabilities of the \textit{Matrix} framework and similar systems:

\begin{description} 

\item[Development of an Automatic Object Placement System:] Future work could include automatic placement of suggested 3D objects within AR environments based on VLM recommendations. This would streamline the customization process and optimize object positioning for improved functionality and visual coherence.

\item[Multimodal Support and Voice Interaction:] Enhancing user interaction by integrating multimodal inputs such as gestures, eye movements, and real-time voice chatting with VLMs will make the system more intuitive and user-friendly. This feature allows dynamic prompting and natural, conversational interactions, providing users with more seamless and engaging experiences.

\item[Adaptive Learning and Personalization:] Implementing reinforcement learning and user feedback mechanisms can enable the system to adapt based on user preferences and behavior. This continuous learning process would result in increasingly personalized and relevant content, improving overall user satisfaction and system performance.

\item[Collaborative and Interactive AR Environments:] Expanding \textit{Matrix} to support multi-user AR experiences could foster collaboration in educational, professional, and recreational contexts. Real-time synchronization and networking capabilities would be essential for these applications, enabling users to interact within shared AR spaces effectively.

\item[User Feedback and Iterative Improvements:] Incorporating user feedback mechanisms will provide valuable insights into system performance and areas needing refinement. This feedback will inform the iterative development of the system, ensuring it meets user expectations and adapts to their evolving needs. \end{description}

\subsection*{Integration of Interactive Optimization Techniques}

Future work could include the application of interactive optimization methodologies, such as the Interactive Design of Experiments (IDoE) approach developed by Splechtna et al.~\cite{DeepAI2024}, to refine 3D object generation in AR environments. This could enable iterative adjustments of generated 3D models based on user-defined goals or real-time environmental feedback, enhancing both the accuracy and contextual relevance of the models.

These future developments aim to push the boundaries of AR technology, deepening the integration of AI-driven solutions and setting new standards for customization, interactivity, and personalization in digital environments. By continuing to innovate and expand upon these findings, we strive to shape the next generation of immersive and adaptive AR experiences.
	\bibliography{VTthesis}
	\bibliographystyle{plainnat}

	% \appendix

	% % Add your appendices here. You must leave the appendices enclosed in the appendices environment in order for the table of contents to be correct.
	% \begin{appendices}
	% 	\chapter{First Appendix} \label{app:appendix_one}
	% 		\section{Section one} \label{ase:app_one_sect_1}
			 
	% 		\section{Section two} \label{ase:app_one_sect_2}
				 
	% 	\chapter{Second Appendix} \label{app:appendix_two}
		 
	% \end{appendices}

\end{document}